\documentclass[]{aastex631}

\usepackage{amsmath}	
\usepackage{amssymb}	
\usepackage{upgreek}

\newcommand{\swift}{{\small \it Swift}}

\newcommand{\uvot}{{\small {\it Swift}/UVOT}}
\usepackage{comment}
\usepackage{tikz}
\usetikzlibrary{matrix}

\shorttitle{{\it WISE} view of CLAGNs}
\shortauthors{Lyu et al.}
\graphicspath{{./}{figures/}}
\graphicspath{{./}{pic/}}

\begin{document}

\title{{\it WISE} view of changing-look AGNs: evidence for a transitional stage of AGNs}

\correspondingauthor{Qingwen Wu}
\email{qwwu@hust.edu.cn}

\author[0000-0001-8879-368X]{Bing Lyu}
\affiliation{School of Physics, Huazhong University of Science and Technology,
1037 Luoyu Road, 
Wuhan, 430074, China \\}
\affiliation{Shanghai Astronomical Observatory, Chinese Academy of Sciences, 80 Nandan Road,
Shanghai, 200030, China}

\author[0000-0003-4773-4987]{Qingwen Wu}
\affiliation{School of Physics, Huazhong University of Science and Technology,
1037 Luoyu Road,
Wuhan, 430074, China \\}

\author[0000-0002-5385-9586]{Zhen Yan}
\affiliation{Shanghai Astronomical Observatory, Chinese Academy of Sciences, 80 Nandan Road,
Shanghai, 200030, China}

\author[0000-0002-3844-9677]{Wenfei Yu}
\affiliation{Shanghai Astronomical Observatory, Chinese Academy of Sciences, 80 Nandan Road,
Shanghai, 200030, China}

\author{Hao Liu}
\affiliation{University of Science and Technology of China,
No.96, JinZhai Road Baohe District, Hefei, Anhui, 230026, China \\}



\begin{abstract}
The discovery of changing-look active galactic nuclei (CLAGNs) with the significant change of optical broad emission lines (optical CLAGNs) and/or strong variation of line-of-sight column densities (X-ray CLAGNs) challenges the orientation-based AGN unification model. We explore mid-infrared (mid-IR) properties for a sample of 57 optical CLAGNs and 11 X-ray CLAGNs based on the {\it Wide-field Infrared Survey Explorer} ({\it WISE}) archive data. We find that Eddington-scaled mid-IR luminosities of both optical and X-ray CLAGNs stay just between low-luminosity AGNs (LLAGNs) and luminous QSOs. The average Eddington-scaled mid-IR luminosities for optical and X-ray CLAGNs are $\sim 0.4$\% and $\sim 0.5$\%, respectively, which roughly correspond the  bolometric luminosity of transition between a radiatively inefficient accretion flow (RIAF) and Shakura-Sunyaev disk (SSD). We estimate the time lags of the variation in the mid-IR behind that in the optical band for 13 CLAGNs with strong mid-IR variability, where the tight correlation between the time lag and the bolometric luminosity ($\tau - L$) for CLAGNs roughly follows that found in the luminous QSOs.

\end{abstract}




\section{Introduction} \label{sec:intro}

Active galactic nuclei (AGNs) are a special class of galaxies characterized by strong variability and high luminosity of non-stellar origin. The accretion onto the central supermassive black hole (SMBH) is widely accepted as the energy source for the AGN activity. The strong emission lines are another typical feature distinguished from the normal galaxies, where the sources with broad lines (1000-20000 $ \rm{km}\, \rm{s}^{-1}$) are called type 1 AGNs. Based on the relative intensity of the broad and narrow components of the Balmer lines, the type 1 AGNs are further classified into several subclasses \citep[e.g., type 1.5, 1.8, and 1.9, see ][]{1976MNRAS.176P..61O,1981ApJ...249..462O}. The sources observed with only narrow lines (e.g., $<$1000 $ \rm{km}\, \rm{s}^{-1}$) are called type 2 AGNs. The broad lines are observed in some type 2 AGNs in polarized light, which suggest that these sources have hidden broad lines \citep[e.g.,][]{1997Natur.385..700H}. In some low-luminosity LINERs, only low ionization emission lines are detected, where these lines are still ionized by the central point-like weak sources \citep[e.g.,][]{2008ARA&A..46..475H}. About 10 percent of AGNs show strong relativistic jets, where the jets start from the vicinity of BHs and can extend far beyond the host galaxy \citep[e.g., Mpc scale,][]{1989AJ.....98.1195K}.

In the last three decades, it was found that the large diversity properties of observed AGN can be explained by a few parameters (e.g., inclination, jet, accretion rate), which is called AGN unification \citep[e.g.,][and references therein]{1993ARA&A..31..473A,2015ARA&A..53..365N}. The first parameter is inclination angle of putative dusty torus. The suppressed multi-waveband continuum and absence of broad emission lines in type 2 AGNs are caused by obscuration of the dust in the torus. The high column density as constrained from X-ray observations and the hidden broad lines in the optical polarization measurements for some type 2 AGNs support this inclination-dependent unification scheme. The second parameter is collimated relativistic radio jets, where these jets are sometimes observed in optical or even in $\gamma$-ray wavebands. Even though the relativistic jets have been observed for several tens of years, the physical reason behind the radio-loud/quiet dichotomy is still an open issue. The third parameter is accretion rate, where different types of accretion modes may exist in the different types of AGNs. For bright AGNs (e.g., QSOs/Seyferts), the big blue bump in optical/UV bands can be well explained by an optically thick and geometrically thin accretion disk \citep[SSD;][]{1973A&A....24..337S}. The SSD may transit to a geometrically thick, advection-dominated accretion flow when the accretion rate is lower than a critical value \citep[ADAF; e.g.,][for a recent review and references therein]{2014ARA&A..52..529Y}, where most of the accretion energy is advected into SMBH rather than radiated away. The ADAF is much hotter than the SSD, which leads to high-energy emission and can explain many typical features in low-luminosity AGNs \citep[LLAGNs,][]{2008ARA&A..46..475H}. Some LLAGNs (e.g., LINERs, FR I radio galaxies, and BL Lacs) lack the broad emission lines and they do not show evident obscuration, which may be caused by absent clouds in the broad-line region (BLR) or central accretion disk does not provide enough ionization photons.

In recent years, it is found that some AGNs, the so-called ``changing-look" AGNs (CLAGNs), show type transitions within a couple of years or even several months. The term ``changing-look" is firstly used to describe the X-ray CLAGNs, which transit from Compton thick (i.e., hydrogen column density, $N_\mathrm{H}> 10^{24}\,\mathrm{cm}^{-2}$) to Compton thin \citep[i.e., $N_\mathrm{H} < 10^{22-23}\,\mathrm{cm}^{-2}$, e.g.,][]{2003MNRAS.342..422M} and vice-visa. This definition has been extended to optical CLAGNs, where broad lines appear/disappear within several years \citep[i.e., transit from type 1 to type 2 and vice-visa, e.g.,][]{2014ApJ...796..134D,2014ApJ...788...48S,2020ApJ...890L..29A,2020ApJ...901....1W}. There has been some systematic search for CLAGNs using multi-epoch optical spectra \citep[e.g.,][]{2018ApJ...862..109Y,2021MNRAS.503.2583S,2021A&A...650A..33P} and the number of CLAGNs is growing. Based on the orientation-unification AGN model, the broad lines and the torus column density will not change within such a short timescale of years or decades.

The physical mechanism of CLAGNs is not fully understood. One scenario is that disappearance/appearance of the broad lines and/or variations of $N_\mathrm{H}$ are caused by the obscuring material moving in or out from our line of sight \citep[e.g.,][]{2013MNRAS.436.1615M,2014MNRAS.443.2862A,2015ApJ...815...55R,2018MNRAS.481.2470T,2019ApJ...887...15W}. However, this scenario is challenged by the low $N_\mathrm{H}$ during type transition \citep[e.g.,][]{2016A&A...593L...9H} and roughly unchanged polarization measurements \citep[e.g.,][]{2019sf2a.conf..509M} in many CLAGNs. The ``changing look'' of second scenario is caused by the change of accretion rate, which will lead to the change of ionization luminosity and/or appearance/disappearance of the clouds in BLR. The study of CLAGNs has been conducted in radio, X-ray, optical, and IR band \citep[e.g.,][]{2017ApJ...846L...7S,2018ApJ...864...27S,2019ApJ...885...44D,2021MNRAS.503.3886Y,2022arXiv220105152F} and the multi-wavelength variability in CLAGNs does support this scenario \citep[e.g.,][]{2016MNRAS.460..304K,2017ApJ...846L...7S,2018ApJ...864...27S,2018MNRAS.480.3898N,2021MNRAS.502L..61Y}. The strong intrinsic continuum variation is also found in some X-ray CLAGNs with strong $N_\mathrm{H}$ variation, which suggests that the variation of $N_\mathrm{H}$ may be also driven by the change of accretion disk \citep[e.g., disk winds;][]{2021RAA....21..199L}.

The mid-IR emission is less influenced by the dust extinction, and has the potential to bring us the clues about the physical mechanism of CLAGNs (e.g., the variable obscurer or the variation of accretion state). In this work, we analyze the mid-IR variability, color, and Eddington ratio for a sample of both X-ray CLAGNs and optical CLAGNs based on the {\it Wide-field Infrared Survey
Explorer} ({\it WISE}) data. We estimate the mid-IR time lags ($\tau$) for 13 bright CLAGNs using multi-epoch optical and mid-IR observations. We use these observations to understand the basic properties of the central engine and possible differences/similarities between the two types of CLAGNs.  We describe the {\it WISE} data and the CLAGN sample in \autoref{sec:sample}. The results of mid-IR variability, color, and luminosity for CLAGNs are shown in \autoref{sec:mir_var_col_lum}. We present the mid-IR dust reverberation mapping analysis and the $\tau$-$L_\mathrm{bol}$ correlation of CLAGNs in \autoref{sec:tau-L}. Conclusion and discussion are presented in \autoref{sec:dis}. Throughout this work, we adopt a flat $\Lambda-$CDM cosmological model with $H_0$=70 km s$^{-1}$ Mpc $^{-1}$, $\Omega_{m}$=0.27, and $\Omega_{\Lambda}=0.73 $.

\section{{\it WISE} Data and CLAGN Sample} \label{sec:sample}
The \textit{WISE} imaged the full sky approximately every six months in four mid-IR bands, centered at 3.4, 4.6, 12, and 22 $\mu$m \citep[referred to as $W1$, $W2$, $W3$, and $W4$, respectively,][]{2010AJ....140.1868W}, which provides an ideal opportunity to explore the infrared properties of CLAGNs. The {\it WISE} surveyed the full sky 1.2 times in the above four bands from 2010 January to September and cryogen for cooling the $W3$ and $W4$ instruments were exhausted. Then it was placed in hibernation on 2011 February 1. On 2013 October 3, it was reactivated (named as \textit{NEOWISE}) with only $W1$ and $W2$ \citep{2014ApJ...792...30M}. We use the \textit{AllWISE} multi-epoch photometry table and \textit{NEOWISE} single exposure (L1b) source table data retrieved from the NASA/IPAC Infrared Science Archive \footnote{\url{https://irsa.ipac.caltech.edu/Missions/wise.html}}. The \textit{AllWISE} and \textit{NEOWISE} data are screened to exclude the possible bad photometric measurements according to the following criteria:\\
(1) Detection from good-quality frame sets\footnote{\url{http://wise2.ipac.caltech.edu/docs/release/neowise/expsup/sec2_3.html}}, with {frame quality score \texttt{qual\_frame}}$>$0, frame image quality score {\texttt{qi\_fact}}$>$0,
South Atlantic Anomaly separation {\texttt{saa\_sep}}$>0$, and Moon masking
flag {\texttt{moon\_masked}}=0.\\ 
(2) $W1$ $<$15 mag and $W2$ $<$13 mag, which approximately correspond to signal-noise-ratio SNR=10.\\
(3) The number of point-spread-function (PSF) components used in profile fitting (\texttt{nb}$<$3), frames are unaffected (\texttt{cc\textunderscore flags}=`0000') and are not actively de-blended \citep[na$=$0; see also][]{2019MNRAS.483.2362R}.

To explore the mid-IR variability properties of CLAGNs, we collect the reported CLAGNs from the literature. To increase the statistical significance on the variability, the sources with at least 20 valid \textit{WISE} data points are considered. Our sample includes 68 sources, which include 57 optical CLAGNs with the disappearance/appearance of broad lines and 11 X-ray CLAGNs with the variation of $N_\mathrm{H}$. It should be noted that 5 CLAGNs show variations of both broad emission lines and X-ray absorption column density (e.g., ESO 362-G18, NGC 2992, NGC 4151, NGC 4395, and NGC 7582). We adopt the BH mass measurements of these CLAGNs from the literature, which are estimated based on kinematics method \citep[e.g.,][]{2003MNRAS.345.1057M}, reverberation mapping (RM) \citep[e.g.,][]{2011MNRAS.410.1877S,2017ApJ...840...97F}, bulge luminosity \citep[$L_\mathrm{bulge}$, e.g.,][]{2006AJ....131.1236D} and velocity dispersion $\sigma_{*}$ of the host galaxy \citep{2002ApJ...574..740T}. The information of the CLAGN sample is listed in \autoref{source}, where the source name, redshift, the type of CLAGNs, BH mass, infrared magnitude, variability, and color are presented.

\startlongtable
\begin{deluxetable}{lcc ccc ccc cccc }
\tablecaption{The sample of CLAGNs. \label{source}}
\tablewidth{0pt}
\tabletypesize{\scriptsize}
\tablehead{
\colhead{Name} &\colhead{z} & \colhead{Type} & \colhead{Ref} & \colhead{$M_{\rm BH}$} & \colhead{Refs} & \colhead{$\sigma_{m W1}$} & \colhead{$<W1>$} & \colhead{$\sigma_{m W2}$} & \colhead{$<W2>$}  & \colhead{$<W1$-$W2>$} & \colhead{$\Delta\,W1$} & \colhead{$\Delta\,W2$} \\
} 

\decimalcolnumbers
\startdata
1H 0419-577 & 0.1040 & X & 1 & 8.6 & 31 & 0.02 & 10.88$\pm$ 0.03 & 0.00 & 9.86$\pm$ 0.03 & 1.02$\pm$ 0.03 & 0.10 & 0.11 \\
2MASS J16171142+0638333 & 0.2291 & O & 2 & 8.0 & 29 & 0.13 & 13.86$\pm$ 0.14 & 0.05 & 12.85$\pm$ 0.09 & 1.02$\pm$ 0.11 & 0.86 & 0.78 \\
2MASS J22053771-0711147 & 0.2950 & O & 2 & 8.0 & 29 & 0.15 & 13.64$\pm$ 0.16 & 0.13 & 12.77$\pm$ 0.15 & 0.87$\pm$ 0.09 & 0.47 & 0.41 \\
2MASX J09381221+0743398 & 0.0220 & O & 3 & 7.5 & 29 & 0.12 & 11.73$\pm$ 0.12 & 0.22 & 11.58$\pm$ 0.22 & 0.11$\pm$ 0.23 & 0.31 & 0.58 \\
2MASX J09483841+4030436 & 0.0468 & O & 3 & 7.5 & 29 & 0.04 & 11.89$\pm$ 0.05 & 0.07 & 11.50$\pm$ 0.08 & 0.40$\pm$ 0.07 & 0.18 & 0.28 \\
3C 390.3 & 0.0561 & O & 4 & 9.3 & 32 & 0.18 & 9.90$\pm$ 0.18 & 0.16 & 8.85$\pm$ 0.16 & 1.05$\pm$ 0.04 & 0.56 & 0.49 \\
ESO 362-G18 & 0.0124 & O & 5 & 7.7 & 33 & 0.09 & 9.88$\pm$ 0.09 & 0.09 & 9.27$\pm$ 0.10 & 0.60$\pm$ 0.05 & 0.29 & 0.33 \\
Fairall 9 & 0.0461 & O & 4 & 8.4 & 34 & 0.06 & 8.99$\pm$ 0.07 & 0.03 & 7.99$\pm$ 0.04 & 1.00$\pm$ 0.04 & 0.25 & 0.16 \\
HE 1136-2304 & 0.0270 & O & 6 & 7.6 & 6 & 0.36 & 10.84$\pm$ 0.36 & 0.21 & 10.02$\pm$ 0.21 & 0.84$\pm$ 0.39 & 0.90 & 0.96 \\
IC 751 & 0.0315 & X & 7 & 8.5 & 7 & 0.03 & 10.76$\pm$ 0.04 & 0.00 & 9.95$\pm$ 0.03 & 0.82$\pm$ 0.04 & 0.06 & 0.10 \\
IRAS 23226-3843 & 0.0359 & O & 8 & 8.2 & 8 & 0.21 & 11.13$\pm$ 0.22 & 0.14 & 10.88$\pm$ 0.14 & 0.25$\pm$ 0.20 & 0.30 & 0.54 \\
Mrk 1018 & 0.0430 & O & 9 & 7.8 & 35 & 0.14 & 10.87$\pm$ 0.15 & 0.16 & 10.39$\pm$ 0.16 & 0.46$\pm$ 0.08 & 0.88 & 1.27 \\
Mrk 530 & 0.0288 & O & 4 & 8.1 & 36 & 0.10 & 8.50$\pm$ 0.11 & 0.10 & 7.58$\pm$ 0.10 & 0.92$\pm$ 0.03 & 0.48 & 0.45 \\
Mrk 590 & 0.0264 & O & 10 & 7.5 & 37 & 0.14 & 10.22$\pm$ 0.14 & 0.22 & 9.79$\pm$ 0.22 & 0.41$\pm$ 0.10 & 0.42 & 0.69 \\
Mrk 6 & 0.0195 & O & 4 & 8.2 & 38 & 0.57 & 8.56$\pm$ 0.57 & 0.49 & 7.70$\pm$ 0.49 & 0.84$\pm$ 0.10 & 1.74 & 1.52 \\
Mrk 609 & 0.0344 & O & 11 & 7.8 & 39 & 0.09 & 10.43$\pm$ 0.10 & 0.13 & 9.94$\pm$ 0.13 & 0.47$\pm$ 0.06 & 0.48 & 0.64 \\
Mrk 926 & 0.0470 & O & 12 & 8.1 & 40 & 0.20 & 9.56$\pm$ 0.20 & 0.16 & 8.66$\pm$ 0.16 & 0.89$\pm$ 0.06 & 0.57 & 0.46 \\
NGC 1097 & 0.0042 & O & 13 & 8.1 & 41 & 0.09 & 8.22$\pm$ 0.09 & 0.09 & 8.00$\pm$ 0.09 & 0.21$\pm$ 0.13 & 0.14 & 0.22 \\
NGC 1365 & 0.0055 & X & 14 & 6.7 & 42 & 0.09 & 7.72$\pm$ 0.09 & 0.07 & 7.03$\pm$ 0.08 & 0.69$\pm$ 0.07 & 0.25 & 0.26 \\
NGC 1566 & 0.0050 & O & 15 & 6.9 & 43 & 0.28 & 8.82$\pm$ 0.28 & 0.42 & 8.53$\pm$ 0.42 & 0.28$\pm$ 0.17 & 1.09 & 1.57 \\
NGC 2617 & 0.0142 & O & 16 & 7.5 & 44 & 0.15 & 10.12$\pm$ 0.15 & 0.18 & 9.55$\pm$ 0.18 & 0.57$\pm$ 0.06 & 0.79 & 1.03 \\
NGC 2992 & 0.0077 & O & 14 & 7.5 & 45 & 0.19 & 8.57$\pm$ 0.19 & 0.23 & 7.94$\pm$ 0.24 & 0.63$\pm$ 0.09 & 0.73 & 1.04 \\
NGC 3065 & 0.0066 & O & 13 & 8.0 & 46 & 0.06 & 9.75$\pm$ 0.06 & 0.04 & 9.73$\pm$ 0.05 & 0.02$\pm$ 0.07 & 0.07 & 0.12 \\
NGC 3516 & 0.0088 & O & 17 & 7.5 & 37 & 0.16 & 8.83$\pm$ 0.16 & 0.17 & 8.19$\pm$ 0.17 & 0.63$\pm$ 0.06 & 0.63 & 0.60 \\
NGC 4051 & 0.0023 & X & 1 & 6.4 & 47 & 0.12 & 8.90$\pm$ 0.12 & 0.12 & 8.10$\pm$ 0.12 & 0.80$\pm$ 0.04 & 0.41 & 0.43 \\
NGC 4151 & 0.0033 & O & 4 & 7.6 & 37 & 0.29 & 7.32$\pm$ 0.29 & 0.25 & 6.24$\pm$ 0.25 & 1.06$\pm$ 0.13 & 1.13 & 0.79 \\
NGC 4388 & 0.0084 & X & 1 & 6.9 & 48 & 0.16 & 9.29$\pm$ 0.16 & 0.23 & 8.39$\pm$ 0.24 & 0.89$\pm$ 0.10 & 0.47 & 0.68 \\
NGC 4395 & 0.0011 & O & 18 & 5.6 & 49 & 0.20 & 12.44$\pm$ 0.20 & 0.18 & 11.66$\pm$ 0.18 & 0.78$\pm$ 0.06 & 0.66 & 0.56 \\
NGC 4507 & 0.0118 & X & 18 & 7.7 & 50 & 0.09 & 8.69$\pm$ 0.09 & 0.08 & 7.61$\pm$ 0.08 & 1.07$\pm$ 0.03 & 0.35 & 0.31 \\
NGC 454 & 0.0122 & X & 1 & 6.2 & 51 & 0.05 & 12.63$\pm$ 0.06 & 0.05 & 12.45$\pm$ 0.08 & 0.17$\pm$ 0.09 & 0.06 & 0.07 \\
NGC 4939 & 0.0104 & X & 19 & 7.5 & 52 & 0.20 & 9.85$\pm$ 0.21 & 0.19 & 9.43$\pm$ 0.19 & 0.42$\pm$ 0.25 & 0.36 & 0.47 \\
NGC 5548 & 0.0172 & O & 20 & 7.5 & 53 & 0.13 & 8.95$\pm$ 0.13 & 0.09 & 8.11$\pm$ 0.10 & 0.84$\pm$ 0.05 & 0.47 & 0.34 \\
NGC 6300 & 0.0037 & X & 21 & 7.0 & 21 & 0.06 & 9.14$\pm$ 0.07 & 0.07 & 8.24$\pm$ 0.07 & 0.89$\pm$ 0.06 & 0.25 & 0.46 \\
NGC 7582 & 0.0053 & O & 4 & 7.7 & 54 & 0.16 & 7.81$\pm$ 0.16 & 0.19 & 6.83$\pm$ 0.19 & 0.98$\pm$ 0.07 & 0.55 & 0.58 \\
NGC 7674 & 0.0290 & X & 22 & 7.6 & 55 & 0.03 & 9.28$\pm$ 0.04 & 0.00 & 8.14$\pm$ 0.02 & 1.14$\pm$ 0.03 & 0.13 & 0.08 \\
SDSS J030510.60-010431.6       & 0.0450 & O & 23 & 7.3 & 23 & 0.06 & 12.17$\pm$ 0.07 & 0.08 & 11.95$\pm$ 0.09 & 0.21$\pm$ 0.07 & 0.24 & 0.38 \\
SDSS J080020.98+263648.8       & 0.0267 & O & 23 & 7.1 & 23 & 0.21 & 10.09$\pm$ 0.22 & 0.24 & 9.33$\pm$ 0.24 & 0.75$\pm$ 0.06 & 0.80 & 0.88 \\
SDSS J081319.34+460849.5 & 0.0538 & O & 24 & 7.6 & 29 & 0.15 & 12.71$\pm$ 0.15 & 0.24 & 12.38$\pm$ 0.23 & 0.28$\pm$ 0.11 & 0.45 & 0.67 \\
SDSS J081726.41+101210.1 & 0.0458 & O & 25 & 7.5 & 25 & 0.36 & 12.51$\pm$ 0.35 & 0.50 & 11.91$\pm$ 0.45 & 0.47$\pm$ 0.13 & 0.97 & 1.25 \\
SDSS J082323.89+422048.3 & 0.1515 & O & 26 & 8.1 & 26 & 0.08 & 13.59$\pm$ 0.09 & 0.09 & 12.90$\pm$ 0.12 & 0.69$\pm$ 0.14 & 0.57 & 0.82 \\
SDSS J082942.67+415436.9 & 0.1263 & O & 26 & 8.5 & 26 & 0.22 & 11.91$\pm$ 0.22 & 0.16 & 11.00$\pm$ 0.16 & 0.92$\pm$ 0.08 & 0.73 & 0.58 \\
SDSS J090902.35+133019.4 & 0.0499 & O & 24 & 7.3 & 29 & 0.09 & 13.09$\pm$ 0.10 & 0.37 & 12.60$\pm$ 0.30 & 0.27$\pm$ 0.37 & 0.83 & 1.16 \\
SDSS J091531.04+481407.7 & 0.1005 & O & 25 & 7.8 & 56 & 0.13 & 13.25$\pm$ 0.14 & 0.15 & 12.73$\pm$ 0.17 & 0.50$\pm$ 0.09 & 0.64 & 0.85 \\
SDSS J111536.57+054449.7       & 0.0900 & O & 27 & 7.6 & 56 & 0.20 & 13.32$\pm$ 0.20 & 0.25 & 12.50$\pm$ 0.24 & 0.77$\pm$ 0.13 & 1.06 & 1.64 \\
SDSS J113355.93+670107.0 & 0.0397 & O & 25 & 8.2 & 56 & 0.16 & 11.76$\pm$ 0.16 & 0.20 & 11.27$\pm$ 0.20 & 0.47$\pm$ 0.07 & 0.54 & 0.77 \\
SDSS J120447.91+170256.8        & 0.2979 & O & 28 & 8.0 & 28 & 0.21 & 13.88$\pm$ 0.21 & 0.14 & 12.70$\pm$ 0.16 & 1.21$\pm$ 0.11 & 1.03 & 0.70 \\
SDSS J122550.30+510846.3 & 0.1679 & O & 25 & 8.6 & 56 & 0.08 & 13.01$\pm$ 0.09 & 0.12 & 12.29$\pm$ 0.13 & 0.72$\pm$ 0.08 & 0.65 & 0.73 \\
SDSS J125403.78+491452.8 & 0.0670 & O & 25 & 8.3 & 56 & 0.08 & 12.74$\pm$ 0.09 & 0.12 & 12.43$\pm$ 0.13 & 0.30$\pm$ 0.08 & 0.34 & 0.45 \\
SDSS J131615.95+301552.2       & 0.0492 & O & 23 & 7.1 & 23 & 0.07 & 11.64$\pm$ 0.07 & 0.11 & 11.39$\pm$ 0.12 & 0.25$\pm$ 0.09 & 0.24 & 0.41 \\
SDSS J132457.29+480241.2       & 0.2716 & O & 26 & 8.0 & 29 & 0.09 & 13.64$\pm$ 0.11 & 0.01 & 12.87$\pm$ 0.07 & 0.77$\pm$ 0.11 & 0.30 & 0.23 \\
SDSS J141324.27+530527.0       & 0.4559 & O & 29 & 8.2 & 29 & 0.30 & 13.45$\pm$ 0.29 & 0.38 & 12.44$\pm$ 0.35 & 0.94$\pm$ 0.11 & 1.45 & 1.72 \\
SDSS J153308.02+443208.4 & 0.0367 & O & 25 & 7.6 & 25 & 0.12 & 11.98$\pm$ 0.13 & 0.24 & 11.85$\pm$ 0.24 & 0.08$\pm$ 0.16 & 0.42 & 0.82 \\
SDSS J155440.25+362952.0       & 0.2368 & O & 24 & 8.0 & 24 & 0.07 & 13.73$\pm$ 0.09 & 0.01 & 12.91$\pm$ 0.08 & 0.82$\pm$ 0.11 & 0.84 & 1.24 \\
SDSS J162501.43+241547.3       & 0.0503 & O & 23 & 6.7 & 23 & 0.12 & 12.55$\pm$ 0.13 & 0.17 & 12.19$\pm$ 0.18 & 0.34$\pm$ 0.08 & 0.39 & 0.55 \\
SDSS J163629.66+410222.4       & 0.0474 & O & 23 & 7.0 & 23 & 0.02 & 12.95$\pm$ 0.05 & 0.00 & 12.88$\pm$ 0.07 & 0.07$\pm$ 0.07 & 0.06 & 0.07 \\
SDSS J172322.31+550413.8 & 0.2947 & O & 26 & 8.6 & 26 & 0.11 & 13.45$\pm$ 0.12 & 0.16 & 12.58$\pm$ 0.17 & 0.83$\pm$ 0.21 & 0.30 & 0.28 \\
UGC 3223 & 0.0156 & O & 30 & 8.0 & 30 & 0.08 & 10.82$\pm$ 0.08 & 0.14 & 10.59$\pm$ 0.14 & 0.21$\pm$ 0.08 & 0.25 & 0.46 \\
UGC 4203 & 0.0135 & X & 14 & 6.8 & 42 & 0.03 & 9.96$\pm$ 0.04 & 0.02 & 8.60$\pm$ 0.03 & 1.36$\pm$ 0.03 & 0.14 & 0.09 \\
WISEA J035301.02-062326.2 & 0.0762 & O & 3 & 7.6 & 29 & 0.07 & 13.09$\pm$ 0.08 & 0.08 & 12.54$\pm$ 0.10 & 0.54$\pm$ 0.09 & 0.24 & 0.20 \\
WISEA J084748.28+182440.0 & 0.0848 & O & 3 & 7.7 & 29 & 0.20 & 12.88$\pm$ 0.20 & 0.24 & 12.25$\pm$ 0.24 & 0.58$\pm$ 0.10 & 0.70 & 0.79 \\
WISEA J100323.46+352503.8 & 0.1189 & O & 27 & 8.0 & 56 & 0.30 & 13.27$\pm$ 0.30 & 0.29 & 12.44$\pm$ 0.28 & 0.81$\pm$ 0.11 & 1.14 & 1.33 \\
WISEA J110057.70-005304.4 & 0.3790 & O & 29 & 8.2 & 29 & 0.09 & 13.55$\pm$ 0.10 & 0.06 & 12.59$\pm$ 0.09 & 0.96$\pm$ 0.11 & 0.40 & 0.40 \\
WISEA J113229.14+035729.1 & 0.0909 & O & 27 & 7.9 & 56 & 0.12 & 13.28$\pm$ 0.13 & 0.12 & 12.70$\pm$ 0.14 & 0.58$\pm$ 0.09 & 0.67 & 0.80 \\
WISEA J131930.75+675355.4 & 0.1664 & O & 27 & 7.5 & 56 & 0.08 & 13.61$\pm$ 0.09 & 0.07 & 12.85$\pm$ 0.10 & 0.76$\pm$ 0.08 & 0.25 & 0.33 \\
WISEA J144754.23+283324.1 & 0.1634 & O & 27 & 8.0 & 56 & 0.13 & 12.89$\pm$ 0.14 & 0.15 & 12.17$\pm$ 0.16 & 0.72$\pm$ 0.08 & 0.56 & 0.67 \\
WISEA J154507.52+170950.8 & 0.0483 & O & 29 & 7.4 & 29 & 0.18 & 12.27$\pm$ 0.18 & 0.21 & 11.65$\pm$ 0.21 & 0.60$\pm$ 0.06 & 0.71 & 0.80 \\
WISEA J154529.63+251127.9 & 0.1170 & O & 27 & 8.1 & 56 & 0.14 & 13.18$\pm$ 0.14 & 0.17 & 12.56$\pm$ 0.17 & 0.60$\pm$ 0.10 & 0.52 & 0.58 \\
WISEA J155258.27+273728.5 & 0.0865 & O & 27 & 8.1 & 56 & 0.06 & 13.51$\pm$ 0.08 & 0.00 & 12.92$\pm$ 0.06 & 0.58$\pm$ 0.09 & 0.32 & 0.49 \\
\enddata
\end{deluxetable}
Note. The table lists source name, redshift, CLAGN type (``O'' for optical CLAGN and ``X'' for X-ray CLAGN), BH mass (log$M_{\rm BH}/M_{\odot}$), references for BH mass, the variability ($\sigma_m$) of $W1$ and $W2$, the mean magnitude of $W1$ and $W2$, mean color of $W1$-$W2$, and maximum magnitude variation ($\Delta\,W1$ and $\Delta\,W2$) of \textit{WISE} data.\\

References. (1)~\citet{2012MNRAS.421.1803M} (2)~\citet{2019ApJ...874....8M} (3)~\citet{2016ApJ...821...33R} (4)~\citet{2019sf2a.conf..509M} (5)~\citet{2018Galax...6...52A} (6)~\citet{2018A&A...619A.168K} (7)~\citet{2016ApJ...820....5R} (8)~\citet{2020A&A...638A..91K} (9)~\citet{2020A&A...644L...5H} (10)~\citet{2020MNRAS.491.4615K} (11)~\citet{2020MNRAS.499.1005W} (12)~\citet{2021IAUS..356..122W} (13)~\citet{2021MNRAS.503.2583S} (14)~\citet{2003MNRAS.342..422M} (15)~\citet{2020MNRAS.498..718O} (16)~\citet{2017OAP....30..117O} (17)~\citet{2021ApJ...909...18F} (18)~\citet{2021MNRAS.507..687J} (19)~\citet{2005MNRAS.356..295G} (20)~\citet{2018ATel11915....1O} (21)~\citet{2020MNRAS.499.5396J} (22)~\citet{2005A&A...442..185B} (23)~\citet{2020MNRAS.498.3985Y} (24)~\citet{2017ApJ...846L...7S} (25)~\citet{2019ApJ...883...31F} (26)~\citet{2021A&A...650A..33P} (27)~\citet{2018ApJ...862..109Y} (28)~\citet{2019ApJ...887...15W} (29)~\citet{2020MNRAS.491.4925G} (30)~\citet{2020ApJ...901....1W} (31)~\citet{2014A&A...563A..95D} (32)~\citet{2011MNRAS.410.1877S} (33)~\citet{2014MNRAS.443.2862A} (34)~\citet{2017MNRAS.466.1777P} (35)~\citet{2018MNRAS.480.3898N} (36)~\citet{2018MNRAS.478.4214E} (37)~\citet{2013ApJ...773...90G} (38)~\citet{2018MNRAS.481.4542S} (39)~\citet{2010ApJ...720..786L} (40)~\citet{2010A&A...522A..36K} (41)~\citet{2015ApJ...800...63S} (42)~\citet{2017MNRAS.468L..97O} (43)~\citet{2002ApJ...579..530W} (44)~\citet{2019arXiv190904676R} (45)~\citet{2020MNRAS.496.3412M} (46)~\citet{2001ApJ...554..240E} (47)~\citet{2019MNRAS.487..667C} (48)~\citet{2019ApJ...884..106M} (49)~\citet{2005ApJ...632..799P} (50)~\citet{2011AAS...21822816M} (51)~\citet{2009ApJ...690.1322W} (52)~\citet{2017ApJ...850...74K} (53)~\citet{2014MNRAS.445.3073P} (54)~\citet{2018MNRAS.473.5334R} (55)~\citet{2017NatAs...1..727K} (56)~\citet{2021ApJ...907L..21D} 

\section{Mid-IR Eddington ratio, variability and color}\label{sec:mir_var_col_lum}
\begin{figure}[h]
\centering
	\includegraphics[width=0.6\textwidth]{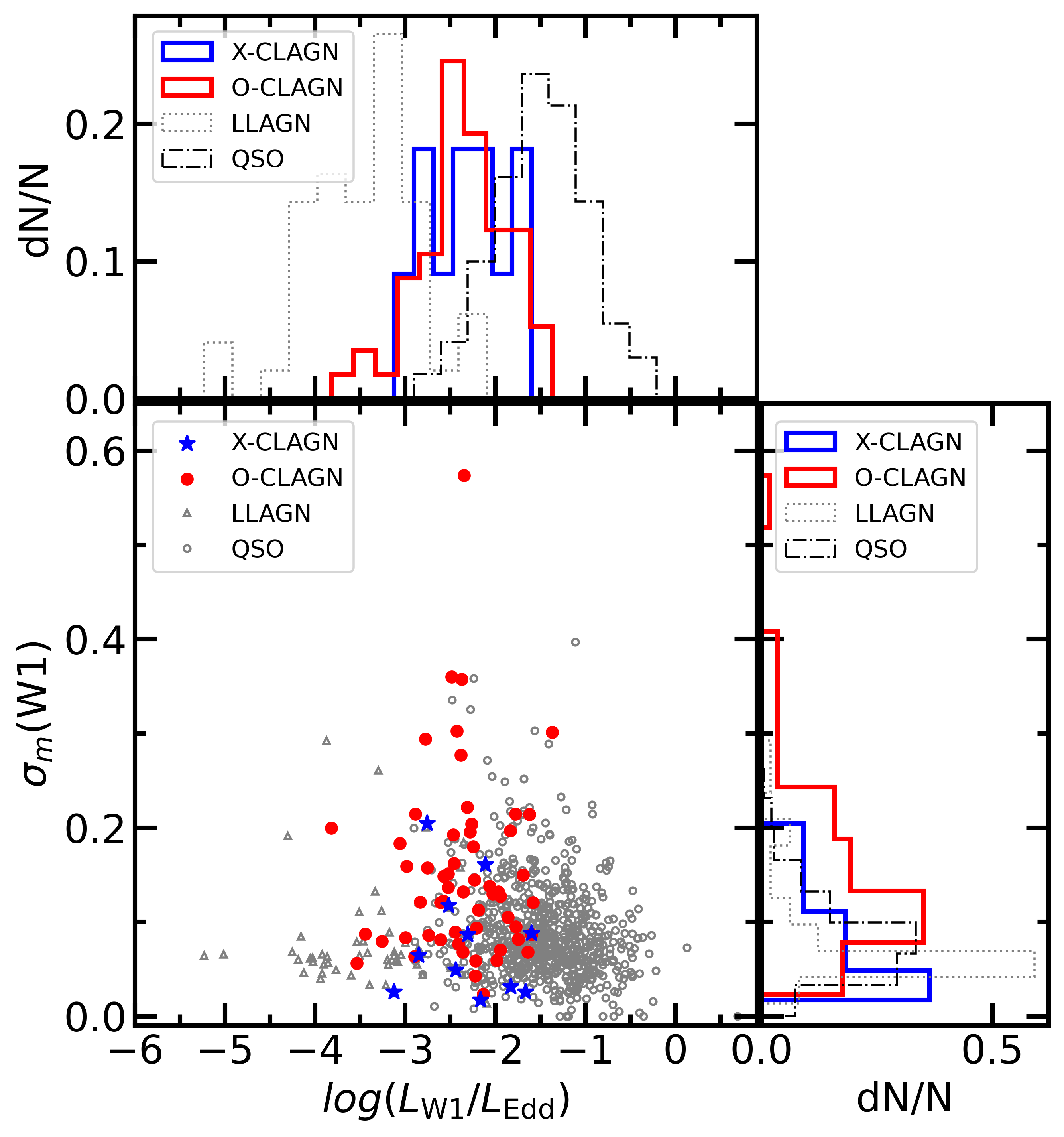}
    \caption{The correlation between the variability of $W1$ band and the mean Eddington-scaled $W1$ band luminosity for LLAGNs \citep{2009MNRAS.399..349G}, CLAGNs, and QSOs \citep{2007ApJ...667..131G} and their distributions. }
    \label{fig:var_ledd_hist}
\end{figure}

We calculate the mid-IR luminosity for CLAGNs from the \textit{WISE} magnitude, and the distribution of the mid-IR Eddington ratio is presented in the top panel of \autoref{fig:var_ledd_hist}. For comparison, we also include the distributions for a sample of 49 LLAGNs from \citet{2009MNRAS.399..349G} and 732 luminous QSOs from \citet{2007ApJ...667..131G}, where the sources with at least 20 valid \textit{WISE} data points are considered. The average log$L_\mathrm{W1}/L_\mathrm{Edd}$\, are $-2.35$, $-3.46$, and $-1.48$ for CLAGNs, LLAGNs, and QSOs, respectively. It can be found that both optical CLAGNs and X-ray CLAGNs have a moderate mean Eddington ratio of mid-IR luminosity between LLAGNs and QSOs. The X-ray CLAGNs show a little bit higher Eddington ratio than optical CLAGNs, where average ratios of log$L_\mathrm{W1}/L_\mathrm{Edd}$\ are $-2.31$ and $-2.36$ respectively (see \autoref{table_MIR_var_cor_lum}). We carry out two parametric statistical test, namely Mann-Whitney U Test (hereafter referred as the U-test), to explore whether two distributions differ significantly. In general, two distributions are considered as different when $p$-value is below 0.05. We find that distributions of the Eddington ratio are clearly different for LLAGNs, QSOs and CLAGNs, where the $p$-value are all $< 10^{-4}$ (see Table 3). The X-ray CLAGNs have a little bit higher Eddington ratio than the optical CLAGNs, but their distributions are quit similar ($p>0.5$ in U-test, see Table 3).

To estimate the mid-IR variability amplitude of CLAGNs, we firstly calculate the maximum variation magnitude $\Delta\,W1$ and $\Delta\,W2$ for each selected CLAGN. We find that the majority of CLAGNs change more than $0.3$ mag for $\Delta\,W1$ (48/68) and $\Delta\,W2$ (54/68), which include 4 X-ray CLAGNs (NGC 4051, NGC 4388, NGC 4507, and NGC 4939). 

We further calculate the intrinsic variability based on the standard deviation about the mean offset for each CLAGN ($\sigma_m$) following \citet{2019MNRAS.483.2362R} \citep[see also][etc]{2007AJ....134.2236S,2012ApJ...759L..31J}.
The magnitude standard deviation $\Sigma$ is given as
\begin{equation}
\Sigma=\sqrt{\frac{1}{n-1}\sum_{i=1}^{N}(m_i - <m>)^2},
\end{equation}
where $m_i$ is the magnitude at $i$-th point and $<m>$ is the error-weighted average.
The amplitude of variability $\sigma_m$ is  
\[\sigma_m  =
  \begin{cases}
    \sqrt{\Sigma^2 - \epsilon^2},  & \quad \text{if } \Sigma>\epsilon,\\
     0,                            & \quad  \text{otherwise.}\\
  \end{cases}
\]	               
where the error $\epsilon$ is calculated from the individual errors as
\begin{equation}
\epsilon^2=\frac{1}{N}\sum_{i=i}^{N}{\epsilon_{i}^{2} + \epsilon_{s}^2}, 
\end{equation}
where $\epsilon_{i}$ is the measurement uncertainty of $i$-th point and $\epsilon_{s}$ is the systematic uncertainty. The systematic uncertainties for $W1$ and $W2$ are 0.024 mag and 0.028 mag \citep{2011ApJ...735..112J}, respectively. We find that the majority of CLAGNs show strong mid-IR variability with $\sigma_{m \rm{W1}}$ (40/68)  and $\sigma_{m \rm{W2}}$ (43/68) greater than 0.1 mag, and 3 of them are X-ray CLAGNs (NGC 4051, NGC 4388, and NGC 4939). In \autoref{fig:var_ledd_hist}, we present the correlation between the variability and the mid-IR Eddington ratio and their distributions. It can be found that the average mid-IR variability is almost two times stronger in optical CLAGNs than that in both LLAGNs and QSOs, where average values of $\sigma_{m \rm{W1}}$ are 0.15, 0.08, and 0.09 for optical CLAGNs, LLAGNs, and QSOs, respectively. The U-test results support that the distribution of $\sigma_{m}$ of optical CLAGNs is much different from those of LLAGNs and QSOs, where all the $p$ are less than $10^{-7}$ in the U-test. It should be noted that the distribution of $\sigma_{m}$ of X-ray CLAGNs of our sample is not much different from that of LLAGNs and QSOs ($p>0.3$ in U-test.)

\begin{figure}[h]
\centering
	\includegraphics[width=0.6\textwidth]{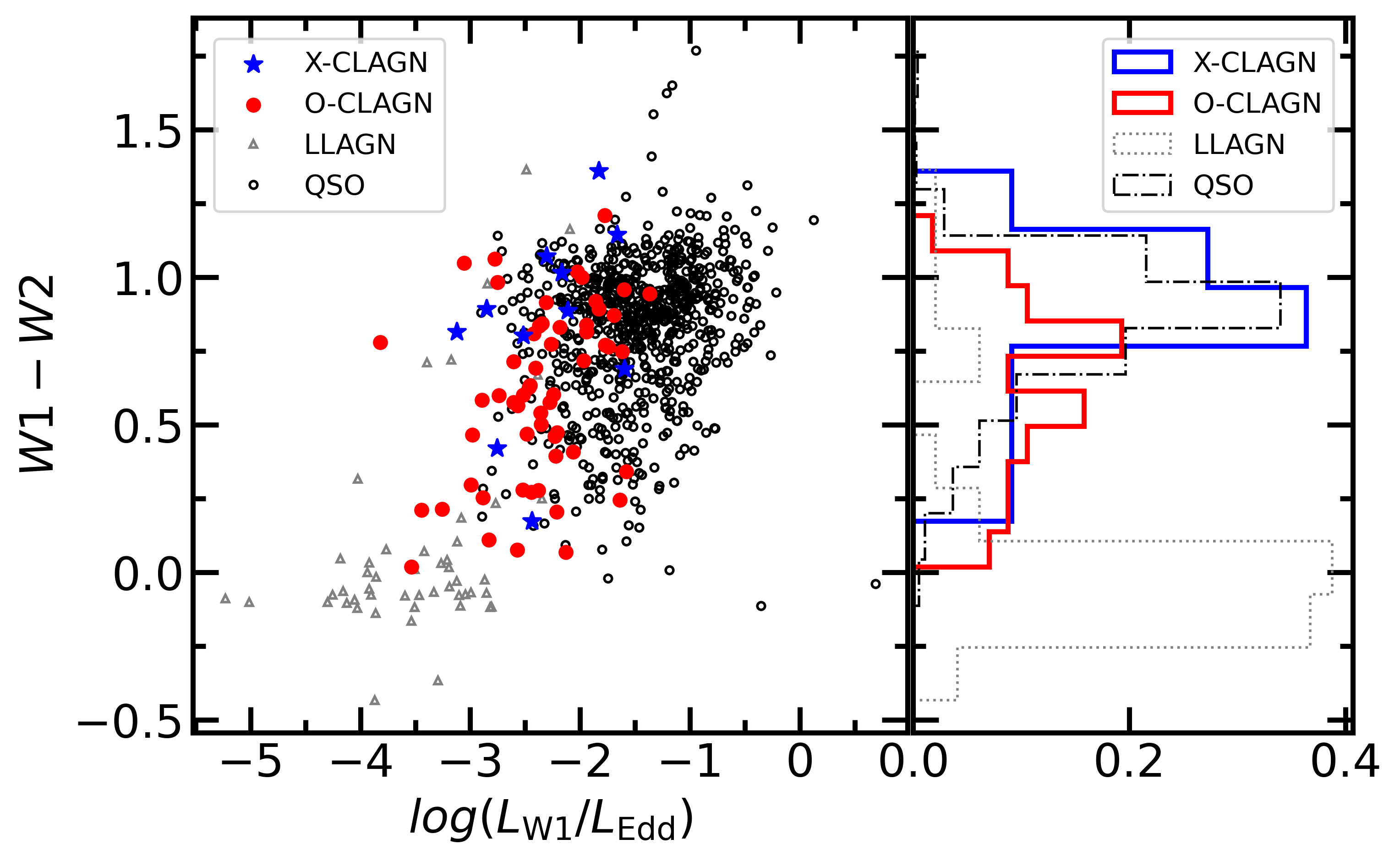}
    \caption{Distribution of the color for $W1-W2$ and its correlation with the Eddington-scaled $W1$ band luminosity for LLAGNs\citep{2009MNRAS.399..349G}, CLAGNs, and QSOs \citep{2007ApJ...667..131G}. }
    \label{fig:color_ledd}
\end{figure}

We calculate the average color $<W1-W2>$ of CLAGNs and also the sample (LLAGNs and QSOs) for comparison. We present the color distribution of $W1$-$W2$ and the correlation between the color and the mid-IR luminosity Eddington ratio in \autoref{fig:color_ledd}. The sources with higher Eddington ratios normally have higher values of $W1$-$W2$. The color $W1$-$W2$ varies from 0 to around 1.4 for CLAGNs with an average value of 0.65, while the average values are 0.08 and 0.83 for LLAGNs and QSOs respectively. The color of brighter/fainter CLAGNs is similar to the QSOs/LLAGNs, respectively (see \autoref{fig:color_ledd}). The color distribution of optical CLAGNs is much different from both QSOs and LLAGNs ($p<10^{-7}$ in U-test), where X-ray CLAGNs is similar to that of QSOs ($p=0.78$ in U-test).

\begin{table}[h]
 \caption{Mid-IR variability, color, and Eddington ratio for LLAGNs, CLAGNs, and QSOs.
}
 \label{table_MIR_var_cor_lum}
 \begin{center}
 \begin{tabular}{lccccc}
 \hline\hline
Sample             & LLAGN &   \multicolumn{3}{c}{CLAGN}        & QSO  \\ 
                   &        &   O & X   & All-CLAGN         &       \\ \hline
$<\sigma_{m \mathrm{W1}}>$    &  0.08      & 0.15    &  0.08  &   0.14   &   0.09  \\ 
$<\sigma_{m \mathrm{W2}}>$    &  0.07      &   0.17 &  0.08   &    0.15  &   0.08    \\ 
$<W1-W2>$          &   0.08      &  0.61   & 0.84      &    0.65    &   0.83   \\ 
$<\mathrm{log} L_\mathrm{W1}/L_\mathrm{bol}>$ & -3.46  &  -2.36  &  -2.31   &  -2.35   &  -1.48\\ 
$<\mathrm{log} L_\mathrm{W2}/L_\mathrm{bol}>$ & -3.81  &   -2.50   &  -2.35   &  -2.47   &   -1.53  \\  
\hline\hline
\end{tabular}
\end{center}
\end{table}

\begin{table}
 \caption{The $p$-value of Mann-Whitney U Test for QSO, LLAGN, optical CLAGN(O), and X-ray CLAGN(X) samples.
}
 \label{table_mann_whitney_U}
 \begin{center}
 \begin{tabular}{lccccc}
 \hline\hline      
Sample    &   O\&LLAGN  &  O\&QSO & O\&X & X\&LLAGN & X\&QSO  \\  \hline
           
$<\sigma_{m \mathrm{W1}}>$&  1.0e-7  & 4.2e-10  & 6.4e-3& 0.66 &  0.33       \\ 
$<\sigma_{m \mathrm{W2}}>$& 3.4e-9   & 8.2e-12 & 4.9e-3  & 0.91  & 0.66 \\ 
$<W1-W2>$                &  4.7e-12   &  2.6e-8   & 2.9e-2 & 1.2e-5  & 0.78    \\ 
$<\mathrm{log} L_\mathrm{W1}/L_\mathrm{bol}>$& 2.7e-13 & 4.7e-23  & 0.78 & 1.1e-5 & 1.6e-5 \\ 
$<\mathrm{log} L_\mathrm{W2}/L_\mathrm{bol}>$& 1.1e-13 &2.9e-23 & 0.54 & 3.4e-6 & 5.8e-5 \\  
\hline\hline
\end{tabular}
\end{center}
\end{table}

\section{Dust reverberation mapping in CLAGNs}\label{sec:tau-L}
Dust reverberation mapping provides a method to explore the properties of AGN torus, where the changes of the disk emission will lead to the variation of infrared emission from the torus but with a time lag due to the signal travel. The continued observations of {\it WISE}, combined with several ground-based optical transient surveys (e.g., All-sky Automated Survey for Supernovae, ASAS-SN) and space-telescope observations (e.g., \swift\,), offer a good opportunity to explore the possible dust structures in these CLAGNs. We adopt the data of $V$ band from the ASAS-SN \citep[][]{2014ApJ...788...48S,2017PASP..129j4502K,2019MNRAS.485..961J}, \uvot\, and mid-IR data from the {\it WISE} to estimate the mid-IR time lag ($\tau$) for CLAGNs. We select 22 bright CLAGNs (e.g., $<W1>$ less than 11 mag) with strong mid-IR variation ($\Delta W1>0.3$ mag), which have good $V$ band and mid-IR band monitorings. For NGC 1566, we adopt the photometric $V$ band data of NGC 1566 from the ASAS-SN and \uvot\,, and the re-binned $W1$/$W2$ band data from the {\it WISE} in each visit to estimate the time lag. We use the tool \textit{uvotmaghist} to do the aperture photometry for V-band data of \uvot\,. The source aperture radius is 3$\arcsec$ and the background is a blank region with a much larger radius. We find that the flux of NGC 1566 in ASAS-SN is systematically higher than that of \uvot\,, which is probably contaminated by the host galaxy due to the poor angular resolution \citep[see ][]{2017PASP..129j4502K}. We then subtract the average offset of V-band data from ASAS-SN data, which is determined from the quasi-simultaneous (within 1 day of the interval) observations of ASAS-SN and \uvot\,. For other sources, we only use the $V$ band data from ASAS-SN and the re-binned $W1$/$W2$ data in each visit with good observational overlaps to estimate the time lags. The $V$ band data are binned in 30 days for some sources with complex short-term variability.

\begin{figure}[b]
\begin{tikzpicture}
    \matrix[matrix of nodes]{
    \includegraphics[width=0.45\textwidth]{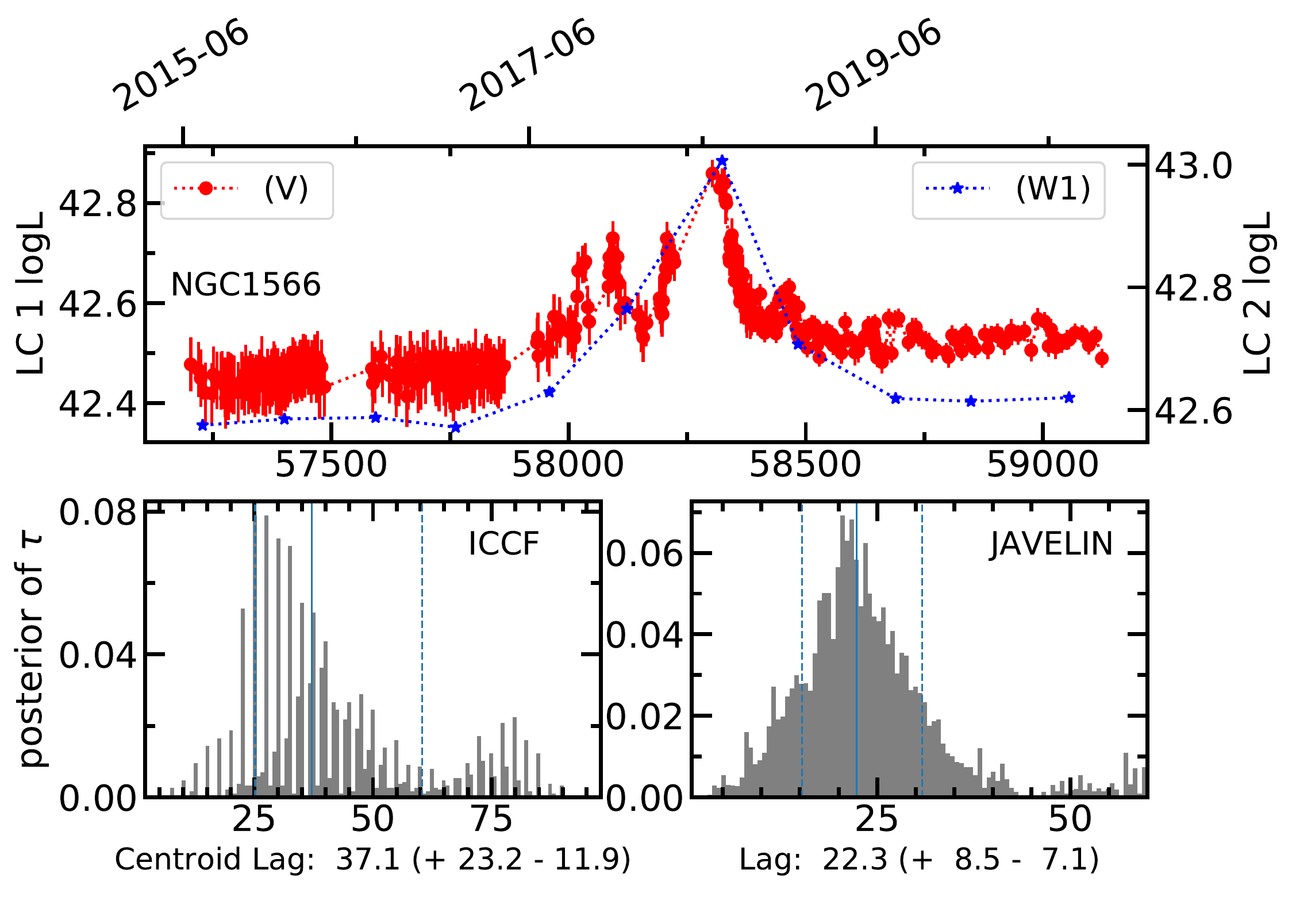} & \hspace{1em} & \includegraphics[width=0.45\textwidth]{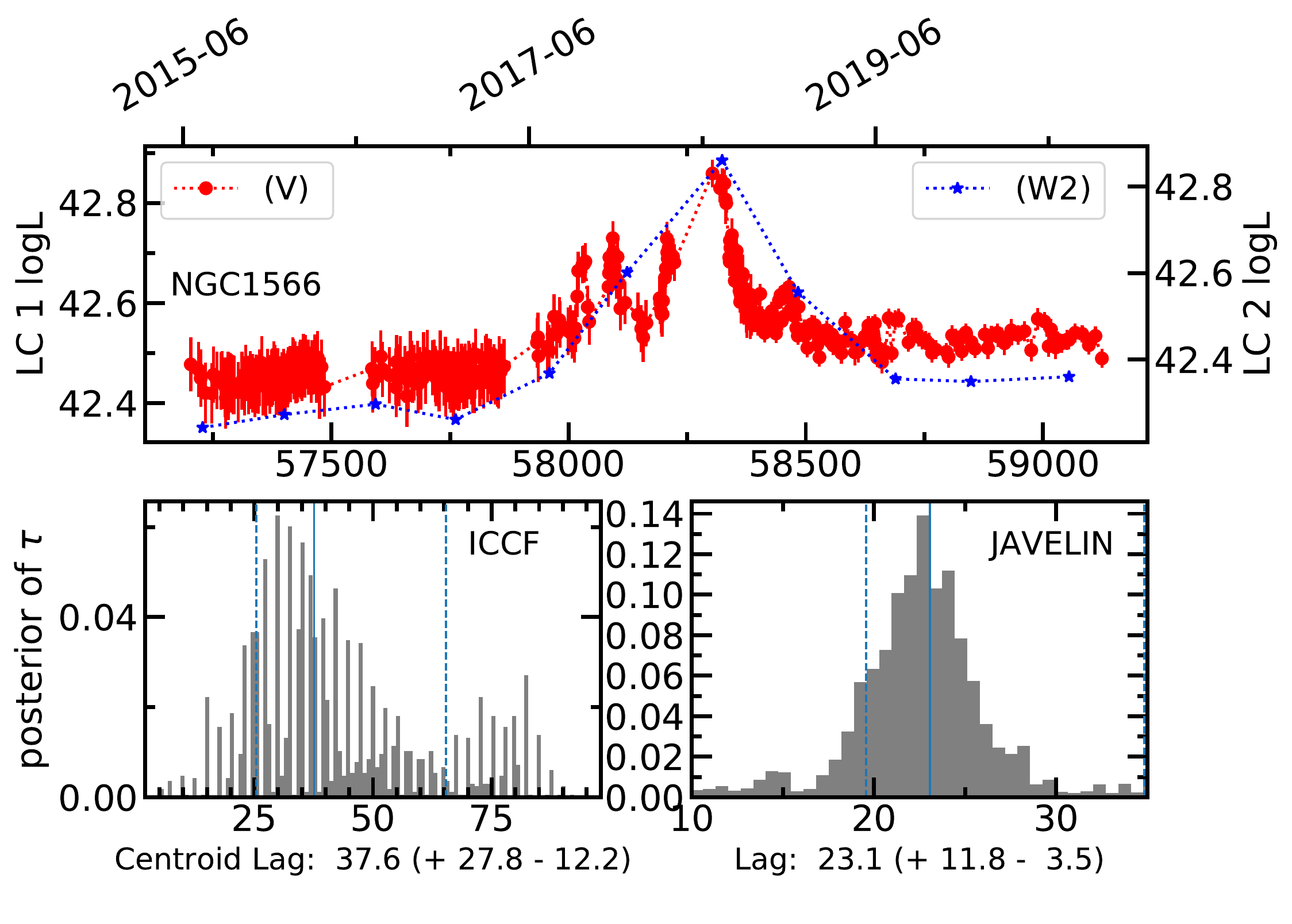} \\
    };
\end{tikzpicture}
\caption{Dust reverberation mapping analysis result for NGC 1566 as an example, and other sources are presented in Appendix. Left panel shows the light curves and the posterior of time lag between $V$ and $W1$ band ($\tau_\mathrm{W1}$). Right panel shows the light curves and the posterior of time lag between $V$ and $W2$ band ($\tau_\mathrm{W2}$).}
\label{fig:lag_NGC1566}
\end{figure}

We use the interpolation cross-correlation function \citep[ICCF;][]{1998PASP..110..660P,2018ascl.soft05032S} in a range of 0--400 days (corresponding to $\sim 70000 R_g$ for $M_\mathrm{BH}=10^8 M_{\odot}$) for the first attempt to estimate $\tau$ and then further limited the range according to the posterior of $\tau$. The interpolation time steps of 5 days are applied to the $V$ and the $W1$/$W2$ band light curves. The flux randomization and random subset selection methods are employed with 50000 realizations in the Monte Carlo simulation to estimate the centroid time lags and the uncertainties\footnote{The code \texttt{pyCCF} is available in \url{http://ascl.net/code/v/1868}}. We also use {\sc javelin} algorithm to further test the results of the ICCF method, where {\sc javelin} algorithm fits the light curves using a damped random walk (DRW) model with amplitude and time scale of the variability. A top-hat transfer function (TF) is convolved with the driving light curve and the best-fit model parameters such as time lag $\tau$ are found through the Markov Chain Monte Carlo method. We restrict the lag and width in the range of 0--400 days for the first attempt. Finally,  we get time lag measurements of 13 sources with maximum cross-correlation coefficient $r>0.6$, which are shown in \autoref{fig:lag_NGC1566} and \autoref{lag_analysis}. In most cases, two methods give the roughly consistent results. Later on, we adopt the time lags from the {\sc javelin} method as primary results in the following analysis, where the time lag $\tau_\mathrm{W1}$ between $V$ and $W1$ band,  $\tau_\mathrm{W2}$ between $V$ and $W2$ band are listed in \autoref{table_lag}. 

\begin{table}
\renewcommand{\arraystretch}{1.5}
 \caption{Mid-IR time lag results of CLAGNs.
}
 \label{table_lag}
 \begin{center}
 \begin{tabular}{lccccccc}
 \hline\hline
Name & log$L_\mathrm{bol}$ & \multicolumn{2}{c}{$\tau_\mathrm{W1}$} & \multicolumn{2}{c}{$\tau_\mathrm{W2}$} &Type \\ 
      &                      &   ICCF     &  {\sc javelin}   &    ICCF & {\sc javelin} &   \\ \hline 
3C 390.3 & 45.79 & $180.5_{-16.6}^{+18.9}$ & $186.5_{-9.5}^{+11.3}$ & $184.8_{-21.3}^{+25.0}$ & $200.1_{-10.6}^{+26.4}$ & O \\
HE 1136-2304 & 44.31 & $127.6_{-43.9}^{+68.3}$ & $178.7_{-47.6}^{+54.9}$ & $177.0_{-62.6}^{+70.6}$ & $196.8_{-28.9}^{+50.8}$ & O \\
Mrk 530 & 44.90 & $212.2_{-39.7}^{+15.2}$ & $215.2_{-6.5}^{+6.1}$ & $215.0_{-20.3}^{+12.9}$ & $216.8_{-6.5}^{+11.4}$ & O \\
Mrk 590 & 44.13 & $57.9_{-20.0}^{+17.8}$ & $97.0_{-29.2}^{+30.6}$ & $56.6_{-21.3}^{+13.8}$ & $115.0_{-36.5}^{+30.1}$ & O \\
Mrk 6 & 44.59 & $141.6_{-11.0}^{+31.7}$ & $170.3_{-15.4}^{+13.5}$ & $180.4_{-15.1}^{+16.6}$ & $222.8_{-34.8}^{+9.1}$ & O \\
Mrk 609 & 44.43 & $173.0_{-37.9}^{+43.6}$ & $239.9_{-19.0}^{+33.0}$ & $187.0_{-60.7}^{+49.5}$ & $262.3_{-18.8}^{+19.5}$ & O \\
Mrk 926 & 45.66 & $213.8_{-31.0}^{+25.3}$ & $242.1_{-9.8}^{+11.7}$ & $252.5_{-38.3}^{+27.5}$ & $273.1_{-13.3}^{+14.2}$ & O \\
NGC 1566 & 42.94 & $37.1_{-11.9}^{+23.2}$ & $22.3_{-7.1}^{+8.5}$ & $37.6_{-12.2}^{+27.8}$ & $23.1_{-3.5}^{+11.8}$ & O \\
NGC 2617 & 43.77 & $93.8_{-17.6}^{+17.0}$ & $69.8_{-12.6}^{+11.9}$ & $93.5_{-17.0}^{+19.1}$ & $73.9_{-16.0}^{+9.7}$ & O \\
NGC 3516 & 44.19 & $91.1_{-31.4}^{+64.0}$ & $88.5_{-34.1}^{+48.5}$ & $100.7_{-28.5}^{+54.4}$ & $119.8_{-37.5}^{+43.3}$ & O \\
NGC 4051 & 42.61 & $22.3_{-10.0}^{+10.4}$ & $26.2_{-11.4}^{+17.0}$ & $24.8_{-12.2}^{+10.2}$ & $29.0_{-12.3}^{+20.9}$ & X \\
NGC 4151 & 44.08 & $74.4_{-19.5}^{+40.8}$ & $77.6_{-12.9}^{+26.1}$ & $99.3_{-32.6}^{+40.7}$ & $89.4_{-12.2}^{+17.0}$ & O \\
NGC 5548 & 44.66 & $162.7_{-31.7}^{+37.2}$ & $132.0_{-4.8}^{+5.7}$ & $207.3_{-29.0}^{+25.7}$ & $236.3_{-5.7}^{+4.7}$ & O \\
\hline\hline
\end{tabular}
\end{center}
Note. The table lists source name, bolometric luminosity (log$L_\mathrm{bol}$), time lag between $V$ band and $W1$/$W2$ band ($\tau_\mathrm{W1}$/$\tau_\mathrm{W2}$) from the ICCF and {\sc javelin} method , CLAGN type (``O'' for optical CLAGN and ``X'' for X-ray CLAGN). 
\end{table}

\begin{figure}[h!]
\centering
	\includegraphics[width=0.6\textwidth]{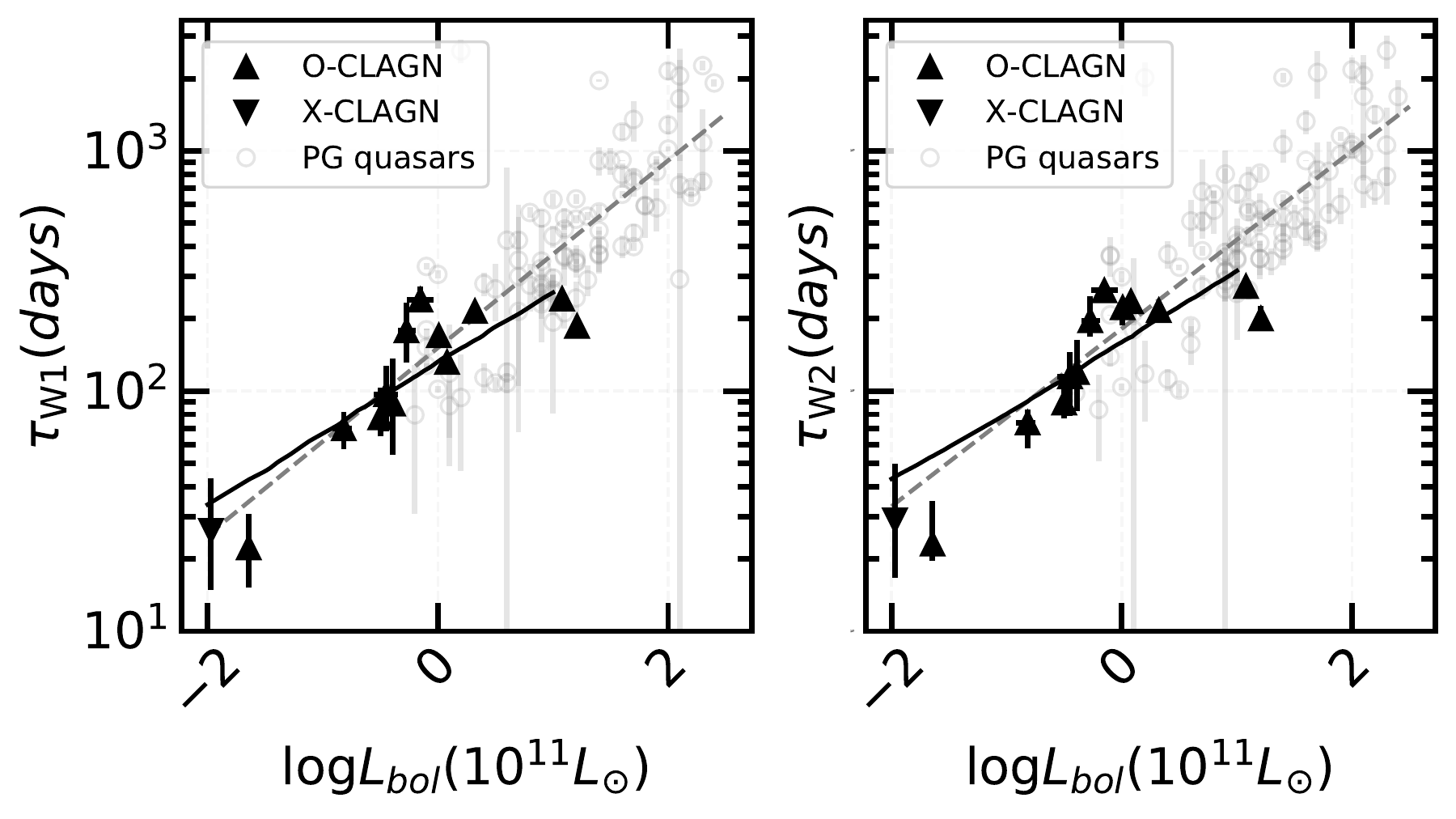}
    \caption{The correlation between the time lag $\tau$ and the bolometric $L_{\rm bol}$ for CLAGNs, where $\tau$ is the time lag between $V$ band and $W1$/$W2$ band. The solid line is the best fitting for CLAGNs. Grey circles represent PG quasars \citep{2019ApJ...886...33L} for comparison. The dashed line represents the best fitting for PG quasars in \citet{2019ApJ...886...33L} and the CLAGNs in this work. } 
    \label{fig:tau_L}
\end{figure}

To explore the possible correlation between time lag ($\tau$) and bolometric luminosity ($L_\mathrm{bol}$), we estimate the bolometric luminosities of CLAGNs from the X-ray luminosity through a bolometric correction of $L_{\mathrm{bol}}=8 \times L_{\mathrm{14-195 keV }}$ \citep[e.g.,][]{2009MNRAS.392.1124V}, where the X-ray data are obtained from the 105-Month Swift-BAT All-sky Hard X-Ray Survey reported in \citet{2018ApJS..235....4O}.  The correlation between $\tau_{\rm{W1}}$/$\tau_{\rm{W2}}$ and $L_\mathrm{bol}$ for CLAGNs is shown in \autoref{fig:tau_L}.


It can be found that the brighter CLAGNs have longer time lags (Spearman's rank correlation coefficient $r=0.88$ with a probability of $p=7.5\times10^{-5}$ for $\tau_{\rm{W1}}$-$L_\mathrm{bol}$ and $r=0.86$ with a probability of $p=1.5\times10^{-4}$ for $\tau_{\rm{W2}}$-$L_\mathrm{bol}$). We use UltraNest\footnote{\url{https://johannesbuchner.github.io/UltraNest/}} package, which implements nested sampling to constrain the model parameters \citep{2021JOSS....6.3001B}, to fit the linear log$\tau$-log($L_{\mathrm{bol}}$) correlation for the 13 CLAGNs using the following equation: 
\begin{equation}
\mathrm{log}(\tau/\mathrm{day})=a\times\mathrm{log}(L_{\mathrm{bol}}/10^{11} L_{\odot})+b.
\end{equation}

We get \begin{equation}
\mathrm{log}(\tau_{\rm{W1}}/\mathrm{day})= 0.28^{+0.08}_{-0.06}\times \mathrm{log}(L_{\mathrm{bol}}/10^{11} L_{\odot})+ 2.12^{+0.05}_{-0.05}
\end{equation} for the $W1$ band, and 
\begin{equation}
\mathrm{log}(\tau_{\rm{W2}}/\mathrm{day})= 0.29^{+0.08}_{-0.07}\times \mathrm{log}(L_{\mathrm{bol}}/10^{11} L_{\odot})+ 2.21^{+0.05}_{-0.06}
\end{equation}for the $W2$ band. The CLAGNs still follow a consistent $\tau$-$L_{\mathrm{bol}}$ correlation with that found for the 87 luminous Palomar--Green quasars \citep[usually with $L_{\mathrm{bol}} \ge 10^{11}L_{\odot}$, see ][]{2019ApJ...886...33L} within the scatter, and the best fitting results are listed in \autoref{table_tau_L_fit}.

\begin{table}[h!]
\renewcommand{\arraystretch}{1.5}
 \caption{The best fitting results of $\tau$-$L_{\mathrm{bol}}$ correlation for CLAGNs and QSOs through UltraNest package.
}
 \label{table_tau_L_fit}
 \begin{center}
 \begin{tabular}{lcccccc}
 \hline\hline
Sample    & \multicolumn{3}{c}{$W1$} &   \multicolumn{3}{c}{$W2$}      \\ 
           &    $a$    & $b$   &    scatter   &  $a$            &     $b$    & scatter        \\ \hline
CLAGNs      &    $0.28^{+0.08}_{-0.06}$ & $2.12^{+0.05}_{-0.05}$ & $0.16^{+0.06}_{-0.04}$  & $0.29^{+0.08}_{-0.07}$  &  $2.21^{+0.05}_{-0.06}$     &  $0.17^{+0.06}_{-0.04}$  \\ 
PG quasars    &    $0.39^{+0.04}_{-0.04}$ & $2.19^{+0.05}_{-0.06}$  &   $0.23^{+0.02}_{-0.02}$   & $0.36^{+0.04}_{-0.04}$  & $2.29^{+0.06}_{-0.05}$    & $0.22^{+0.02}_{-0.02}$  \\
CLAGNs \& PG quasars &   $0.39^{+0.03}_{-0.03}$ & $2.17^{+0.04}_{-0.04}$ & $0.22^{+0.02}_{-0.02}$ & $0.37^{+0.02}_{-0.03}$   & $2.27^{+0.04}_{-0.04}$        &$0.21^{+0.02}_{-0.02}$   \\ 

\hline\hline
\end{tabular}
\end{center}
Note. The slope (a), intercept (b) and scatter of the best fitting for $\tau$-$L_{\mathrm{bol}}$ correlation. The data of PG quasars are adopted from \citet{2019ApJ...886...33L}.
\end{table}

\section{Conclusion and Discussion} \label{sec:dis}
We perform an analysis on the {\it WISE} mid-IR data for a sample of both X-ray and optical CLAGNs to explore the possible physical mechanism behind them. We find that the average Eddington scaled mid-IR luminosity of both X-ray and optical CLAGNs is lower than that of QSOs but higher than that of LLAGNs. The corresponding bolometric luminosity Eddington ratio at a few percent is around the critical luminosity where AGNs may easily suffer the accretion-mode transition between RIAF and SSD. CLAGNs show a wide distribution of mid-IR color $W1-W2$, where brighter and fainter CLAGNs are similar to QSOs and LLAGNs, respectively. Above results support that the CLAGNs possibly stay in a transitional stage. Based on the mid-IR and optical variability, we estimate the dust echo time lags of 13 CLAGNs, which roughly follow the correlation of $\tau-L_\mathrm{bol}$ as found in the QSOs \citep{2019ApJ...886...33L}.

\subsection{Transitional stage of CLAGNs}
The study of the intrinsic physics of CLAGNs will help us to understand the nature of accretion flow and the evolution of galaxies. Since the mid-IR emission is mainly produced by the hot dust heated by the UV radiation of the accretion disk, the mid-IR properties of CLAGNs can shed light on their central engines. We find that the most of the CLAGNs show strong variation in the mid-IR band (e.g., more than 70\% sources have maximum variation magnitude $\Delta W1>0.3$). The variation suggests that most of the optical and X-ray CLAGNs cannot be simply explained by the motion of absorbing clouds moving in/out of our line of sight, where the ``changing-look'' should be correlated with the variation of the accretion disk (e.g., change of disk structure, disk wind, etc). 

The Eddington ratio of mid-IR luminosity for CLAGNs just stays between QSOs and LLAGNs, where the SSD and ADAF (or RIAF) are widely believed to be dominant in QSOs and LLAGNs, respectively. The average ratio of $L_\mathrm{W1}/L_\mathrm{Edd}$ for CLAGNs is $\sim $0.5\%, which roughly corresponds to $L_\mathrm{bol}/L_\mathrm{Edd}\sim$ several percent by assuming a bolometric correction factor of $\sim 10$ \citep[e.g.,][]{2012MNRAS.426.2677R}. The transition between IRAF and SSD is expected at a critical accretion rate of $\sim$1\% Eddington ratio. So the distribution of mid-IR luminosity Eddington ratio of CLAGNs supports the possible accretion-mode transition in CLAGNs. The CLAGNs will show galaxy-like mid-IR color (e.g., $W1-W2\leq0.5$) when the sources enter into the faint state, while they will show AGN-like mid-IR color (e.g., $W1-W2\geq0.5$) when the sources enter into the bright state and vice versa \citep[e.g.,][]{2012ApJ...753...30S,2013AJ....145...55Y,2020ApJ...889...46S}. Therefore, the color distribution of CLAGNs also supports the possible accretion-mode transition in CLAGNs.

We find that the optical CLAGNs show stronger mid-IR variation compared to LLAGNs and QSOs (see \autoref{fig:var_ledd_hist}), which may be triggered by the strong variation of radiative efficiency of accretion flow \citep[see ][]{2012MNRAS.427.1580X} due to the transition of accretion mode. In the SSD case, the radiative efficiency is roughly unchanged for a given BH spin. However, in the ADAF case, the radiative efficiency drops quickly with accretion rate when $\dot{M}/\dot{M}_{Edd}<1\%$. Therefore, the central ionization luminosity will vary a lot at the critical accretion rate even though the accretion rate itself only changes a little bit \citep[see ][]{2012MNRAS.427.1580X}, which can lead to the stronger mid-IR variability in optical CLAGNs. As the accretion rate decreases below the critical value, the inner cold disk might transit into ADAF and the intrinsic luminosities will drop, which will reduce the ionizing photons for exciting the broad emission lines. The AGNs will change their optical type from type 1 to type 1.5-1.8 or even type 2 as seen in the optical CLAGNs. The sources will go back to type 1 state when the SSD is reformed as the accretion rate increases. The multi-epoch observations of several optical CLAGNs also show that they cross a critical bolometric Eddington ratio of $L_\mathrm{bol}/L_\mathrm{Edd}\sim$1\% during their dramatic variations \citep[e.g.,][]{2019ApJ...874....8M,2019ApJ...883...76R,2021MNRAS.508..144G,2021MNRAS.506.4188L,2021MNRAS.507..687J}, which is similar to the state transition in black hole X-ray binaries \citep[e.g.,][]{2008ApJ...682..212W}.

In this work, we find several X-ray CLAGNs also show strong mid-IR variability (4/11 with $\Delta W1>0.3$) and they have average mid-IR Eddington ratio at $\sim 0.5$\%, which is not much different from the optical ones ($\sim 0.4$\%). It should be noted that the variability and color of X-ray CLAGNs are more consistent with QSOs, even though their Eddington-scaled mid-IR luminosities are much lower than those of QSOs. The variation of $N_\mathrm{H}$ in X-ray CLAGNs may be caused by the variation of disk wind which will be observed at a moderate inclination angle (e.g., $40^{\rm o}$-$60^{\rm o}$). The X-ray spectral evolution in a X-ray CLAGN of NGC 1365 does support this scenario \citep[e.g.,][]{2014MNRAS.440.3503C,2021RAA....21..199L}. The formation of disk wind and its opening angle are closely correlated to the underlying SSD \citep[e.g.,][]{2017PASA...34...42Y,2020MNRAS.492.5540M}. In the ADAF case, the plasma is fully ionized and the wind will not affect the absorption. Therefore, if the disk for those X-ray CLAGNs has transited to SSD, the variability and color of X-ray CLAGNs would be more similar to those of QSOs. We note that the X-ray CLAGNs are still limited in our sample, and more sources are wished to further test this issue.


\subsection{Dust echo in CLAGNs}
Mid-IR emission at $W1$ and $W2$ band mainly comes from the hot dust heated by central engine of AGNs. The dust reverberation mapping method can be applied when the source has strong optical and mid-IR variability. We explore the dust echo time lag $\tau$ for 13 CLAGNs based on mid-IR and optical multi-epoch observations. Several CLAGNs are famous nearby Seyferts, where the infrared reverberation mappings have been explored in the previous works. The time lags between $K$ band (2.19 $\mu$m) and $V$ band of 5 CLAGNs (Mrk 590, NGC 3516, NGC 4051, NGC 4151, and NGC 5548) in a sample of 17 Seyfert 1 AGNs have been reported \citep[see][]{2014ApJ...788..159K,2019ApJ...886...33L}. Adopting the time-lag ratio $\tau_{\mathrm{K}}$: $\tau_{\mathrm{W1}}$=$0.6$: $1$ \citep[see][]{2019ApJ...886...33L}, the derived time lags for the 5 CLAGNs from \citet[]{2014ApJ...788..159K} are roughly consistent with our results. The time-lag ratio $\tau_{\mathrm{W1}}$: $\tau_{\mathrm{W2}}$=$1$: $1.2$ for the 13 CLAGNs is also consistent with that of the luminous QSOs \citep[see][]{2019ApJ...886...33L}. The positive correlation between mid-IR time lag and bolometric luminosity ($\tau$-$L$) for AGNs has been widely investigated \citep[e.g.,][]{2014ApJ...788..159K,2019ApJ...886...33L,2019ApJ...886..150M,2020MNRAS.495.2921N,2020AJ....159..259S,2021MNRAS.501.3905M}. In CLAGNs, there is also a tight $\tau$-$L$ correlation, which is roughly consistent with that of the luminous QSOs \citep{2019ApJ...886...33L} within the errorbar (see \autoref{fig:tau_L}). The time lag of mid-IR band suggests the similar circumnuclear dust structure existing both in CLAGNs and QSOs. The $\tau$-$L$ correlation for CLAGNs is slightly shallower than expected $\tau \propto L^{0.5}$ assuming the similar dust sublimation temperature \citep{2013peag.book.....N} and a face-on viewing angle \citep{2019ApJ...886...33L}. The physical reason for this shallower slope is unclear, which might be influenced by the structure of the torus or the inclination angle.

\begin{acknowledgments}
We thank the anonymous referee for useful comments which help to
improve the manuscript. BL and QW were supported in part by the Natural Science Foundation of China (grant U1931203) and the science research grants from the China Manned Space Project with NO. CMS-CSST-2021-A06; ZY was supported in part by the Natural Science Foundation of China (grants U1938114 and 11773055), the Youth Innovation Promotion Association of CAS (id 2020265); and WY would like to acknowledge the support in part by the National Program on Key Research and Development Project (grant 2016YFA0400804) and the National Natural Science Foundation of China (grants 11333005 and U1838203).
This research has made use of WISE and NEOWISE data products from the NASA/IPAC Infrared Science Archive, which is a joint project of the University of California, Los Angeles, and the Jet Propulsion Laboratory/California Institute of Technology, funded by the National Aeronautics and Space Administration. 


\end{acknowledgments}

%

\facilities{\textit{WISE}, \uvot\,, ASAS-SN}%


\software{astropy \citep{2013A&A...558A..33A,2018AJ....156..123A},  
         \sc{javelin}\citep{2011ApJ...735...80Z,2013ApJ...765..106Z}, 
         \sc{pyccf}\citep{1998PASP..110..660P,2018ascl.soft05032S},
          }




\bibliographystyle{aasjournal}
\bibliography{ref.bib}

\begin{thebibliography}{}
\expandafter\ifx\csname natexlab\endcsname\relax\def\natexlab#1{#1}\fi
\providecommand{\url}[1]{\href{#1}{#1}}
\providecommand{\dodoi}[1]{doi:~\href{http://doi.org/#1}{\nolinkurl{#1}}}
\providecommand{\doeprint}[1]{\href{http://ascl.net/#1}{\nolinkurl{http://ascl.net/#1}}}
\providecommand{\doarXiv}[1]{\href{https://arxiv.org/abs/#1}{\nolinkurl{https://arxiv.org/abs/#1}}}

\bibitem[{{Ag{\'\i}s-Gonz{\'a}lez} {et~al.}(2018){Ag{\'\i}s-Gonz{\'a}lez},
  {Hutsem{\'e}kers}, \& {Miniutti}}]{2018Galax...6...52A}
{Ag{\'\i}s-Gonz{\'a}lez}, B., {Hutsem{\'e}kers}, D., \& {Miniutti}, G. 2018,
  Galaxies, 6, 52, \dodoi{10.3390/galaxies6020052}

\bibitem[{{Ag{\'\i}s-Gonz{\'a}lez} {et~al.}(2014){Ag{\'\i}s-Gonz{\'a}lez},
  {Miniutti}, {Kara}, {Fabian}, {Sanfrutos}, {Risaliti}, {Bianchi},
  {Strotjohann}, {Saxton}, \& {Parker}}]{2014MNRAS.443.2862A}
{Ag{\'\i}s-Gonz{\'a}lez}, B., {Miniutti}, G., {Kara}, E., {et~al.} 2014,
  \mnras, 443, 2862, \dodoi{10.1093/mnras/stu1358}

\bibitem[{{Ai} {et~al.}(2020){Ai}, {Dou}, {Yang}, {Sun}, {Xie}, {Yao}, {Wu},
  {Wang}, {Shu}, \& {Jiang}}]{2020ApJ...890L..29A}
{Ai}, Y., {Dou}, L., {Yang}, C., {et~al.} 2020, \apjl, 890, L29,
  \dodoi{10.3847/2041-8213/ab7306}

\bibitem[{{Antonucci}(1993)}]{1993ARA&A..31..473A}
{Antonucci}, R. 1993, Annual Review of Astronomy and Astrophysics, 31, 473,
  \dodoi{10.1146/annurev.aa.31.090193.002353}

\bibitem[{{Astropy Collaboration} {et~al.}(2013){Astropy Collaboration},
  {Robitaille}, {Tollerud}, {Greenfield}, {Droettboom}, {Bray}, {Aldcroft},
  {Davis}, {Ginsburg}, {Price-Whelan}, {Kerzendorf}, {Conley}, {Crighton},
  {Barbary}, {Muna}, {Ferguson}, {Grollier}, {Parikh}, {Nair}, {Unther},
  {Deil}, {Woillez}, {Conseil}, {Kramer}, {Turner}, {Singer}, {Fox}, {Weaver},
  {Zabalza}, {Edwards}, {Azalee Bostroem}, {Burke}, {Casey}, {Crawford},
  {Dencheva}, {Ely}, {Jenness}, {Labrie}, {Lim}, {Pierfederici}, {Pontzen},
  {Ptak}, {Refsdal}, {Servillat}, \& {Streicher}}]{2013A&A...558A..33A}
{Astropy Collaboration}, {Robitaille}, T.~P., {Tollerud}, E.~J., {et~al.} 2013,
  \aap, 558, A33, \dodoi{10.1051/0004-6361/201322068}

\bibitem[{{Astropy Collaboration} {et~al.}(2018){Astropy Collaboration},
  {Price-Whelan}, {Sip{\H{o}}cz}, {G{\"u}nther}, {Lim}, {Crawford}, {Conseil},
  {Shupe}, {Craig}, {Dencheva}, {Ginsburg}, {VanderPlas}, {Bradley},
  {P{\'e}rez-Su{\'a}rez}, {de Val-Borro}, {Aldcroft}, {Cruz}, {Robitaille},
  {Tollerud}, {Ardelean}, {Babej}, {Bach}, {Bachetti}, {Bakanov}, {Bamford},
  {Barentsen}, {Barmby}, {Baumbach}, {Berry}, {Biscani}, {Boquien}, {Bostroem},
  {Bouma}, {Brammer}, {Bray}, {Breytenbach}, {Buddelmeijer}, {Burke},
  {Calderone}, {Cano Rodr{\'\i}guez}, {Cara}, {Cardoso}, {Cheedella}, {Copin},
  {Corrales}, {Crichton}, {D'Avella}, {Deil}, {Depagne}, {Dietrich}, {Donath},
  {Droettboom}, {Earl}, {Erben}, {Fabbro}, {Ferreira}, {Finethy}, {Fox},
  {Garrison}, {Gibbons}, {Goldstein}, {Gommers}, {Greco}, {Greenfield},
  {Groener}, {Grollier}, {Hagen}, {Hirst}, {Homeier}, {Horton}, {Hosseinzadeh},
  {Hu}, {Hunkeler}, {Ivezi{\'c}}, {Jain}, {Jenness}, {Kanarek}, {Kendrew},
  {Kern}, {Kerzendorf}, {Khvalko}, {King}, {Kirkby}, {Kulkarni}, {Kumar},
  {Lee}, {Lenz}, {Littlefair}, {Ma}, {Macleod}, {Mastropietro}, {McCully},
  {Montagnac}, {Morris}, {Mueller}, {Mumford}, {Muna}, {Murphy}, {Nelson},
  {Nguyen}, {Ninan}, {N{\"o}the}, {Ogaz}, {Oh}, {Parejko}, {Parley}, {Pascual},
  {Patil}, {Patil}, {Plunkett}, {Prochaska}, {Rastogi}, {Reddy Janga},
  {Sabater}, {Sakurikar}, {Seifert}, {Sherbert}, {Sherwood-Taylor}, {Shih},
  {Sick}, {Silbiger}, {Singanamalla}, {Singer}, {Sladen}, {Sooley},
  {Sornarajah}, {Streicher}, {Teuben}, {Thomas}, {Tremblay}, {Turner},
  {Terr{\'o}n}, {van Kerkwijk}, {de la Vega}, {Watkins}, {Weaver}, {Whitmore},
  {Woillez}, {Zabalza}, \& {Astropy Contributors}}]{2018AJ....156..123A}
{Astropy Collaboration}, {Price-Whelan}, A.~M., {Sip{\H{o}}cz}, B.~M., {et~al.}
  2018, \aj, 156, 123, \dodoi{10.3847/1538-3881/aabc4f}

\bibitem[{{Bianchi} {et~al.}(2005){Bianchi}, {Guainazzi}, {Matt}, {Chiaberge},
  {Iwasawa}, {Fiore}, \& {Maiolino}}]{2005A&A...442..185B}
{Bianchi}, S., {Guainazzi}, M., {Matt}, G., {et~al.} 2005, \aap, 442, 185,
  \dodoi{10.1051/0004-6361:20053389}

\bibitem[{{Buchner}(2021)}]{2021JOSS....6.3001B}
{Buchner}, J. 2021, The Journal of Open Source Software, 6, 3001,
  \dodoi{10.21105/joss.03001}

\bibitem[{{Chainakun} {et~al.}(2019){Chainakun}, {Watcharangkool}, {Young}, \&
  {Hancock}}]{2019MNRAS.487..667C}
{Chainakun}, P., {Watcharangkool}, A., {Young}, A.~J., \& {Hancock}, S. 2019,
  \mnras, 487, 667, \dodoi{10.1093/mnras/stz1319}

\bibitem[{{Connolly} {et~al.}(2014){Connolly}, {McHardy}, \&
  {Dwelly}}]{2014MNRAS.440.3503C}
{Connolly}, S.~D., {McHardy}, I.~M., \& {Dwelly}, T. 2014, \mnras, 440, 3503,
  \dodoi{10.1093/mnras/stu546}

\bibitem[{{Denney} {et~al.}(2014){Denney}, {De Rosa}, {Croxall}, {Gupta},
  {Bentz}, {Fausnaugh}, {Grier}, {Martini}, {Mathur}, {Peterson}, {Pogge}, \&
  {Shappee}}]{2014ApJ...796..134D}
{Denney}, K.~D., {De Rosa}, G., {Croxall}, K., {et~al.} 2014, \apj, 796, 134,
  \dodoi{10.1088/0004-637X/796/2/134}

\bibitem[{{Dexter} {et~al.}(2019){Dexter}, {Xin}, {Shen}, {Grier}, {Liu},
  {Gezari}, {McGreer}, {Brand t}, {Hall}, {Horne}, {Simm}, {Merloni}, {Green},
  {Vivek}, {Trump}, {Homayouni}, {Peterson}, {Schneider}, {Kinemuchi}, {Pan},
  \& {Bizyaev}}]{2019ApJ...885...44D}
{Dexter}, J., {Xin}, S., {Shen}, Y., {et~al.} 2019, \apj, 885, 44,
  \dodoi{10.3847/1538-4357/ab4354}

\bibitem[{{Di Gesu} {et~al.}(2014){Di Gesu}, {Costantini}, {Piconcelli},
  {Ebrero}, {Mehdipour}, \& {Kaastra}}]{2014A&A...563A..95D}
{Di Gesu}, L., {Costantini}, E., {Piconcelli}, E., {et~al.} 2014, \aap, 563,
  A95, \dodoi{10.1051/0004-6361/201322916}

\bibitem[{{Dodd} {et~al.}(2021){Dodd}, {Law-Smith}, {Auchettl}, {Ramirez-Ruiz},
  \& {Foley}}]{2021ApJ...907L..21D}
{Dodd}, S.~A., {Law-Smith}, J. A.~P., {Auchettl}, K., {Ramirez-Ruiz}, E., \&
  {Foley}, R.~J. 2021, \apjl, 907, L21, \dodoi{10.3847/2041-8213/abd852}

\bibitem[{{Dong} \& {De Robertis}(2006)}]{2006AJ....131.1236D}
{Dong}, X.~Y., \& {De Robertis}, M.~M. 2006, \aj, 131, 1236,
  \dodoi{10.1086/499334}

\bibitem[{{Ehler} {et~al.}(2018){Ehler}, {Gonzalez}, \&
  {Gallo}}]{2018MNRAS.478.4214E}
{Ehler}, H.~J.~S., {Gonzalez}, A.~G., \& {Gallo}, L.~C. 2018, \mnras, 478,
  4214, \dodoi{10.1093/mnras/sty1306}

\bibitem[{{Eracleous} \& {Halpern}(2001)}]{2001ApJ...554..240E}
{Eracleous}, M., \& {Halpern}, J.~P. 2001, \apj, 554, 240,
  \dodoi{10.1086/321331}

\bibitem[{{Fausnaugh} {et~al.}(2017){Fausnaugh}, {Grier}, {Bentz}, {Denney},
  {De Rosa}, {Peterson}, {Kochanek}, {Pogge}, {Adams}, {Barth}, {Beatty},
  {Bhattacharjee}, {Borman}, {Boroson}, {Bottorff}, {Brown}, {Brown},
  {Brotherton}, {Coker}, {Crawford}, {Croxall}, {Eftekharzadeh}, {Eracleous},
  {Joner}, {Henderson}, {Holoien}, {Horne}, {Hutchison}, {Kaspi}, {Kim},
  {King}, {Li}, {Lochhaas}, {Ma}, {MacInnis}, {Manne-Nicholas}, {Mason},
  {Montuori}, {Mosquera}, {Mudd}, {Musso}, {Nazarov}, {Nguyen}, {Okhmat},
  {Onken}, {Ou-Yang}, {Pancoast}, {Pei}, {Penny}, {Poleski}, {Rafter},
  {Romero-Colmenero}, {Runnoe}, {Sand}, {Schimoia}, {Sergeev}, {Shappee},
  {Simonian}, {Somers}, {Spencer}, {Starkey}, {Stevens}, {Tayar}, {Treu},
  {Valenti}, {Van Saders}, {Villanueva}, {Villforth}, {Weiss}, {Winkler}, \&
  {Zhu}}]{2017ApJ...840...97F}
{Fausnaugh}, M.~M., {Grier}, C.~J., {Bentz}, M.~C., {et~al.} 2017, \apj, 840,
  97, \dodoi{10.3847/1538-4357/aa6d52}

\bibitem[{{Feng} {et~al.}(2021){Feng}, {Hu}, {Li}, {Liu}, {Bai}, {Xing},
  {Wang}, {Yang}, {Xiao}, \& {Lu}}]{2021ApJ...909...18F}
{Feng}, H.-C., {Hu}, C., {Li}, S.-S., {et~al.} 2021, \apj, 909, 18,
  \dodoi{10.3847/1538-4357/abd851}

\bibitem[{{Fernandez} {et~al.}(2022){Fernandez}, {Secrest}, {Johnson},
  {Schmitt}, {Fischer}, {Cigan}, \& {Dorland}}]{2022arXiv220105152F}
{Fernandez}, L.~C., {Secrest}, N.~J., {Johnson}, M.~C., {et~al.} 2022, arXiv
  e-prints, arXiv:2201.05152.
\newblock \doarXiv{2201.05152}

\bibitem[{{Frederick} {et~al.}(2019){Frederick}, {Gezari}, {Graham}, {Cenko},
  {van Velzen}, {Stern}, {Blagorodnova}, {Kulkarni}, {Yan}, {De}, {Fremling},
  {Hung}, {Kara}, {Shupe}, {Ward}, {Bellm}, {Dekany}, {Duev}, {Feindt},
  {Giomi}, {Kupfer}, {Laher}, {Masci}, {Miller}, {Neill}, {Ngeow}, {Patterson},
  {Porter}, {Rusholme}, {Sollerman}, \& {Walters}}]{2019ApJ...883...31F}
{Frederick}, S., {Gezari}, S., {Graham}, M.~J., {et~al.} 2019, \apj, 883, 31,
  \dodoi{10.3847/1538-4357/ab3a38}

\bibitem[{{Graham} {et~al.}(2020){Graham}, {Ross}, {Stern}, {Drake},
  {McKernan}, {Ford}, {Djorgovski}, {Mahabal}, {Glikman}, {Larson}, \&
  {Christensen}}]{2020MNRAS.491.4925G}
{Graham}, M.~J., {Ross}, N.~P., {Stern}, D., {et~al.} 2020, \mnras, 491, 4925,
  \dodoi{10.1093/mnras/stz3244}

\bibitem[{{Greene} \& {Ho}(2007)}]{2007ApJ...667..131G}
{Greene}, J.~E., \& {Ho}, L.~C. 2007, \apj, 667, 131, \dodoi{10.1086/520497}

\bibitem[{{Grier} {et~al.}(2013){Grier}, {Martini}, {Watson}, {Peterson},
  {Bentz}, {Dasyra}, {Dietrich}, {Ferrarese}, {Pogge}, \&
  {Zu}}]{2013ApJ...773...90G}
{Grier}, C.~J., {Martini}, P., {Watson}, L.~C., {et~al.} 2013, \apj, 773, 90,
  \dodoi{10.1088/0004-637X/773/2/90}

\bibitem[{{Gu} \& {Cao}(2009)}]{2009MNRAS.399..349G}
{Gu}, M., \& {Cao}, X. 2009, \mnras, 399, 349,
  \dodoi{10.1111/j.1365-2966.2009.15277.x}

\bibitem[{{Guainazzi} {et~al.}(2005){Guainazzi}, {Fabian}, {Iwasawa}, {Matt},
  \& {Fiore}}]{2005MNRAS.356..295G}
{Guainazzi}, M., {Fabian}, A.~C., {Iwasawa}, K., {Matt}, G., \& {Fiore}, F.
  2005, \mnras, 356, 295, \dodoi{10.1111/j.1365-2966.2004.08448.x}

\bibitem[{{Guolo} {et~al.}(2021){Guolo}, {Ruschel-Dutra}, {Grupe}, {Peterson},
  {Storchi-Bergmann}, {Schimoia}, {Nemmen}, \&
  {Robinson}}]{2021MNRAS.508..144G}
{Guolo}, M., {Ruschel-Dutra}, D., {Grupe}, D., {et~al.} 2021, \mnras, 508, 144,
  \dodoi{10.1093/mnras/stab2550}

\bibitem[{{Heisler} {et~al.}(1997){Heisler}, {Lumsden}, \&
  {Bailey}}]{1997Natur.385..700H}
{Heisler}, C.~A., {Lumsden}, S.~L., \& {Bailey}, J.~A. 1997, \nat, 385, 700,
  \dodoi{10.1038/385700a0}

\bibitem[{{Ho}(2008)}]{2008ARA&A..46..475H}
{Ho}, L.~C. 2008, \araa, 46, 475,
  \dodoi{10.1146/annurev.astro.45.051806.110546}

\bibitem[{{Husemann} {et~al.}(2016){Husemann}, {Urrutia}, {Tremblay}, {Krumpe},
  {Dexter}, {Busch}, {Combes}, {Croom}, {Davis}, {Eckart}, {McElroy},
  {Perez-Torres}, {Powell}, \& {Scharw{\"a}chter}}]{2016A&A...593L...9H}
{Husemann}, B., {Urrutia}, T., {Tremblay}, G.~R., {et~al.} 2016, Astronomy and
  Astrophysics, 593, L9, \dodoi{10.1051/0004-6361/201629245}

\bibitem[{{Hutsem{\'e}kers} {et~al.}(2020){Hutsem{\'e}kers}, {Ag{\'\i}s
  Gonz{\'a}lez}, {Marin}, \& {Sluse}}]{2020A&A...644L...5H}
{Hutsem{\'e}kers}, D., {Ag{\'\i}s Gonz{\'a}lez}, B., {Marin}, F., \& {Sluse},
  D. 2020, \aap, 644, L5, \dodoi{10.1051/0004-6361/202039760}

\bibitem[{{Jana} {et~al.}(2020){Jana}, {Chatterjee}, {Kumari}, {Nandi}, {Naik},
  \& {Patra}}]{2020MNRAS.499.5396J}
{Jana}, A., {Chatterjee}, A., {Kumari}, N., {et~al.} 2020, \mnras, 499, 5396,
  \dodoi{10.1093/mnras/staa2552}

\bibitem[{{Jana} {et~al.}(2021){Jana}, {Kumari}, {Nandi}, {Naik}, {Chatterjee},
  {Jaisawal}, {Hayasaki}, \& {Ricci}}]{2021MNRAS.507..687J}
{Jana}, A., {Kumari}, N., {Nandi}, P., {et~al.} 2021, \mnras, 507, 687,
  \dodoi{10.1093/mnras/stab2155}

\bibitem[{{Jarrett} {et~al.}(2011){Jarrett}, {Cohen}, {Masci}, {Wright},
  {Stern}, {Benford}, {Blain}, {Carey}, {Cutri}, {Eisenhardt}, {Lonsdale},
  {Mainzer}, {Marsh}, {Padgett}, {Petty}, {Ressler}, {Skrutskie}, {Stanford},
  {Surace}, {Tsai}, {Wheelock}, \& {Yan}}]{2011ApJ...735..112J}
{Jarrett}, T.~H., {Cohen}, M., {Masci}, F., {et~al.} 2011, \apj, 735, 112,
  \dodoi{10.1088/0004-637X/735/2/112}

\bibitem[{{Jayasinghe} {et~al.}(2019){Jayasinghe}, {Stanek}, {Kochanek},
  {Shappee}, {Holoien}, {Thompson}, {Prieto}, {Dong}, {Pawlak}, {Pejcha},
  {Shields}, {Pojmanski}, {Otero}, {Hurst}, {Britt}, \&
  {Will}}]{2019MNRAS.485..961J}
{Jayasinghe}, T., {Stanek}, K.~Z., {Kochanek}, C.~S., {et~al.} 2019, \mnras,
  485, 961, \dodoi{10.1093/mnras/stz444}

\bibitem[{{Jiang} {et~al.}(2012){Jiang}, {Zhou}, {Ho}, {Yuan}, {Wang}, {Dong},
  {Jiang}, {Ji}, \& {Tian}}]{2012ApJ...759L..31J}
{Jiang}, N., {Zhou}, H.-Y., {Ho}, L.~C., {et~al.} 2012, \apjl, 759, L31,
  \dodoi{10.1088/2041-8205/759/2/L31}

\bibitem[{{Kellermann} {et~al.}(1989){Kellermann}, {Sramek}, {Schmidt},
  {Shaffer}, \& {Green}}]{1989AJ.....98.1195K}
{Kellermann}, K.~I., {Sramek}, R., {Schmidt}, M., {Shaffer}, D.~B., \& {Green},
  R. 1989, \aj, 98, 1195, \dodoi{10.1086/115207}

\bibitem[{{Kharb} {et~al.}(2017){Kharb}, {Lal}, \&
  {Merritt}}]{2017NatAs...1..727K}
{Kharb}, P., {Lal}, D.~V., \& {Merritt}, D. 2017, Nature Astronomy, 1, 727,
  \dodoi{10.1038/s41550-017-0256-4}

\bibitem[{{Koay} {et~al.}(2016){Koay}, {Vestergaard}, {Bignall}, {Reynolds}, \&
  {Peterson}}]{2016MNRAS.460..304K}
{Koay}, J.~Y., {Vestergaard}, M., {Bignall}, H.~E., {Reynolds}, C., \&
  {Peterson}, B.~M. 2016, \mnras, 460, 304, \dodoi{10.1093/mnras/stw975}

\bibitem[{{Kochanek} {et~al.}(2017){Kochanek}, {Shappee}, {Stanek}, {Holoien},
  {Thompson}, {Prieto}, {Dong}, {Shields}, {Will}, {Britt}, {Perzanowski}, \&
  {Pojma{\'n}ski}}]{2017PASP..129j4502K}
{Kochanek}, C.~S., {Shappee}, B.~J., {Stanek}, K.~Z., {et~al.} 2017, \pasp,
  129, 104502, \dodoi{10.1088/1538-3873/aa80d9}

\bibitem[{{Kokubo} \& {Minezaki}(2020)}]{2020MNRAS.491.4615K}
{Kokubo}, M., \& {Minezaki}, T. 2020, \mnras, 491, 4615,
  \dodoi{10.1093/mnras/stz3397}

\bibitem[{{Kollatschny} {et~al.}(2018){Kollatschny}, {Ochmann}, {Zetzl},
  {Haas}, {Chelouche}, {Kaspi}, {Pozo Nu{\~n}ez}, \&
  {Grupe}}]{2018A&A...619A.168K}
{Kollatschny}, W., {Ochmann}, M.~W., {Zetzl}, M., {et~al.} 2018, \aap, 619,
  A168, \dodoi{10.1051/0004-6361/201833727}

\bibitem[{{Kollatschny} \& {Zetzl}(2010)}]{2010A&A...522A..36K}
{Kollatschny}, W., \& {Zetzl}, M. 2010, \aap, 522, A36,
  \dodoi{10.1051/0004-6361/200913463}

\bibitem[{{Kollatschny} {et~al.}(2020){Kollatschny}, {Grupe}, {Parker},
  {Ochmann}, {Schartel}, {Herwig}, {Komossa}, {Romero-Colmenero}, \&
  {Santos-Lleo}}]{2020A&A...638A..91K}
{Kollatschny}, W., {Grupe}, D., {Parker}, M.~L., {et~al.} 2020, \aap, 638, A91,
  \dodoi{10.1051/0004-6361/202037897}

\bibitem[{{Koshida} {et~al.}(2014){Koshida}, {Minezaki}, {Yoshii}, {Kobayashi},
  {Sakata}, {Sugawara}, {Enya}, {Suganuma}, {Tomita}, {Aoki}, \&
  {Peterson}}]{2014ApJ...788..159K}
{Koshida}, S., {Minezaki}, T., {Yoshii}, Y., {et~al.} 2014, \apj, 788, 159,
  \dodoi{10.1088/0004-637X/788/2/159}

\bibitem[{{Koss} {et~al.}(2017){Koss}, {Trakhtenbrot}, {Ricci}, {Lamperti},
  {Oh}, {Berney}, {Schawinski}, {Balokovi{\'c}}, {Baronchelli}, {Crenshaw},
  {Fischer}, {Gehrels}, {Harrison}, {Hashimoto}, {Hogg}, {Ichikawa}, {Masetti},
  {Mushotzky}, {Sartori}, {Stern}, {Treister}, {Ueda}, {Veilleux}, \&
  {Winter}}]{2017ApJ...850...74K}
{Koss}, M., {Trakhtenbrot}, B., {Ricci}, C., {et~al.} 2017, \apj, 850, 74,
  \dodoi{10.3847/1538-4357/aa8ec9}

\bibitem[{{LaMassa} {et~al.}(2010){LaMassa}, {Heckman}, {Ptak}, {Martins},
  {Wild}, \& {Sonnentrucker}}]{2010ApJ...720..786L}
{LaMassa}, S.~M., {Heckman}, T.~M., {Ptak}, A., {et~al.} 2010, \apj, 720, 786,
  \dodoi{10.1088/0004-637X/720/1/786}

\bibitem[{{Liu} {et~al.}(2021){Liu}, {Wu}, {Xue}, {Wang}, {Yang}, {Guo}, \&
  {He}}]{2021RAA....21..199L}
{Liu}, H., {Wu}, Q.-W., {Xue}, Y.-Q., {et~al.} 2021, Research in Astronomy and
  Astrophysics, 21, 199, \dodoi{10.1088/1674-4527/21/8/199}

\bibitem[{{Lyu} {et~al.}(2021){Lyu}, {Yan}, {Yu}, \&
  {Wu}}]{2021MNRAS.506.4188L}
{Lyu}, B., {Yan}, Z., {Yu}, W., \& {Wu}, Q. 2021, \mnras, 506, 4188,
  \dodoi{10.1093/mnras/stab1581}

\bibitem[{{Lyu} {et~al.}(2019){Lyu}, {Rieke}, \& {Smith}}]{2019ApJ...886...33L}
{Lyu}, J., {Rieke}, G.~H., \& {Smith}, P.~S. 2019, \apj, 886, 33,
  \dodoi{10.3847/1538-4357/ab481d}

\bibitem[{{MacLeod} {et~al.}(2019){MacLeod}, {Green}, {Anderson}, {Bruce},
  {Eracleous}, {Graham}, {Homan}, {Lawrence}, {LeBleu}, {Ross}, {Ruan},
  {Runnoe}, {Stern}, {Burgett}, {Chambers}, {Kaiser}, {Magnier}, \&
  {Metcalfe}}]{2019ApJ...874....8M}
{MacLeod}, C.~L., {Green}, P.~J., {Anderson}, S.~F., {et~al.} 2019, \apj, 874,
  8, \dodoi{10.3847/1538-4357/ab05e2}

\bibitem[{{Mainzer} {et~al.}(2014){Mainzer}, {Bauer}, {Cutri}, {Grav},
  {Masiero}, {Beck}, {Clarkson}, {Conrow}, {Dailey}, {Eisenhardt}, {Fabinsky},
  {Fajardo-Acosta}, {Fowler}, {Gelino}, {Grillmair}, {Heinrichsen}, {Kendall},
  {Kirkpatrick}, {Liu}, {Masci}, {McCallon}, {Nugent}, {Papin}, {Rice},
  {Royer}, {Ryan}, {Sevilla}, {Sonnett}, {Stevenson}, {Thompson}, {Wheelock},
  {Wiemer}, {Wittman}, {Wright}, \& {Yan}}]{2014ApJ...792...30M}
{Mainzer}, A., {Bauer}, J., {Cutri}, R.~M., {et~al.} 2014, \apj, 792, 30,
  \dodoi{10.1088/0004-637X/792/1/30}

\bibitem[{{Mandal} {et~al.}(2021){Mandal}, {Rakshit}, {Stalin}, {Wylezalek},
  {Patig}, {Sagar}, {Mathew}, {Muneer}, \& {Pal}}]{2021MNRAS.501.3905M}
{Mandal}, A.~K., {Rakshit}, S., {Stalin}, C.~S., {et~al.} 2021, \mnras, 501,
  3905, \dodoi{10.1093/mnras/staa3828}

\bibitem[{{Marchese} {et~al.}(2012){Marchese}, {Braito}, {Della Ceca},
  {Caccianiga}, \& {Severgnini}}]{2012MNRAS.421.1803M}
{Marchese}, E., {Braito}, V., {Della Ceca}, R., {Caccianiga}, A., \&
  {Severgnini}, P. 2012, \mnras, 421, 1803,
  \dodoi{10.1111/j.1365-2966.2012.20445.x}

\bibitem[{{Marin} {et~al.}(2019){Marin}, {Hutsem{\'e}kers}, \& {Ag{\'\i}s
  Gonz{\'a}lez}}]{2019sf2a.conf..509M}
{Marin}, F., {Hutsem{\'e}kers}, D., \& {Ag{\'\i}s Gonz{\'a}lez}, B. 2019, in
  SF2A-2019: Proceedings of the Annual meeting of the French Society of
  Astronomy and Astrophysics, ed. P.~{Di Matteo}, O.~{Creevey}, A.~{Crida},
  G.~{Kordopatis}, J.~{Malzac}, J.~B. {Marquette}, M.~{N'Diaye}, \& O.~{Venot},
  Di.
\newblock \doarXiv{1909.02801}

\bibitem[{{Marin} {et~al.}(2013){Marin}, {Porquet}, {Goosmann},
  {Dov{\v{c}}iak}, {Muleri}, {Grosso}, \& {Karas}}]{2013MNRAS.436.1615M}
{Marin}, F., {Porquet}, D., {Goosmann}, R.~W., {et~al.} 2013, \mnras, 436,
  1615, \dodoi{10.1093/mnras/stt1677}

\bibitem[{{Marinucci} {et~al.}(2020){Marinucci}, {Bianchi}, {Braito}, {De
  Marco}, {Matt}, {Middei}, {Nardini}, \& {Reeves}}]{2020MNRAS.496.3412M}
{Marinucci}, A., {Bianchi}, S., {Braito}, V., {et~al.} 2020, \mnras, 496, 3412,
  \dodoi{10.1093/mnras/staa1683}

\bibitem[{{Marinucci} {et~al.}(2011){Marinucci}, {Risaliti}, {Elvis},
  {Bianchi}, \& {Matt}}]{2011AAS...21822816M}
{Marinucci}, A., {Risaliti}, G., {Elvis}, M., {Bianchi}, S., \& {Matt}, G.
  2011, in American Astronomical Society Meeting Abstracts, Vol. 218, American
  Astronomical Society Meeting Abstracts \#218, 228.16

\bibitem[{{Matt} {et~al.}(2003){Matt}, {Guainazzi}, \&
  {Maiolino}}]{2003MNRAS.342..422M}
{Matt}, G., {Guainazzi}, M., \& {Maiolino}, R. 2003, \mnras, 342, 422,
  \dodoi{10.1046/j.1365-8711.2003.06539.x}

\bibitem[{{Matthews} {et~al.}(2020){Matthews}, {Knigge}, {Higginbottom},
  {Long}, {Sim}, {Mangham}, {Parkinson}, \& {Hewitt}}]{2020MNRAS.492.5540M}
{Matthews}, J.~H., {Knigge}, C., {Higginbottom}, N., {et~al.} 2020, \mnras,
  492, 5540, \dodoi{10.1093/mnras/staa136}

\bibitem[{{Merloni} {et~al.}(2003){Merloni}, {Heinz}, \& {di
  Matteo}}]{2003MNRAS.345.1057M}
{Merloni}, A., {Heinz}, S., \& {di Matteo}, T. 2003, \mnras, 345, 1057,
  \dodoi{10.1046/j.1365-2966.2003.07017.x}

\bibitem[{{Miller} {et~al.}(2019){Miller}, {Kammoun}, {Ludlam}, {Gendreau},
  {Arzoumanian}, {Cackett}, \& {Tombesi}}]{2019ApJ...884..106M}
{Miller}, J.~M., {Kammoun}, E., {Ludlam}, R.~M., {et~al.} 2019, \apj, 884, 106,
  \dodoi{10.3847/1538-4357/ab3e05}

\bibitem[{{Minezaki} {et~al.}(2019){Minezaki}, {Yoshii}, {Kobayashi},
  {Sugawara}, {Sakata}, {Enya}, {Koshida}, {Tomita}, {Suganuma}, {Aoki}, \&
  {Peterson}}]{2019ApJ...886..150M}
{Minezaki}, T., {Yoshii}, Y., {Kobayashi}, Y., {et~al.} 2019, \apj, 886, 150,
  \dodoi{10.3847/1538-4357/ab4f7b}

\bibitem[{{Netzer}(2013)}]{2013peag.book.....N}
{Netzer}, H. 2013, {The Physics and Evolution of Active Galactic Nuclei}

\bibitem[{{Netzer}(2015)}]{2015ARA&A..53..365N}
---. 2015, \araa, 53, 365, \dodoi{10.1146/annurev-astro-082214-122302}

\bibitem[{{Noda} \& {Done}(2018)}]{2018MNRAS.480.3898N}
{Noda}, H., \& {Done}, C. 2018, \mnras, 480, 3898,
  \dodoi{10.1093/mnras/sty2032}

\bibitem[{{Noda} {et~al.}(2020){Noda}, {Kawamuro}, {Kokubo}, \&
  {Minezaki}}]{2020MNRAS.495.2921N}
{Noda}, H., {Kawamuro}, T., {Kokubo}, M., \& {Minezaki}, T. 2020, \mnras, 495,
  2921, \dodoi{10.1093/mnras/staa1376}

\bibitem[{{Oh} {et~al.}(2018){Oh}, {Koss}, {Markwardt}, {Schawinski},
  {Baumgartner}, {Barthelmy}, {Cenko}, {Gehrels}, {Mushotzky}, {Petulante},
  {Ricci}, {Lien}, \& {Trakhtenbrot}}]{2018ApJS..235....4O}
{Oh}, K., {Koss}, M., {Markwardt}, C.~B., {et~al.} 2018, \apjs, 235, 4,
  \dodoi{10.3847/1538-4365/aaa7fd}

\bibitem[{{Oknyansky} {et~al.}(2018){Oknyansky}, {Lipunov}, {Gorbovskoy},
  {Winkler}, {van Wyk}, {Tsygankov}, \& {Buckley}}]{2018ATel11915....1O}
{Oknyansky}, V.~L., {Lipunov}, V.~M., {Gorbovskoy}, E.~S., {et~al.} 2018, The
  Astronomer's Telegram, 11915, 1

\bibitem[{{Oknyansky} {et~al.}(2017){Oknyansky}, {Gaskell}, {Huseynov},
  {Mikailov}, {Lipunov}, {Shatsky}, {Tsygankov}, {Gorbovskoy}, {Tatarnikov},
  {Metlov}, {Malanchev}, {Brotherton}, {Kasper}, {Du}, {Chen}, {Burlak},
  {Buckley}, {Rebolo}, {Serra-Ricart}, {Podesta}, \&
  {Levato}}]{2017OAP....30..117O}
{Oknyansky}, V.~L., {Gaskell}, C.~M., {Huseynov}, N.~A., {et~al.} 2017, Odessa
  Astronomical Publications, 30, 117, \dodoi{10.18524/1810-4215.2017.30.114366}

\bibitem[{{Oknyansky} {et~al.}(2020){Oknyansky}, {Winkler}, {Tsygankov},
  {Lipunov}, {Gorbovskoy}, {van Wyk}, {Buckley}, {Jiang}, \&
  {Tyurina}}]{2020MNRAS.498..718O}
{Oknyansky}, V.~L., {Winkler}, H., {Tsygankov}, S.~S., {et~al.} 2020, \mnras,
  498, 718, \dodoi{10.1093/mnras/staa1552}

\bibitem[{{Onori} {et~al.}(2017){Onori}, {Ricci}, {La Franca}, {Bianchi},
  {Bongiorno}, {Brusa}, {Fiore}, {Maiolino}, {Marconi}, {Sani}, \&
  {Vignali}}]{2017MNRAS.468L..97O}
{Onori}, F., {Ricci}, F., {La Franca}, F., {et~al.} 2017, \mnras, 468, L97,
  \dodoi{10.1093/mnrasl/slx032}

\bibitem[{{Osterbrock}(1981)}]{1981ApJ...249..462O}
{Osterbrock}, D.~E. 1981, The Astrophysical Journal, 249, 462,
  \dodoi{10.1086/159306}

\bibitem[{{Osterbrock} \& {Koski}(1976)}]{1976MNRAS.176P..61O}
{Osterbrock}, D.~E., \& {Koski}, A.~T. 1976, Monthly Notices of the Royal
  Astronomical Society, 176, 61P, \dodoi{10.1093/mnras/176.1.61P}

\bibitem[{{Pal} {et~al.}(2017){Pal}, {Dewangan}, {Connolly}, \&
  {Misra}}]{2017MNRAS.466.1777P}
{Pal}, M., {Dewangan}, G.~C., {Connolly}, S.~D., \& {Misra}, R. 2017, \mnras,
  466, 1777, \dodoi{10.1093/mnras/stw3173}

\bibitem[{{Pancoast} {et~al.}(2014){Pancoast}, {Brewer}, {Treu}, {Park},
  {Barth}, {Bentz}, \& {Woo}}]{2014MNRAS.445.3073P}
{Pancoast}, A., {Brewer}, B.~J., {Treu}, T., {et~al.} 2014, \mnras, 445, 3073,
  \dodoi{10.1093/mnras/stu1419}

\bibitem[{{Peterson} {et~al.}(1998){Peterson}, {Wanders}, {Horne}, {Collier},
  {Alexander}, {Kaspi}, \& {Maoz}}]{1998PASP..110..660P}
{Peterson}, B.~M., {Wanders}, I., {Horne}, K., {et~al.} 1998, \pasp, 110, 660,
  \dodoi{10.1086/316177}

\bibitem[{{Peterson} {et~al.}(2005){Peterson}, {Bentz}, {Desroches},
  {Filippenko}, {Ho}, {Kaspi}, {Laor}, {Maoz}, {Moran}, {Pogge}, \&
  {Quillen}}]{2005ApJ...632..799P}
{Peterson}, B.~M., {Bentz}, M.~C., {Desroches}, L.-B., {et~al.} 2005, \apj,
  632, 799, \dodoi{10.1086/444494}

\bibitem[{{Potts} \& {Villforth}(2021)}]{2021A&A...650A..33P}
{Potts}, B., \& {Villforth}, C. 2021, \aap, 650, A33,
  \dodoi{10.1051/0004-6361/202140597}

\bibitem[{{Rakshit} {et~al.}(2019){Rakshit}, {Johnson}, {Stalin}, {Gandhi}, \&
  {Hoenig}}]{2019MNRAS.483.2362R}
{Rakshit}, S., {Johnson}, A., {Stalin}, C.~S., {Gandhi}, P., \& {Hoenig}, S.
  2019, \mnras, 483, 2362, \dodoi{10.1093/mnras/sty3261}

\bibitem[{{Ricci} {et~al.}(2016){Ricci}, {Bauer}, {Arevalo}, {Boggs}, {Brandt},
  {Christensen}, {Craig}, {Gandhi}, {Hailey}, {Harrison}, {Koss}, {Markwardt},
  {Stern}, {Treister}, \& {Zhang}}]{2016ApJ...820....5R}
{Ricci}, C., {Bauer}, F.~E., {Arevalo}, P., {et~al.} 2016, \apj, 820, 5,
  \dodoi{10.3847/0004-637X/820/1/5}

\bibitem[{{Ricci} {et~al.}(2018){Ricci}, {Steiner}, {May}, {Garcia-Rissmann},
  \& {Menezes}}]{2018MNRAS.473.5334R}
{Ricci}, T.~V., {Steiner}, J.~E., {May}, D., {Garcia-Rissmann}, A., \&
  {Menezes}, R.~B. 2018, \mnras, 473, 5334, \dodoi{10.1093/mnras/stx2746}

\bibitem[{{Rivers} {et~al.}(2015){Rivers}, {Balokovi{\'c}}, {Ar{\'e}valo},
  {Bauer}, {Boggs}, {Brandt}, {Brightman}, {Christensen}, {Craig}, {Gandhi},
  {Hailey}, {Harrison}, {Koss}, {Ricci}, {Stern}, {Walton}, \&
  {Zhang}}]{2015ApJ...815...55R}
{Rivers}, E., {Balokovi{\'c}}, M., {Ar{\'e}valo}, P., {et~al.} 2015, \apj, 815,
  55, \dodoi{10.1088/0004-637X/815/1/55}

\bibitem[{{Ruan} {et~al.}(2019{\natexlab{a}}){Ruan}, {Anderson}, {Eracleous},
  {Green}, {Haggard}, {MacLeod}, {Runnoe}, \&
  {Sobolewska}}]{2019arXiv190904676R}
{Ruan}, J.~J., {Anderson}, S.~F., {Eracleous}, M., {et~al.} 2019{\natexlab{a}},
  arXiv e-prints, arXiv:1909.04676.
\newblock \doarXiv{1909.04676}

\bibitem[{{Ruan} {et~al.}(2019{\natexlab{b}}){Ruan}, {Anderson}, {Eracleous},
  {Green}, {Haggard}, {MacLeod}, {Runnoe}, \&
  {Sobolewska}}]{2019ApJ...883...76R}
---. 2019{\natexlab{b}}, \apj, 883, 76, \dodoi{10.3847/1538-4357/ab3c1a}

\bibitem[{{Runco} {et~al.}(2016){Runco}, {Cosens}, {Bennert}, {Scott},
  {Komossa}, {Malkan}, {Lazarova}, {Auger}, {Treu}, \&
  {Park}}]{2016ApJ...821...33R}
{Runco}, J.~N., {Cosens}, M., {Bennert}, V.~N., {et~al.} 2016, \apj, 821, 33,
  \dodoi{10.3847/0004-637X/821/1/33}

\bibitem[{{Runnoe} {et~al.}(2012){Runnoe}, {Brotherton}, \&
  {Shang}}]{2012MNRAS.426.2677R}
{Runnoe}, J.~C., {Brotherton}, M.~S., \& {Shang}, Z. 2012, \mnras, 426, 2677,
  \dodoi{10.1111/j.1365-2966.2012.21644.x}

\bibitem[{{Schimoia} {et~al.}(2015){Schimoia}, {Storchi-Bergmann}, {Grupe},
  {Eracleous}, {Peterson}, {Baldwin}, {Nemmen}, \&
  {Winge}}]{2015ApJ...800...63S}
{Schimoia}, J.~S., {Storchi-Bergmann}, T., {Grupe}, D., {et~al.} 2015, \apj,
  800, 63, \dodoi{10.1088/0004-637X/800/1/63}

\bibitem[{{Senarath} {et~al.}(2021){Senarath}, {Brown}, {Cluver}, {Jarrett},
  {Wolf}, {Ross}, {Lucey}, {Parkash}, \& {Hon}}]{2021MNRAS.503.2583S}
{Senarath}, M.~R., {Brown}, M. J.~I., {Cluver}, M.~E., {et~al.} 2021, \mnras,
  503, 2583, \dodoi{10.1093/mnras/stab393}

\bibitem[{{Sergeev} {et~al.}(2011){Sergeev}, {Klimanov}, {Doroshenko},
  {Efimov}, {Nazarov}, \& {Pronik}}]{2011MNRAS.410.1877S}
{Sergeev}, S.~G., {Klimanov}, S.~A., {Doroshenko}, V.~T., {et~al.} 2011,
  \mnras, 410, 1877, \dodoi{10.1111/j.1365-2966.2010.17569.x}

\bibitem[{{Sesar} {et~al.}(2007){Sesar}, {Ivezi{\'c}}, {Lupton}, {Juri{\'c}},
  {Gunn}, {Knapp}, {DeLee}, {Smith}, {Miknaitis}, {Lin}, {Tucker}, {Doi},
  {Tanaka}, {Fukugita}, {Holtzman}, {Kent}, {Yanny}, {Schlegel}, {Finkbeiner},
  {Padmanabhan}, {Rockosi}, {Bond}, {Lee}, {Stoughton}, {Jester}, {Harris},
  {Harding}, {Brinkmann}, {Schneider}, {York}, {Richmond}, \& {Vanden
  Berk}}]{2007AJ....134.2236S}
{Sesar}, B., {Ivezi{\'c}}, {\v{Z}}., {Lupton}, R.~H., {et~al.} 2007, \aj, 134,
  2236, \dodoi{10.1086/521819}

\bibitem[{{Shakura} \& {Sunyaev}(1973)}]{1973A&A....24..337S}
{Shakura}, N.~I., \& {Sunyaev}, R.~A. 1973, \aap, 500, 33

\bibitem[{{Shappee} {et~al.}(2014){Shappee}, {Prieto}, {Grupe}, {Kochanek},
  {Stanek}, {De Rosa}, {Mathur}, {Zu}, {Peterson}, {Pogge}, {Komossa}, {Im},
  {Jencson}, {Holoien}, {Basu}, {Beacom}, {Szczygie{\l}}, {Brimacombe},
  {Adams}, {Campillay}, {Choi}, {Contreras}, {Dietrich}, {Dubberley},
  {Elphick}, {Foale}, {Giustini}, {Gonzalez}, {Hawkins}, {Howell}, {Hsiao},
  {Koss}, {Leighly}, {Morrell}, {Mudd}, {Mullins}, {Nugent}, {Parrent},
  {Phillips}, {Pojmanski}, {Rosing}, {Ross}, {Sand}, {Terndrup}, {Valenti},
  {Walker}, \& {Yoon}}]{2014ApJ...788...48S}
{Shappee}, B.~J., {Prieto}, J.~L., {Grupe}, D., {et~al.} 2014, \apj, 788, 48,
  \dodoi{10.1088/0004-637X/788/1/48}

\bibitem[{{Sheng} {et~al.}(2017){Sheng}, {Wang}, {Jiang}, {Yang}, {Yan}, {Dou},
  \& {Peng}}]{2017ApJ...846L...7S}
{Sheng}, Z., {Wang}, T., {Jiang}, N., {et~al.} 2017, \apjl, 846, L7,
  \dodoi{10.3847/2041-8213/aa85de}

\bibitem[{{Sheng} {et~al.}(2020){Sheng}, {Wang}, {Jiang}, {Ding}, {Cai}, {Guo},
  {Sun}, {Dou}, \& {Yang}}]{2020ApJ...889...46S}
---. 2020, \apj, 889, 46, \dodoi{10.3847/1538-4357/ab5af9}

\bibitem[{{Smirnova} {et~al.}(2018){Smirnova}, {Moiseev}, \&
  {Dodonov}}]{2018MNRAS.481.4542S}
{Smirnova}, A.~A., {Moiseev}, A.~V., \& {Dodonov}, S.~N. 2018, \mnras, 481,
  4542, \dodoi{10.1093/mnras/sty2569}

\bibitem[{{Sobrino Figaredo} {et~al.}(2020){Sobrino Figaredo}, {Haas},
  {Ramolla}, {Chini}, {Blex}, {Hodapp}, {Murphy}, {Kollatschny}, {Chelouche},
  \& {Kaspi}}]{2020AJ....159..259S}
{Sobrino Figaredo}, C., {Haas}, M., {Ramolla}, M., {et~al.} 2020, \aj, 159,
  259, \dodoi{10.3847/1538-3881/ab89b1}

\bibitem[{{Stern} {et~al.}(2012){Stern}, {Assef}, {Benford}, {Blain}, {Cutri},
  {Dey}, {Eisenhardt}, {Griffith}, {Jarrett}, {Lake}, {Masci}, {Petty},
  {Stanford}, {Tsai}, {Wright}, {Yan}, {Harrison}, \&
  {Madsen}}]{2012ApJ...753...30S}
{Stern}, D., {Assef}, R.~J., {Benford}, D.~J., {et~al.} 2012, \apj, 753, 30,
  \dodoi{10.1088/0004-637X/753/1/30}

\bibitem[{{Stern} {et~al.}(2018){Stern}, {McKernan}, {Graham}, {Ford}, {Ross},
  {Meisner}, {Assef}, {Balokovi{\'c}}, {Brightman}, {Dey}, {Drake},
  {Djorgovski}, {Eisenhardt}, \& {Jun}}]{2018ApJ...864...27S}
{Stern}, D., {McKernan}, B., {Graham}, M.~J., {et~al.} 2018, \apj, 864, 27,
  \dodoi{10.3847/1538-4357/aac726}

\bibitem[{{Sun} {et~al.}(2018){Sun}, {Grier}, \&
  {Peterson}}]{2018ascl.soft05032S}
{Sun}, M., {Grier}, C.~J., \& {Peterson}, B.~M. 2018, {PyCCF: Python Cross
  Correlation Function for reverberation mapping studies}.
\newblock \doeprint{1805.032}

\bibitem[{{Tremaine} {et~al.}(2002){Tremaine}, {Gebhardt}, {Bender}, {Bower},
  {Dressler}, {Faber}, {Filippenko}, {Green}, {Grillmair}, {Ho}, {Kormendy},
  {Lauer}, {Magorrian}, {Pinkney}, \& {Richstone}}]{2002ApJ...574..740T}
{Tremaine}, S., {Gebhardt}, K., {Bender}, R., {et~al.} 2002, \apj, 574, 740,
  \dodoi{10.1086/341002}

\bibitem[{{Turner} {et~al.}(2018){Turner}, {Reeves}, {Braito}, {Lobban},
  {Kraemer}, \& {Miller}}]{2018MNRAS.481.2470T}
{Turner}, T.~J., {Reeves}, J.~N., {Braito}, V., {et~al.} 2018, \mnras, 481,
  2470, \dodoi{10.1093/mnras/sty2447}

\bibitem[{{Vasudevan} \& {Fabian}(2009)}]{2009MNRAS.392.1124V}
{Vasudevan}, R.~V., \& {Fabian}, A.~C. 2009, \mnras, 392, 1124,
  \dodoi{10.1111/j.1365-2966.2008.14108.x}

\bibitem[{{Wang} {et~al.}(2019){Wang}, {Xu}, {Wang}, {Zhang}, {Zheng}, \&
  {Wei}}]{2019ApJ...887...15W}
{Wang}, J., {Xu}, D.~W., {Wang}, Y., {et~al.} 2019, \apj, 887, 15,
  \dodoi{10.3847/1538-4357/ab4d90}

\bibitem[{{Wang} {et~al.}(2020){Wang}, {Xu}, \& {Wei}}]{2020ApJ...901....1W}
{Wang}, J., {Xu}, D.~W., \& {Wei}, J.~Y. 2020, \apj, 901, 1,
  \dodoi{10.3847/1538-4357/abaa48}

\bibitem[{{Winkler}(2021)}]{2021IAUS..356..122W}
{Winkler}, H. 2021, IAU Symposium, 356, 122, \dodoi{10.1017/S1743921320002719}

\bibitem[{{Winter} {et~al.}(2009){Winter}, {Mushotzky}, {Reynolds}, \&
  {Tueller}}]{2009ApJ...690.1322W}
{Winter}, L.~M., {Mushotzky}, R.~F., {Reynolds}, C.~S., \& {Tueller}, J. 2009,
  \apj, 690, 1322, \dodoi{10.1088/0004-637X/690/2/1322}

\bibitem[{{Wolf} {et~al.}(2020){Wolf}, {Golding}, {Hon}, \&
  {Onken}}]{2020MNRAS.499.1005W}
{Wolf}, C., {Golding}, J., {Hon}, W.~J., \& {Onken}, C.~A. 2020, \mnras, 499,
  1005, \dodoi{10.1093/mnras/staa2794}

\bibitem[{{Woo} \& {Urry}(2002)}]{2002ApJ...579..530W}
{Woo}, J.-H., \& {Urry}, C.~M. 2002, \apj, 579, 530, \dodoi{10.1086/342878}

\bibitem[{{Wright} {et~al.}(2010){Wright}, {Eisenhardt}, {Mainzer}, {Ressler},
  {Cutri}, {Jarrett}, {Kirkpatrick}, {Padgett}, {McMillan}, {Skrutskie},
  {Stanford}, {Cohen}, {Walker}, {Mather}, {Leisawitz}, {Gautier}, {McLean},
  {Benford}, {Lonsdale}, {Blain}, {Mendez}, {Irace}, {Duval}, {Liu}, {Royer},
  {Heinrichsen}, {Howard}, {Shannon}, {Kendall}, {Walsh}, {Larsen}, {Cardon},
  {Schick}, {Schwalm}, {Abid}, {Fabinsky}, {Naes}, \&
  {Tsai}}]{2010AJ....140.1868W}
{Wright}, E.~L., {Eisenhardt}, P. R.~M., {Mainzer}, A.~K., {et~al.} 2010, \aj,
  140, 1868, \dodoi{10.1088/0004-6256/140/6/1868}

\bibitem[{{Wu} \& {Gu}(2008)}]{2008ApJ...682..212W}
{Wu}, Q., \& {Gu}, M. 2008, \apj, 682, 212, \dodoi{10.1086/588187}

\bibitem[{{Xie} \& {Yuan}(2012)}]{2012MNRAS.427.1580X}
{Xie}, F.-G., \& {Yuan}, F. 2012, \mnras, 427, 1580,
  \dodoi{10.1111/j.1365-2966.2012.22030.x}

\bibitem[{{Yan} {et~al.}(2013){Yan}, {Donoso}, {Tsai}, {Stern}, {Assef},
  {Eisenhardt}, {Blain}, {Cutri}, {Jarrett}, {Stanford}, {Wright}, {Bridge}, \&
  {Riechers}}]{2013AJ....145...55Y}
{Yan}, L., {Donoso}, E., {Tsai}, C.-W., {et~al.} 2013, \aj, 145, 55,
  \dodoi{10.1088/0004-6256/145/3/55}

\bibitem[{{Yang} {et~al.}(2021{\natexlab{a}}){Yang}, {Paragi}, {Beswick},
  {Chen}, {van Bemmel}, {Wu}, {An}, {Wu}, {Fan}, {Oonk}, {Liu}, \&
  {Wang}}]{2021MNRAS.503.3886Y}
{Yang}, J., {Paragi}, Z., {Beswick}, R.~J., {et~al.} 2021{\natexlab{a}},
  \mnras, 503, 3886, \dodoi{10.1093/mnras/stab706}

\bibitem[{{Yang} {et~al.}(2021{\natexlab{b}}){Yang}, {van Bemmel}, {Paragi},
  {Komossa}, {Yuan}, {Yang}, {An}, {Koay}, {Reynolds}, {Oonk}, {Liu}, \&
  {Wu}}]{2021MNRAS.502L..61Y}
{Yang}, J., {van Bemmel}, I., {Paragi}, Z., {et~al.} 2021{\natexlab{b}},
  \mnras, 502, L61, \dodoi{10.1093/mnrasl/slab005}

\bibitem[{{Yang} {et~al.}(2018){Yang}, {Wu}, {Fan}, {Jiang}, {McGreer},
  {Shangguan}, {Yao}, {Wang}, {Joshi}, {Green}, {Wang}, {Feng}, {Fu}, {Yang},
  \& {Liu}}]{2018ApJ...862..109Y}
{Yang}, Q., {Wu}, X.-B., {Fan}, X., {et~al.} 2018, \apj, 862, 109,
  \dodoi{10.3847/1538-4357/aaca3a}

\bibitem[{{Yong} {et~al.}(2017){Yong}, {Webster}, {King}, {Bate}, {O'Dowd}, \&
  {Labrie}}]{2017PASA...34...42Y}
{Yong}, S.~Y., {Webster}, R.~L., {King}, A.~L., {et~al.} 2017, \pasa, 34, e042,
  \dodoi{10.1017/pasa.2017.37}

\bibitem[{{Yu} {et~al.}(2020){Yu}, {Shi}, {Chen}, {Chen}, {Li}, {Bing}, {Ge},
  {Riffel}, \& {Riffel}}]{2020MNRAS.498.3985Y}
{Yu}, X., {Shi}, Y., {Chen}, Y., {et~al.} 2020, \mnras, 498, 3985,
  \dodoi{10.1093/mnras/staa2627}

\bibitem[{{Yuan} \& {Narayan}(2014)}]{2014ARA&A..52..529Y}
{Yuan}, F., \& {Narayan}, R. 2014, \araa, 52, 529,
  \dodoi{10.1146/annurev-astro-082812-141003}

\bibitem[{{Zu} {et~al.}(2013){Zu}, {Kochanek}, {Koz{\l}owski}, \&
  {Udalski}}]{2013ApJ...765..106Z}
{Zu}, Y., {Kochanek}, C.~S., {Koz{\l}owski}, S., \& {Udalski}, A. 2013, \apj,
  765, 106, \dodoi{10.1088/0004-637X/765/2/106}

\bibitem[{{Zu} {et~al.}(2011){Zu}, {Kochanek}, \&
  {Peterson}}]{2011ApJ...735...80Z}
{Zu}, Y., {Kochanek}, C.~S., \& {Peterson}, B.~M. 2011, \apj, 735, 80,
  \dodoi{10.1088/0004-637X/735/2/80}

\end{thebibliography}


\appendix
\section{time lag analysis}
\label{lag_analysis}
\begin{figure}[h!]
\begin{tikzpicture}
    \matrix[matrix of nodes]{
    \includegraphics[width=0.45\textwidth]{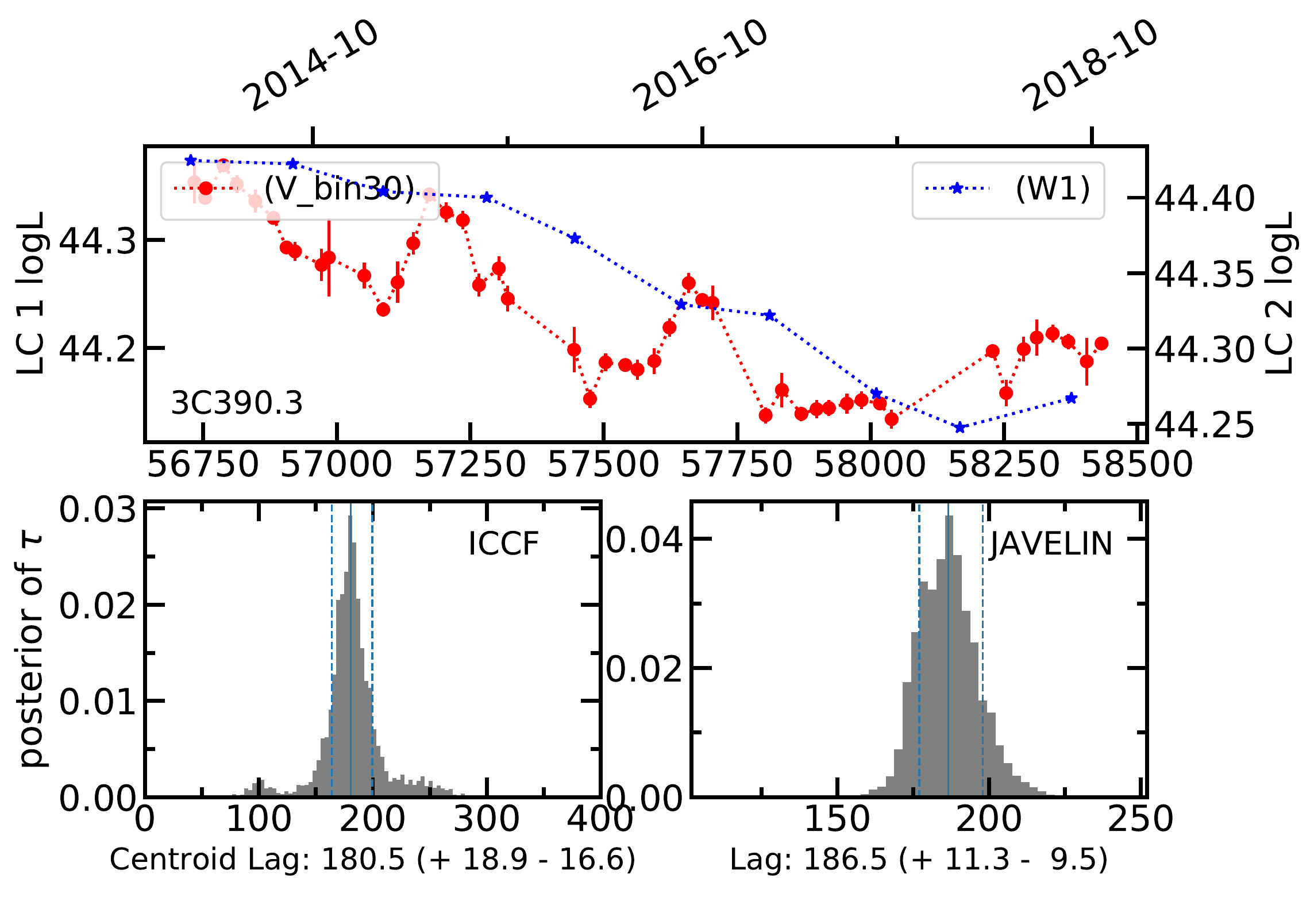}& \hspace{1em}  &
    \includegraphics[width=0.45\textwidth]{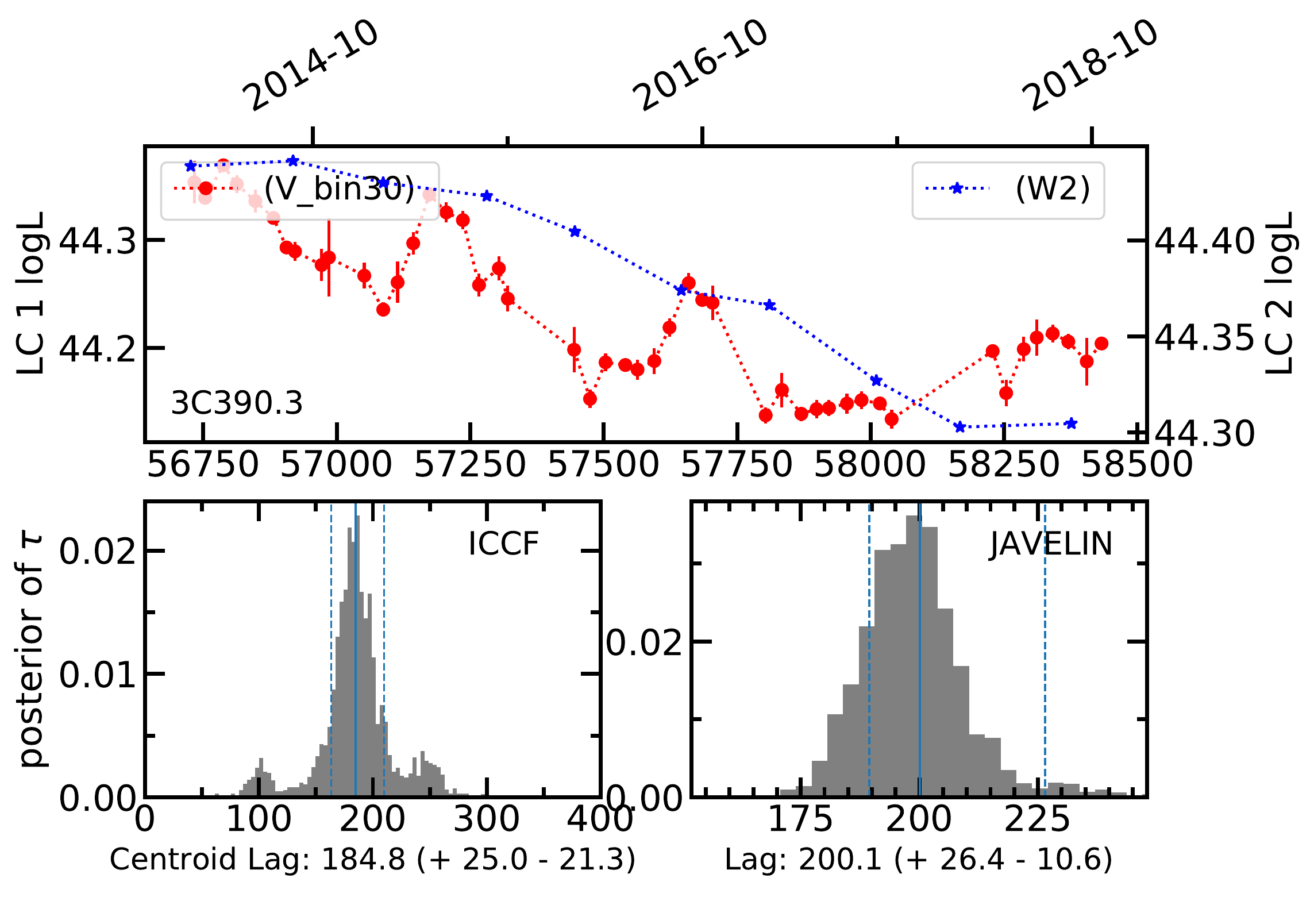} \\
    \includegraphics[width=0.45\textwidth]{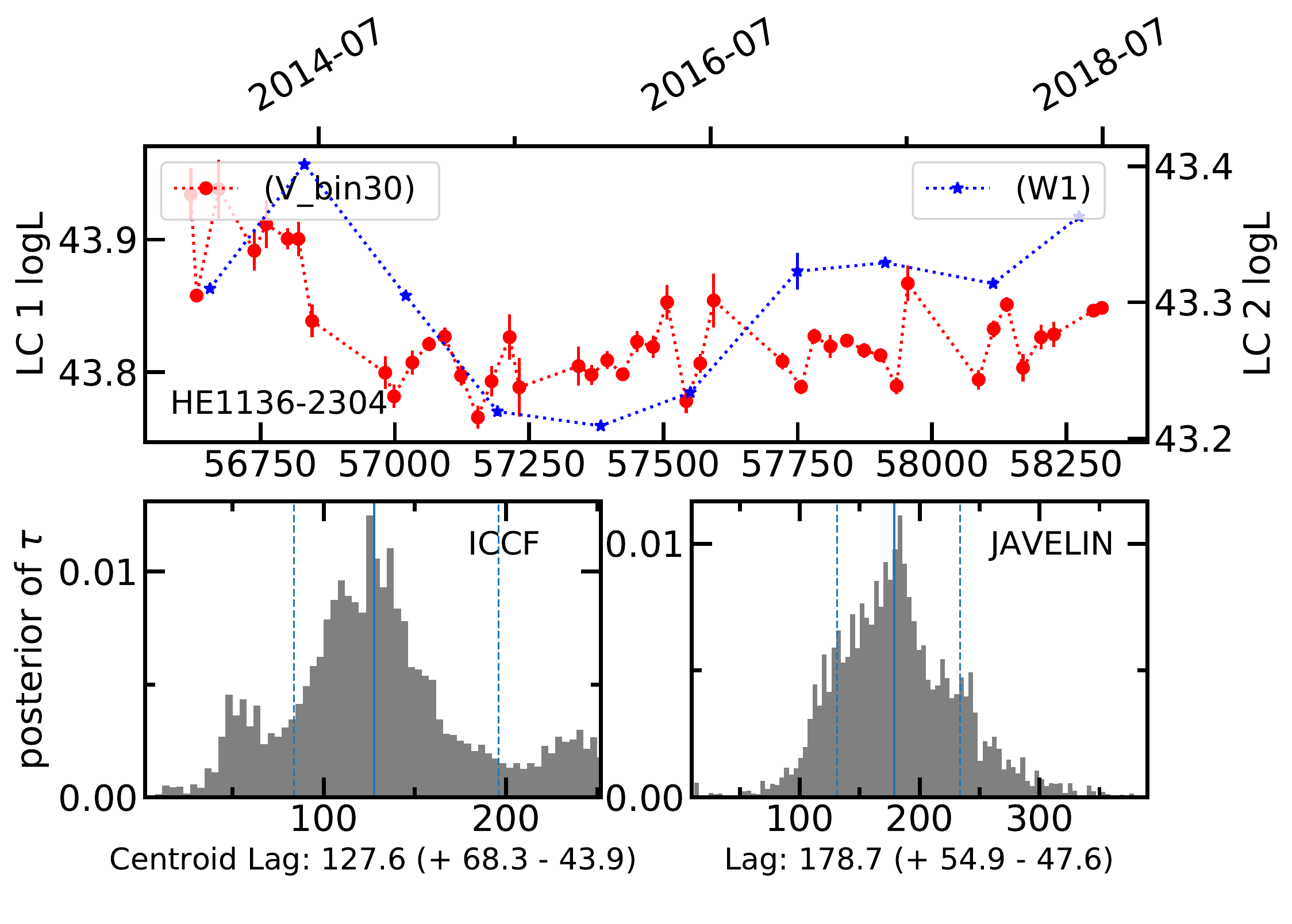}& \hspace{1em}  &
    \includegraphics[width=0.45\textwidth]{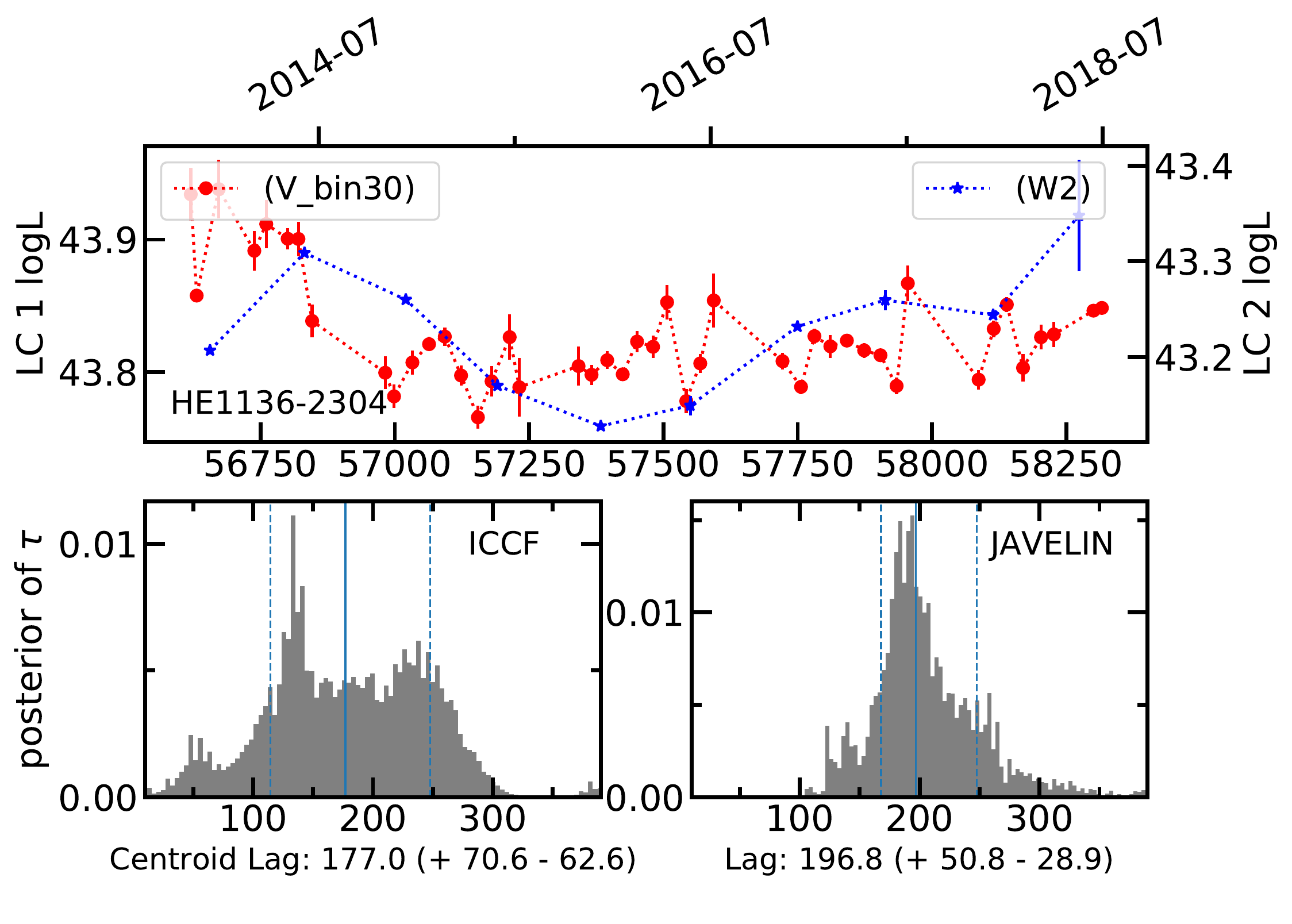} \\
    \includegraphics[width=0.45\textwidth]{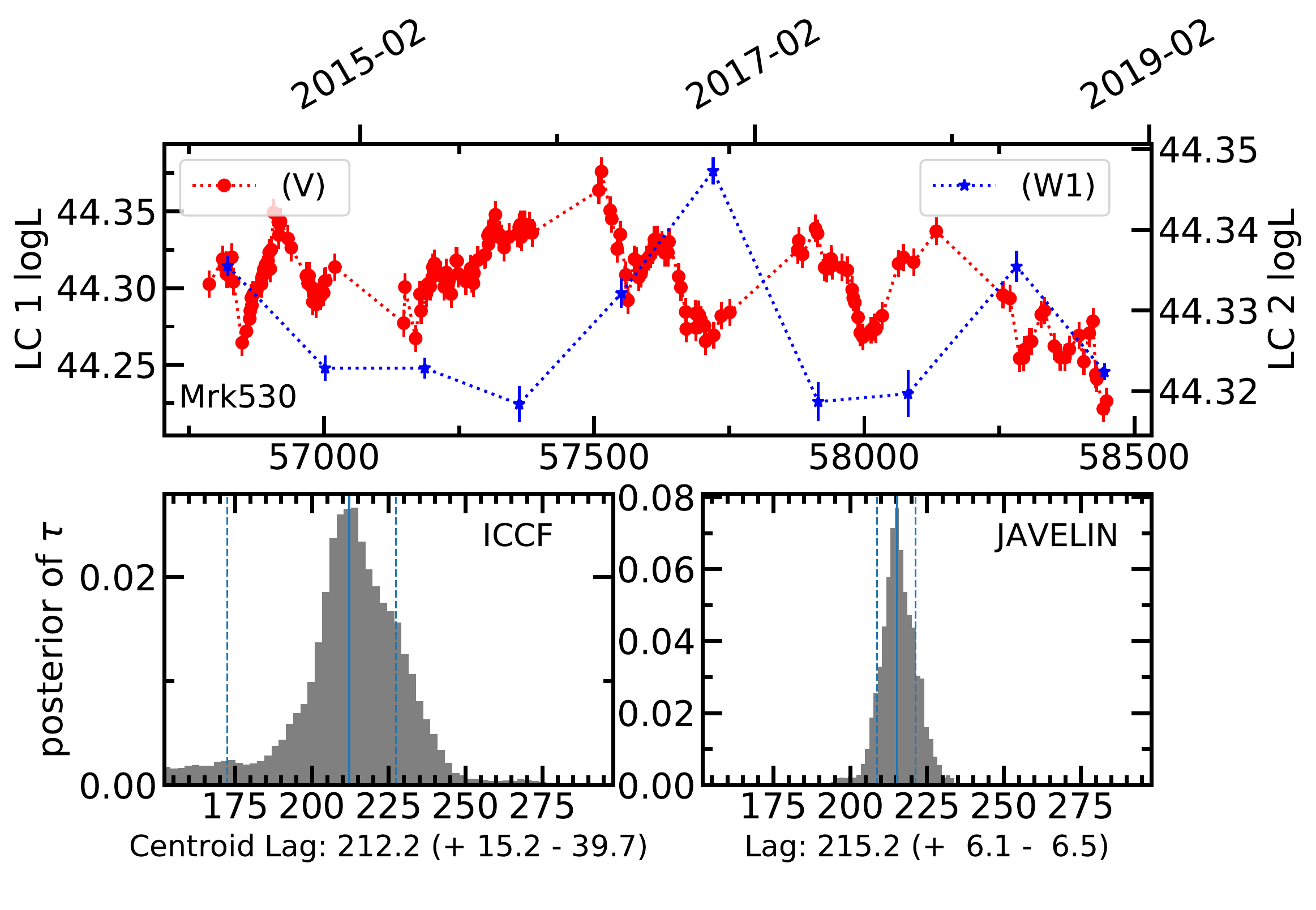}& \hspace{1em}  &
    \includegraphics[width=0.45\textwidth]{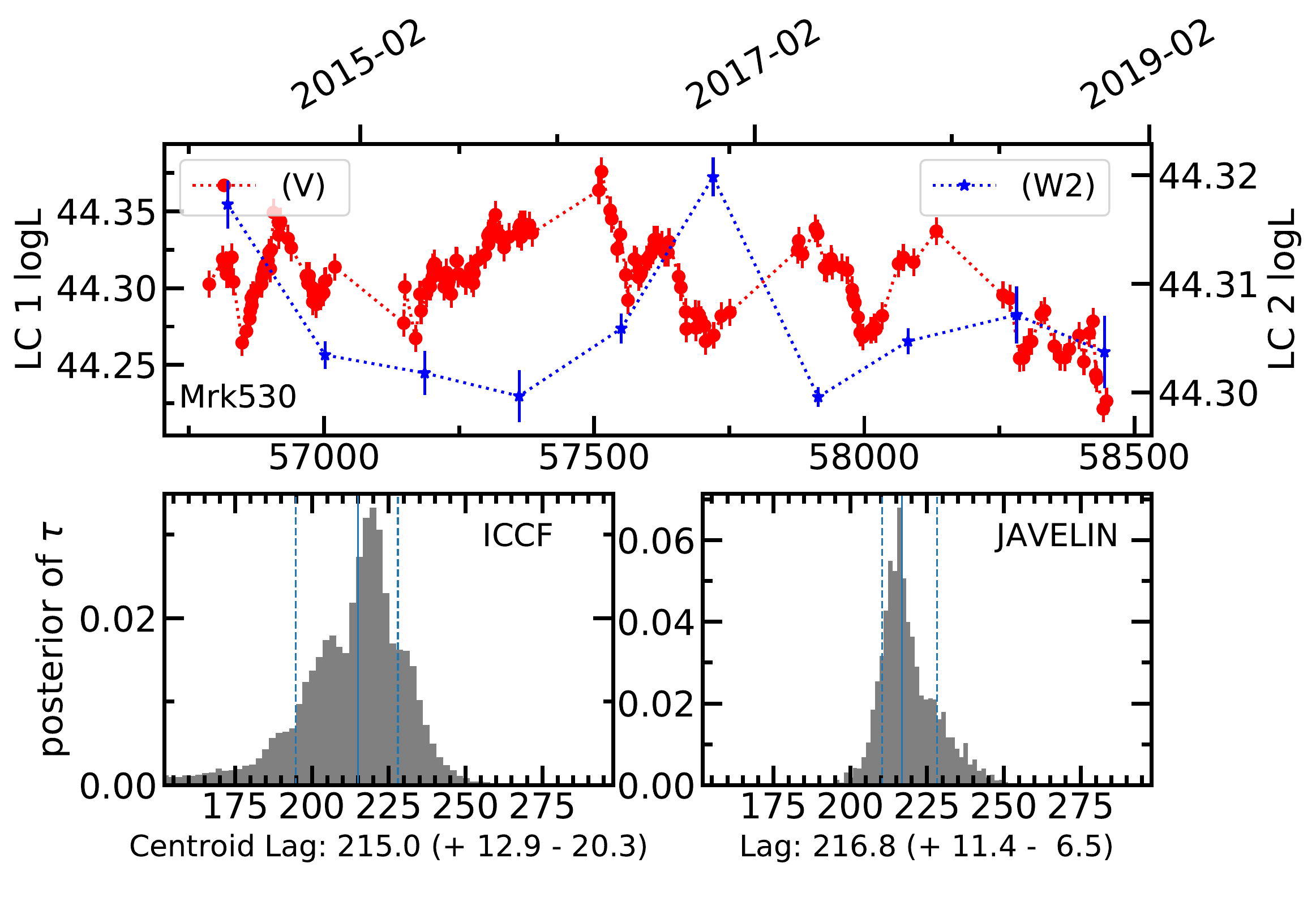} \\
    };
\end{tikzpicture}
\end{figure}  

\begin{figure}[h!]
\begin{tikzpicture}
    \matrix[matrix of nodes]{
    \includegraphics[width=0.45\textwidth]{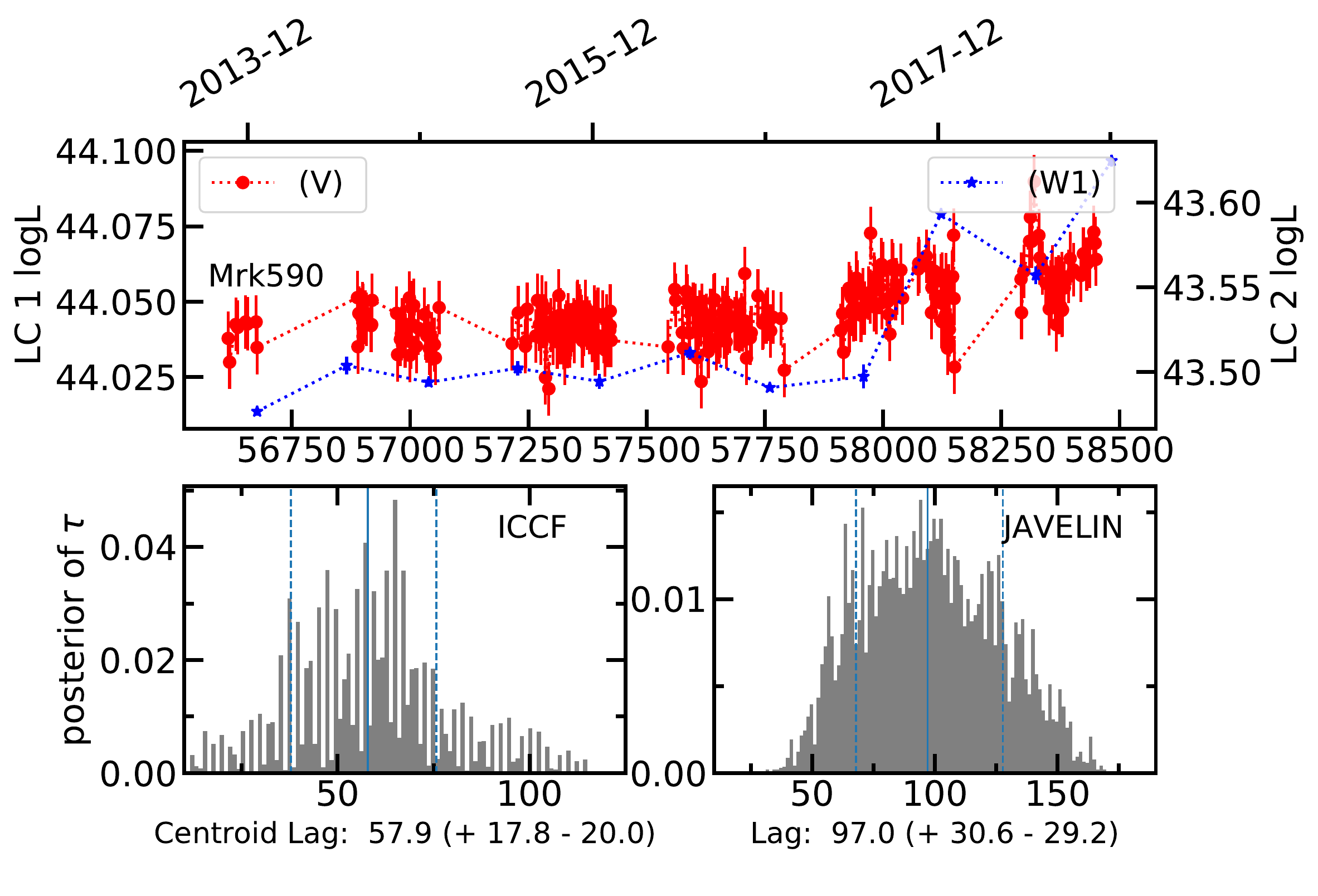}& \hspace{1em}  &
    \includegraphics[width=0.45\textwidth]{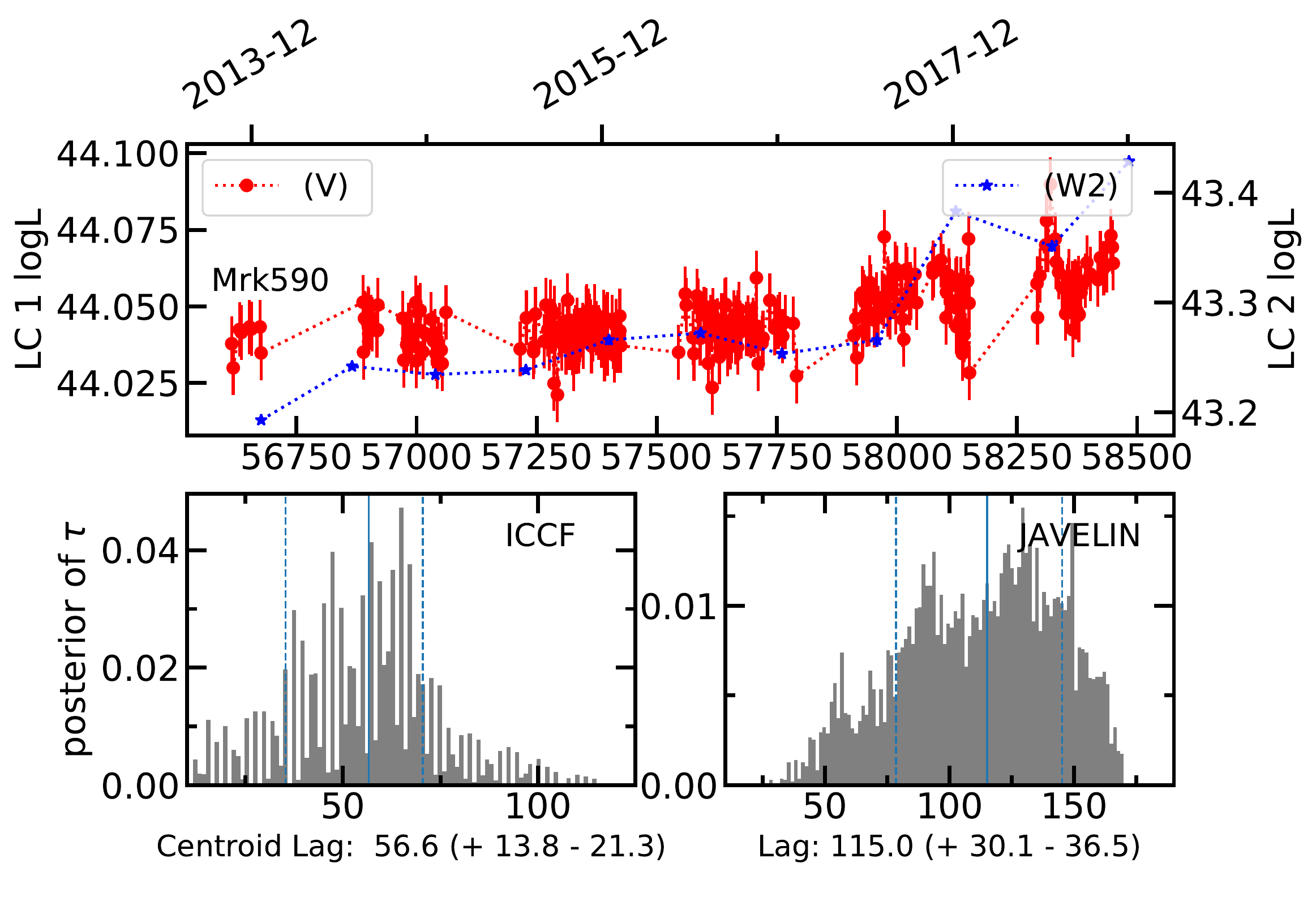} \\
    \includegraphics[width=0.45\textwidth]{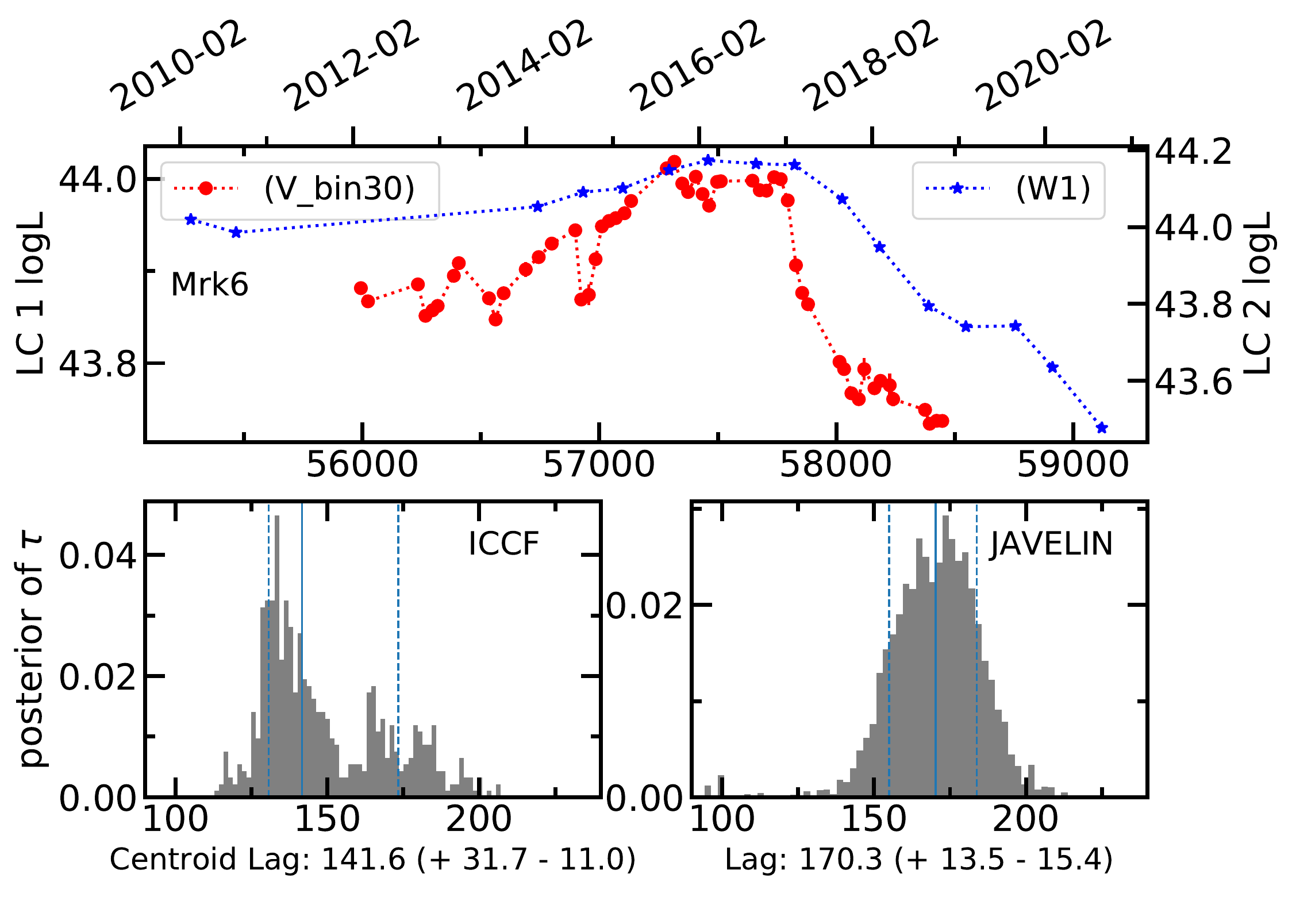}& \hspace{1em}  &
    \includegraphics[width=0.45\textwidth]{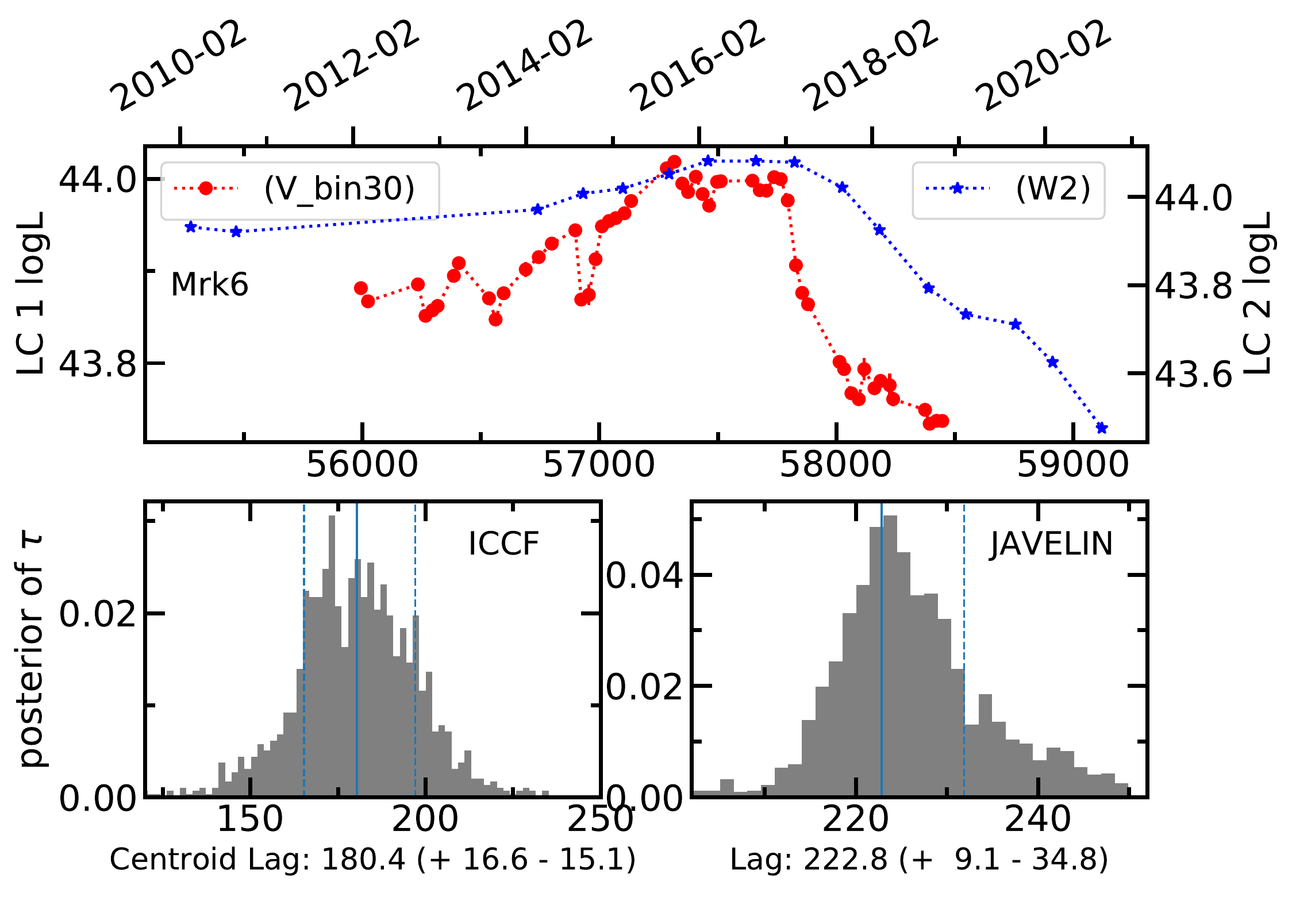} \\
    \includegraphics[width=0.45\textwidth]{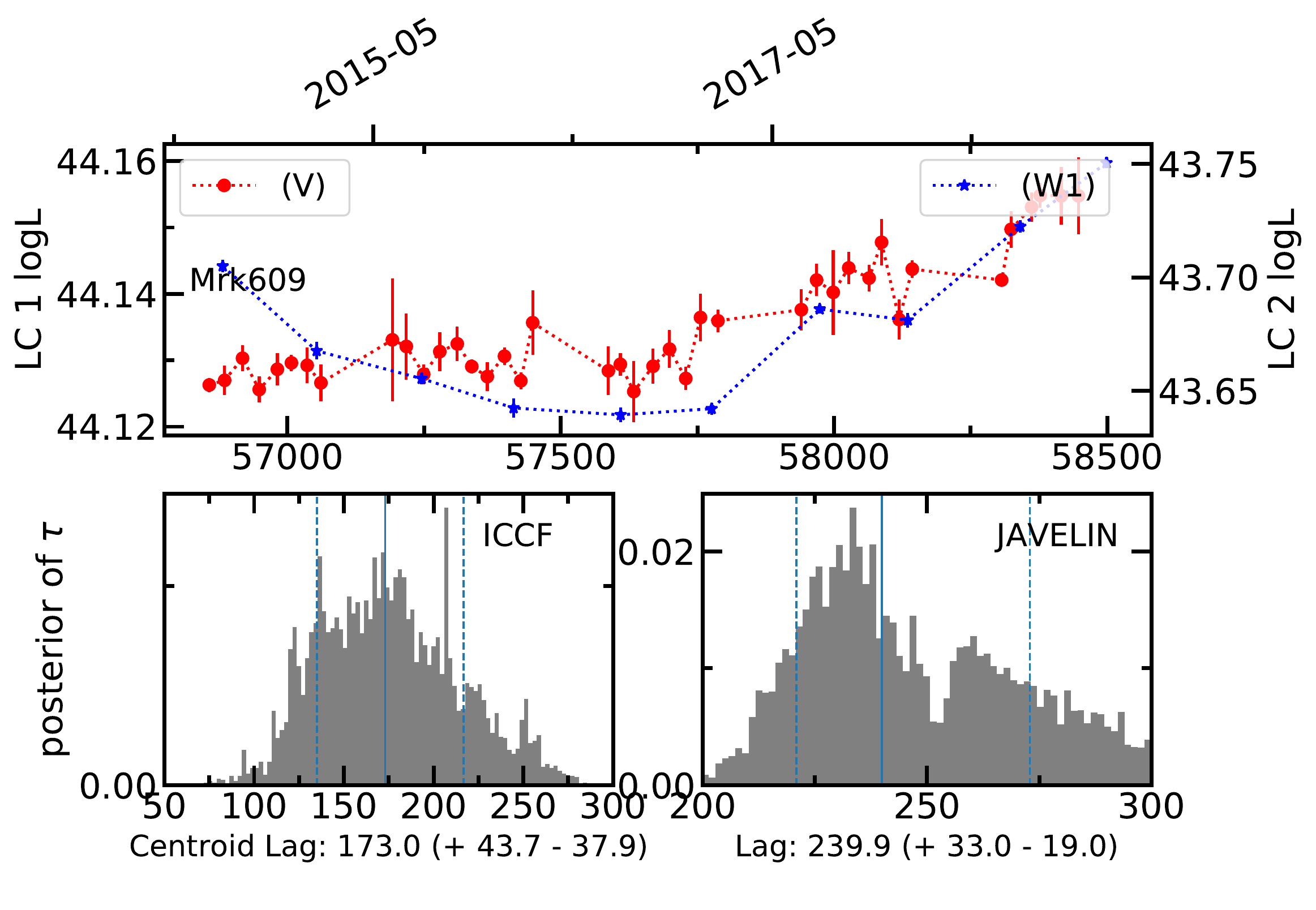}& \hspace{1em}  &
    \includegraphics[width=0.45\textwidth]{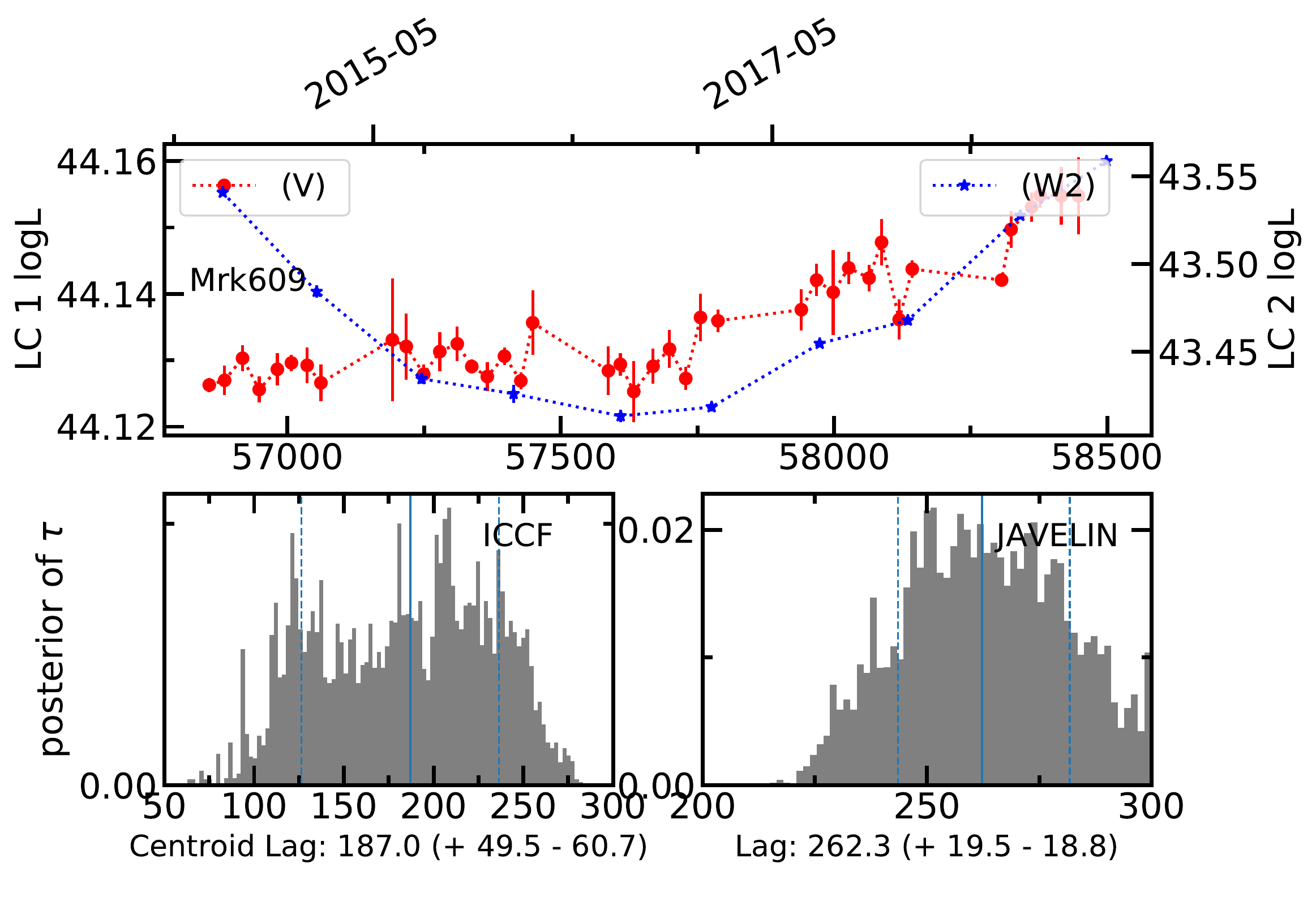} \\    
    \includegraphics[width=0.45\textwidth]{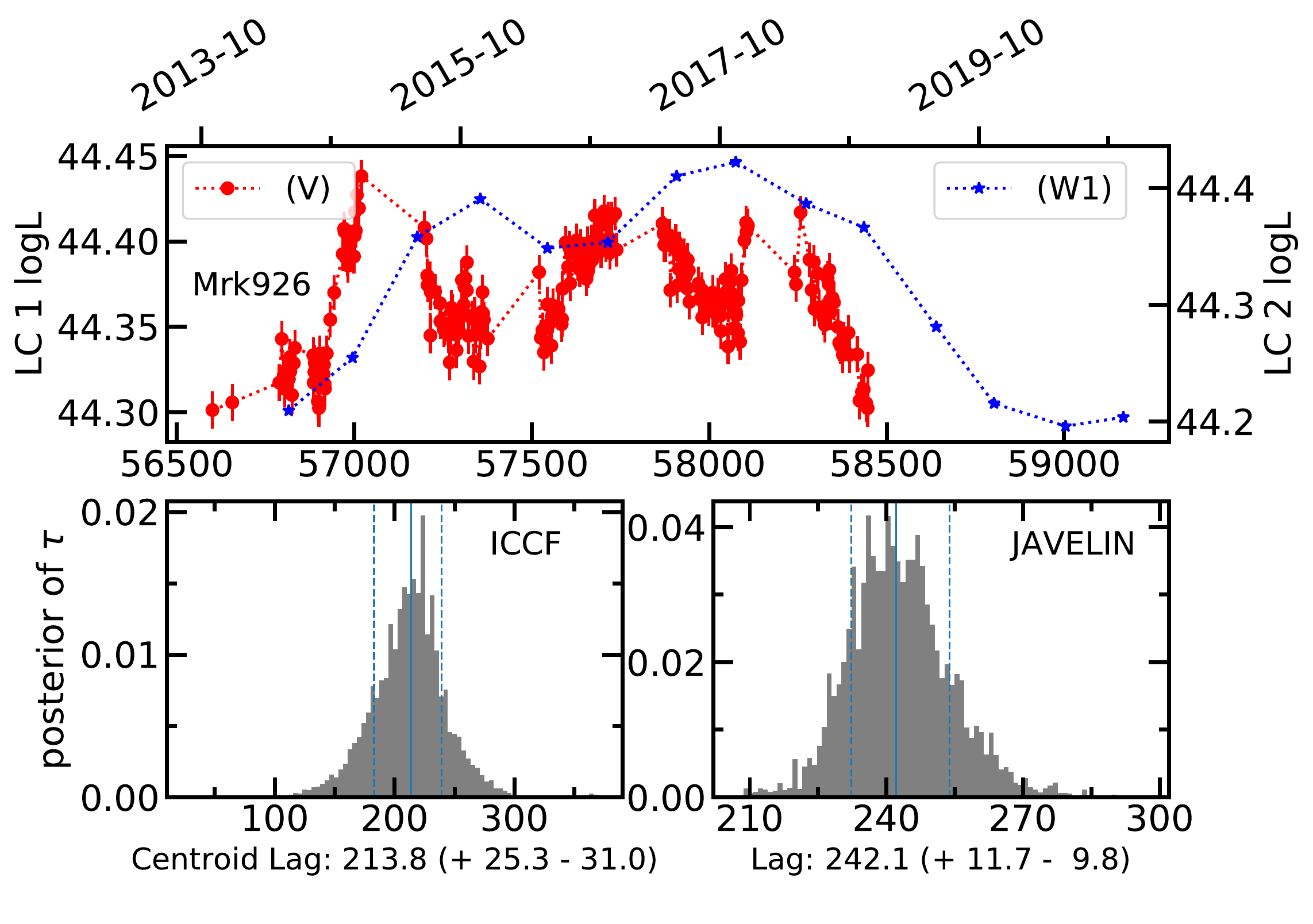}& \hspace{1em}  &
    \includegraphics[width=0.45\textwidth]{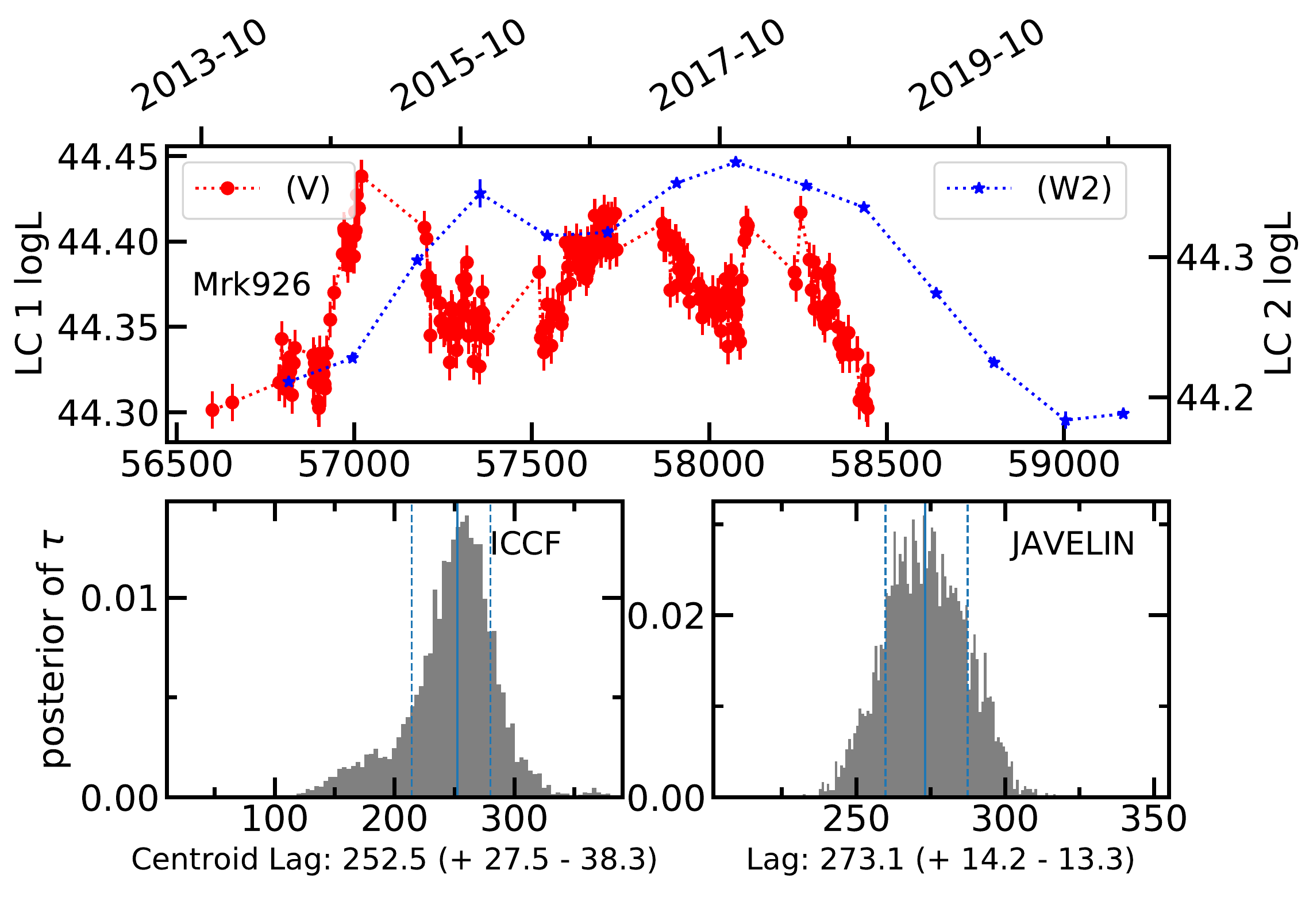} \\
    };
\end{tikzpicture}
\end{figure}  
\begin{figure}[h!]
\begin{tikzpicture}
    \matrix[matrix of nodes]{
    \includegraphics[width=0.45\textwidth]{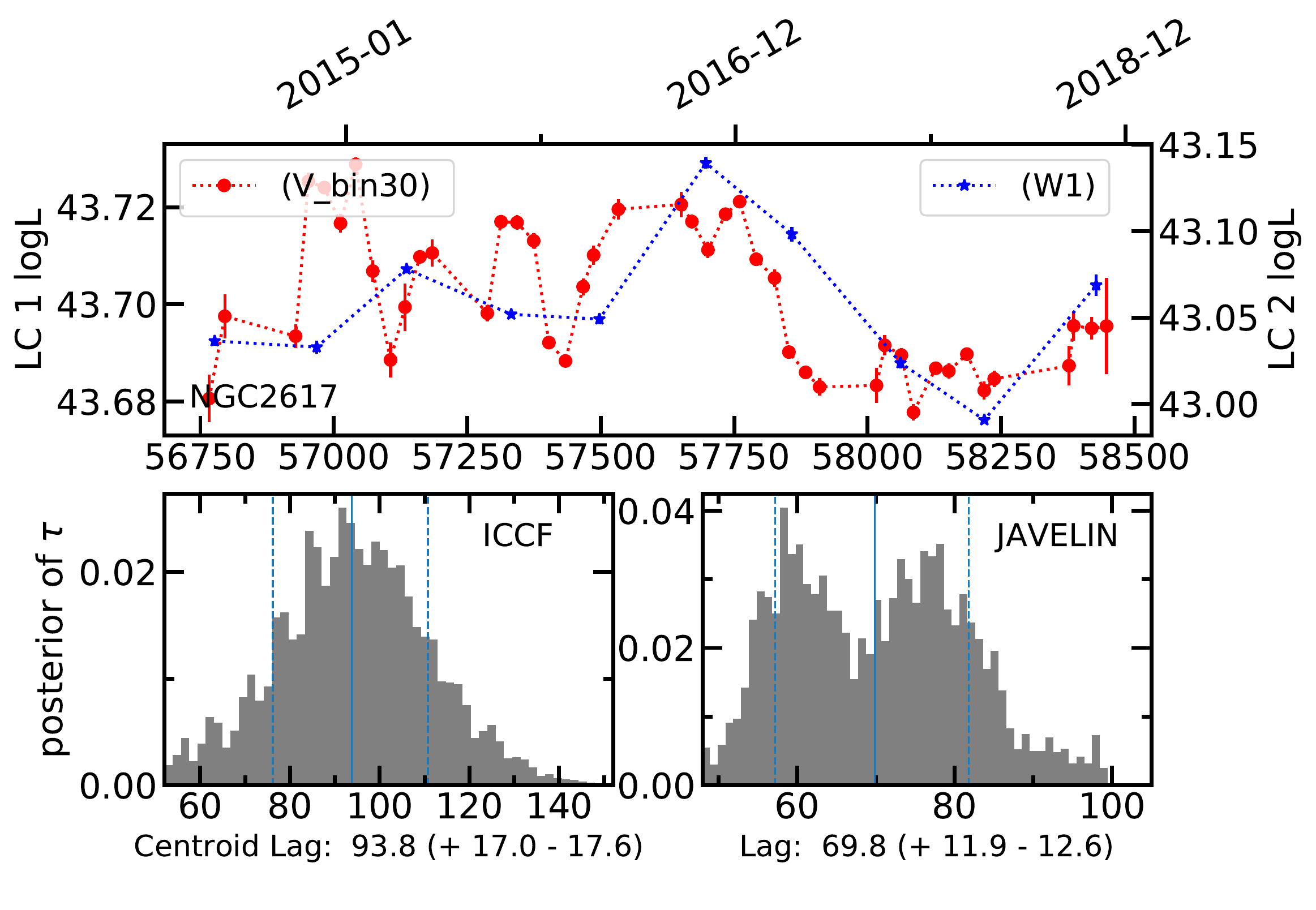}& \hspace{1em}  &
    \includegraphics[width=0.45\textwidth]{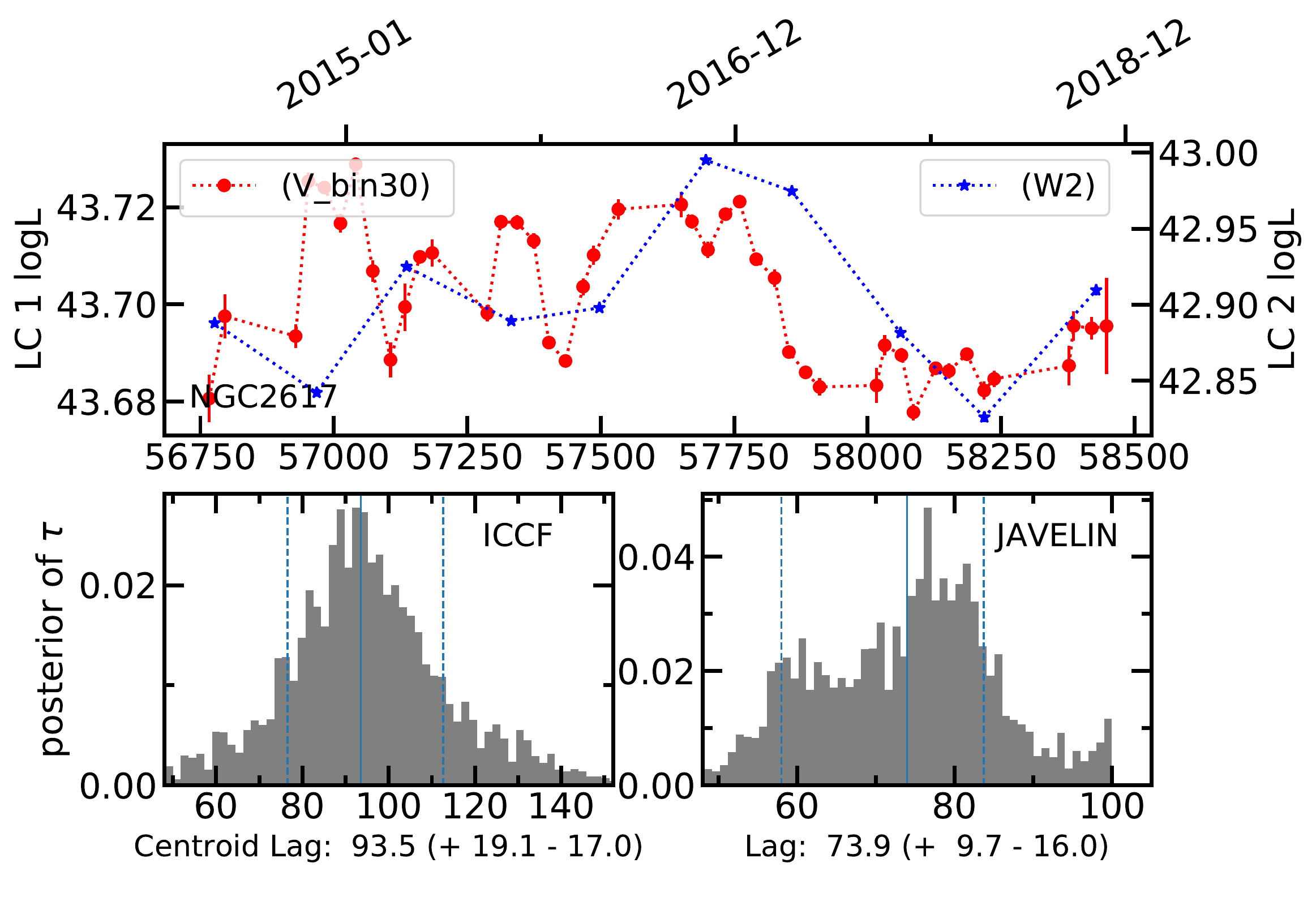} \\   
    \includegraphics[width=0.45\textwidth]{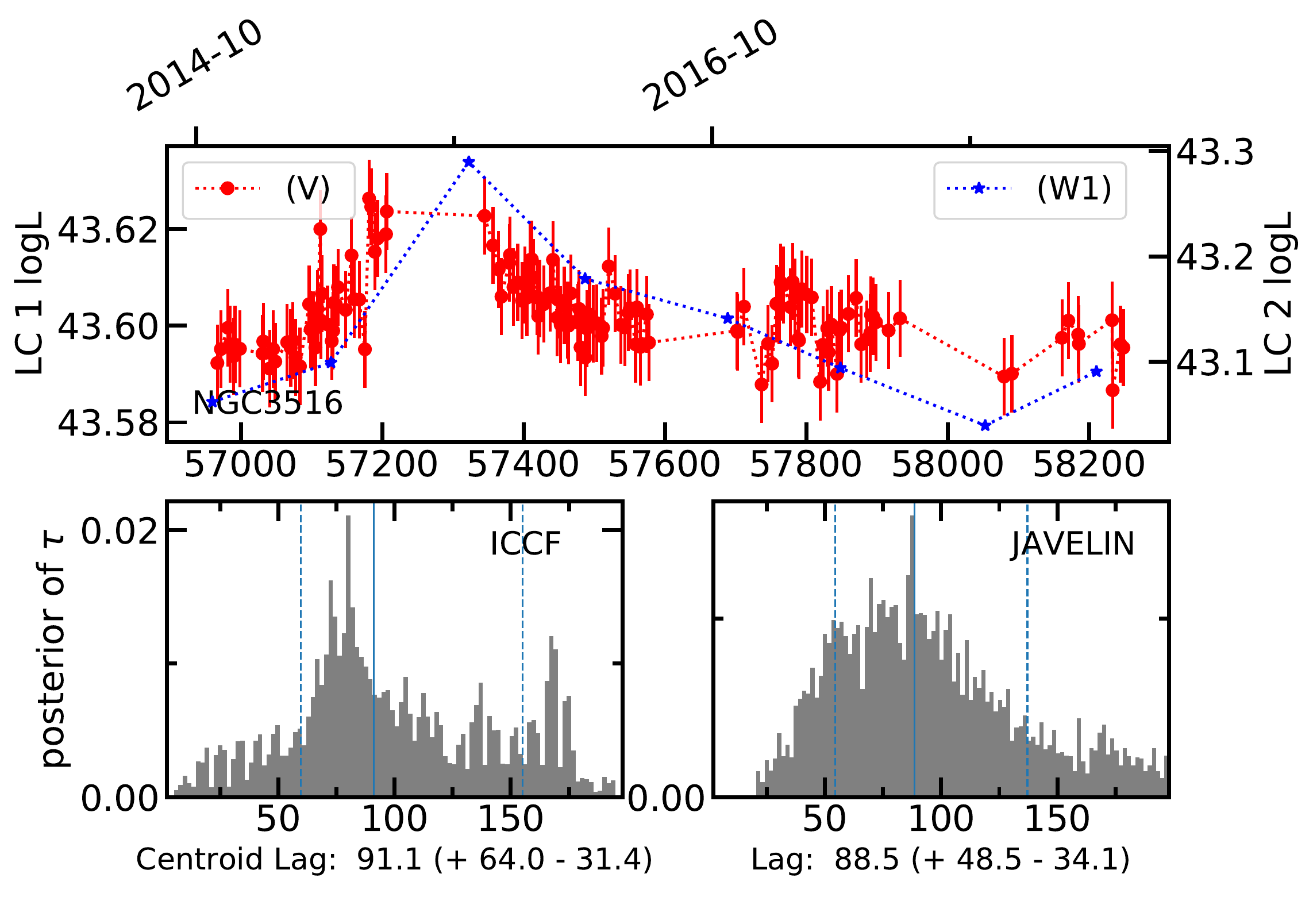}& \hspace{1em}  &
    \includegraphics[width=0.45\textwidth]{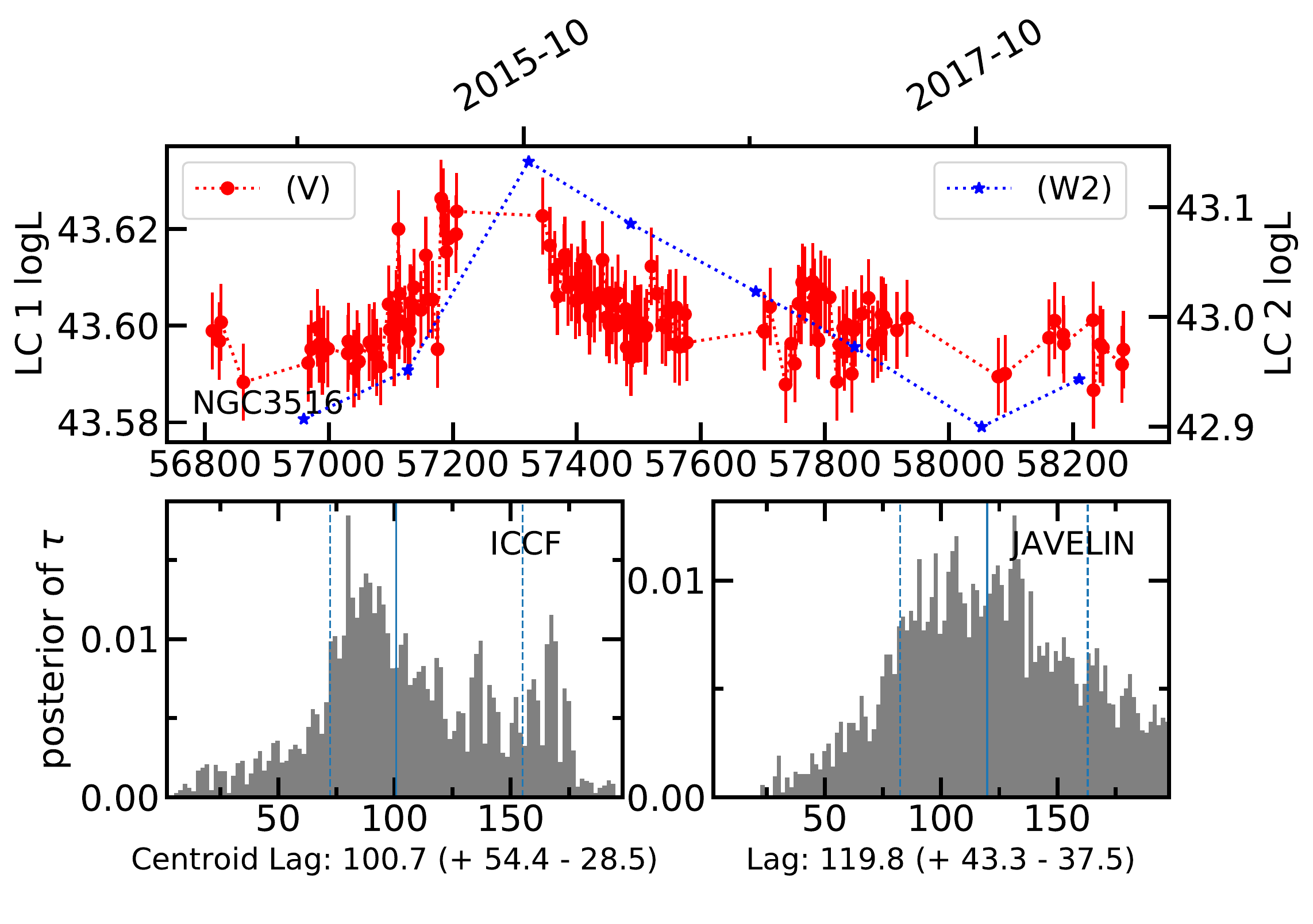} \\
    \includegraphics[width=0.45\textwidth]{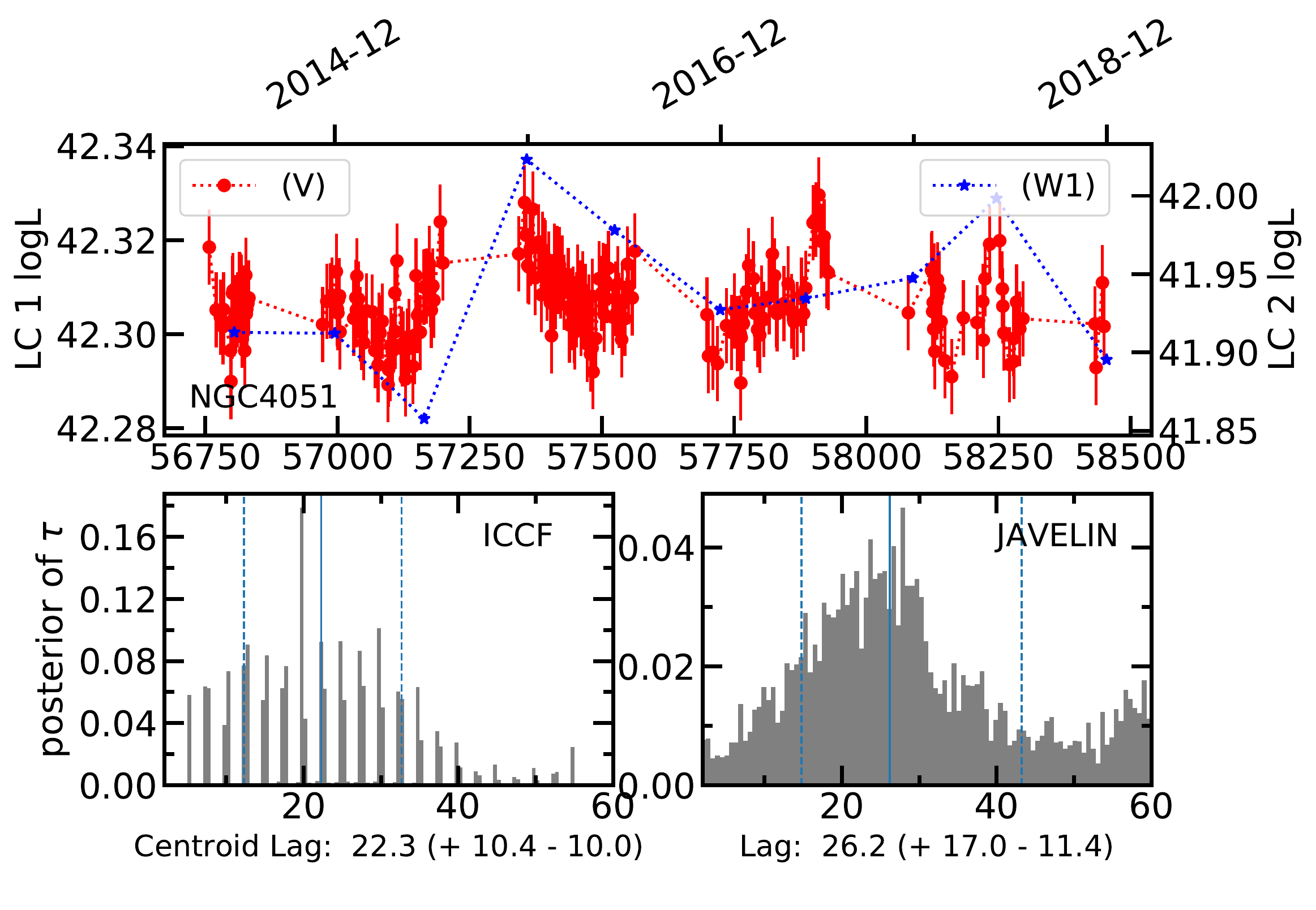}& \hspace{1em}  &
    \includegraphics[width=0.45\textwidth]{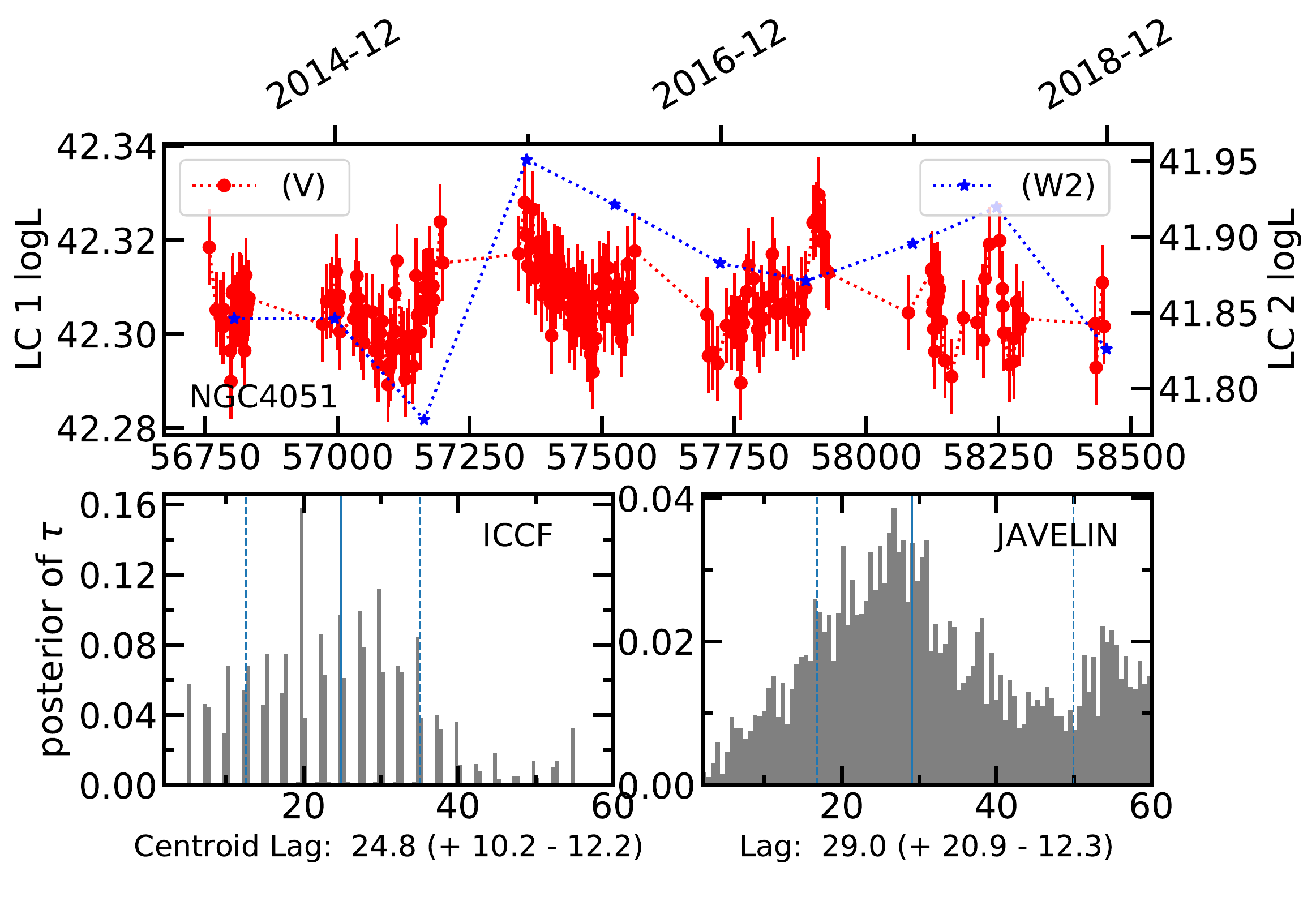} \\
    \includegraphics[width=0.45\textwidth]{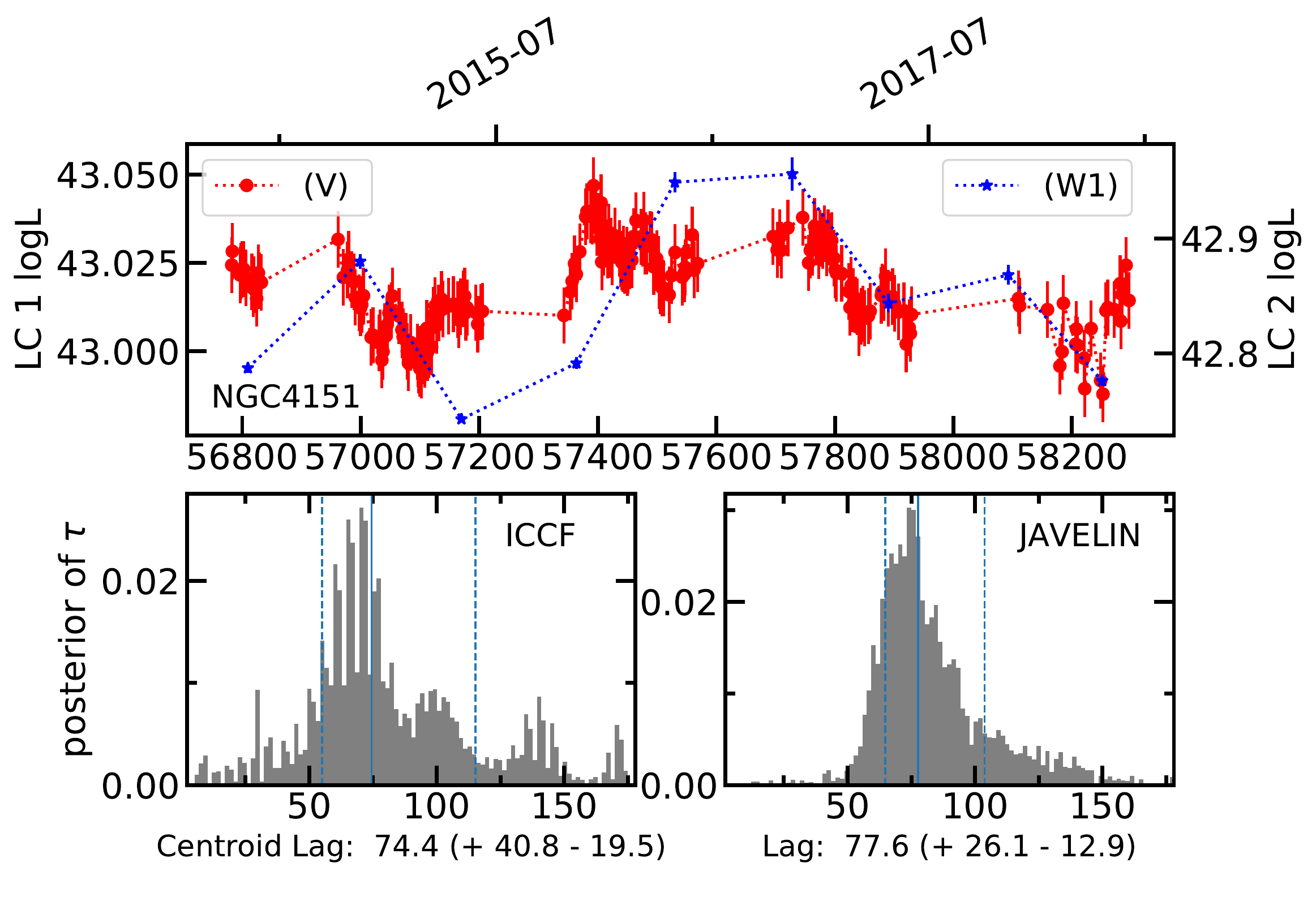}& \hspace{1em}  &
    \includegraphics[width=0.45\textwidth]{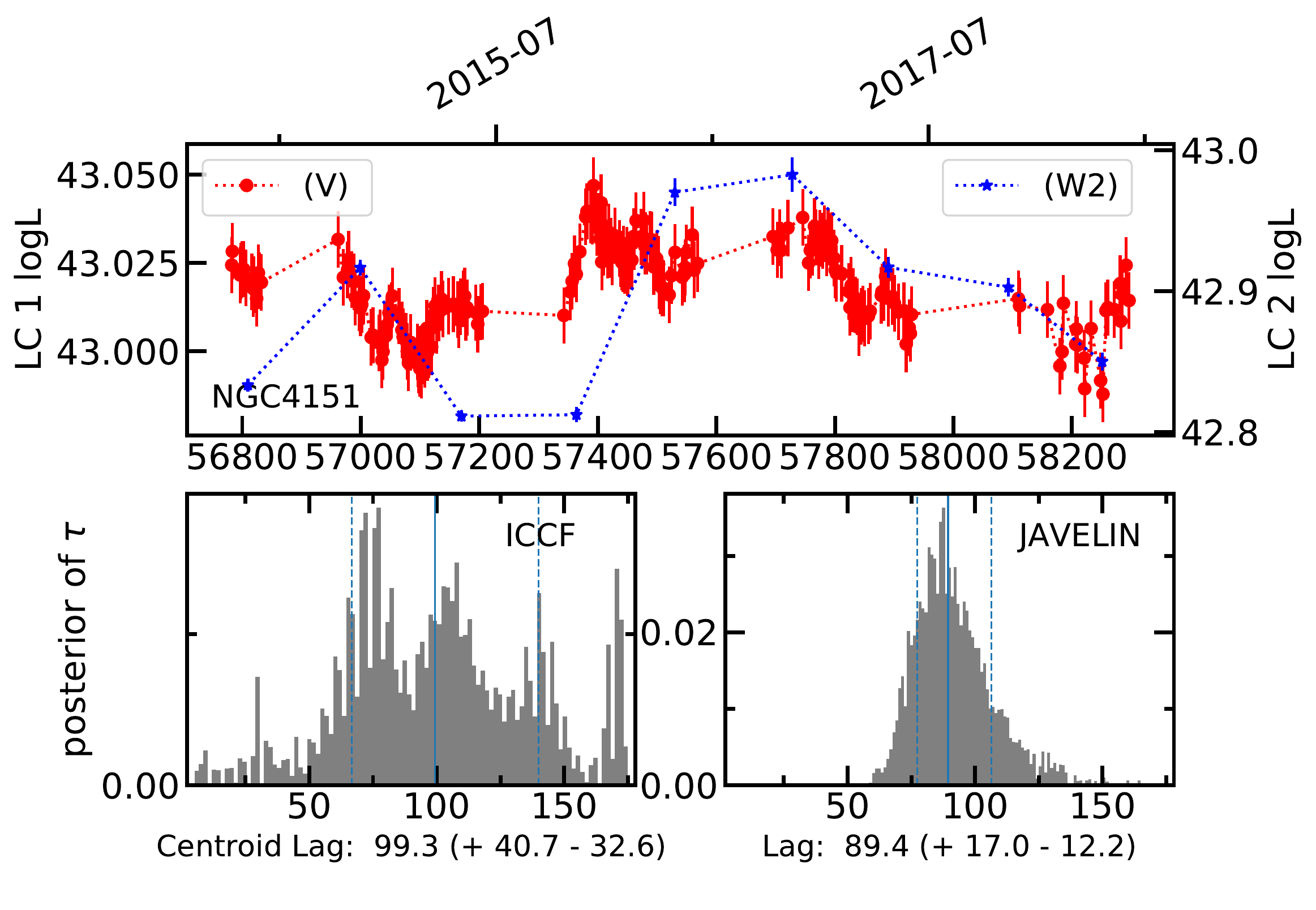} \\
    };
\end{tikzpicture}
\end{figure}

\begin{figure}[h!]
\begin{tikzpicture}
    \matrix[matrix of nodes]{
    \includegraphics[width=0.45\textwidth]{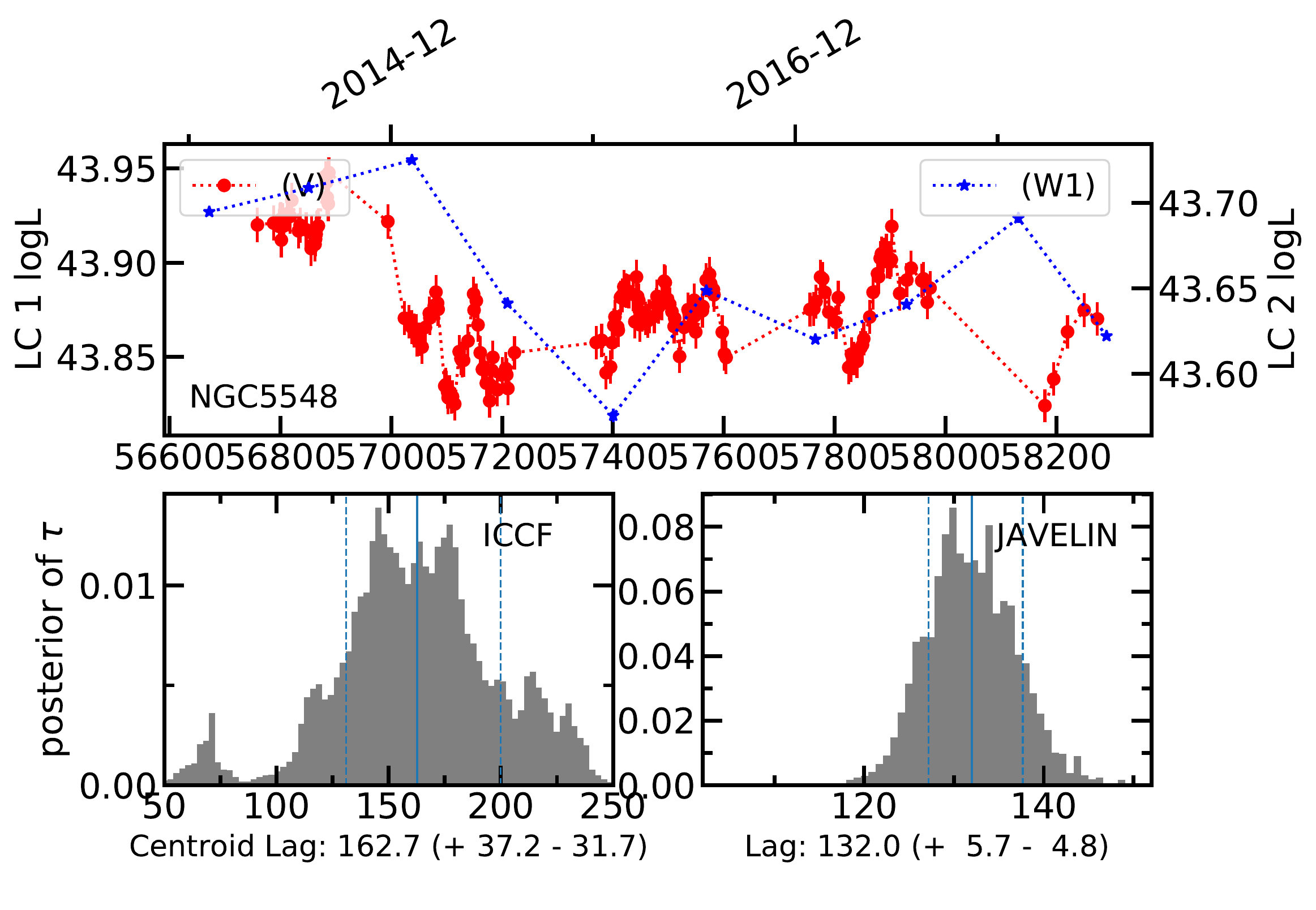}& \hspace{1em}  &
    \includegraphics[width=0.45\textwidth]{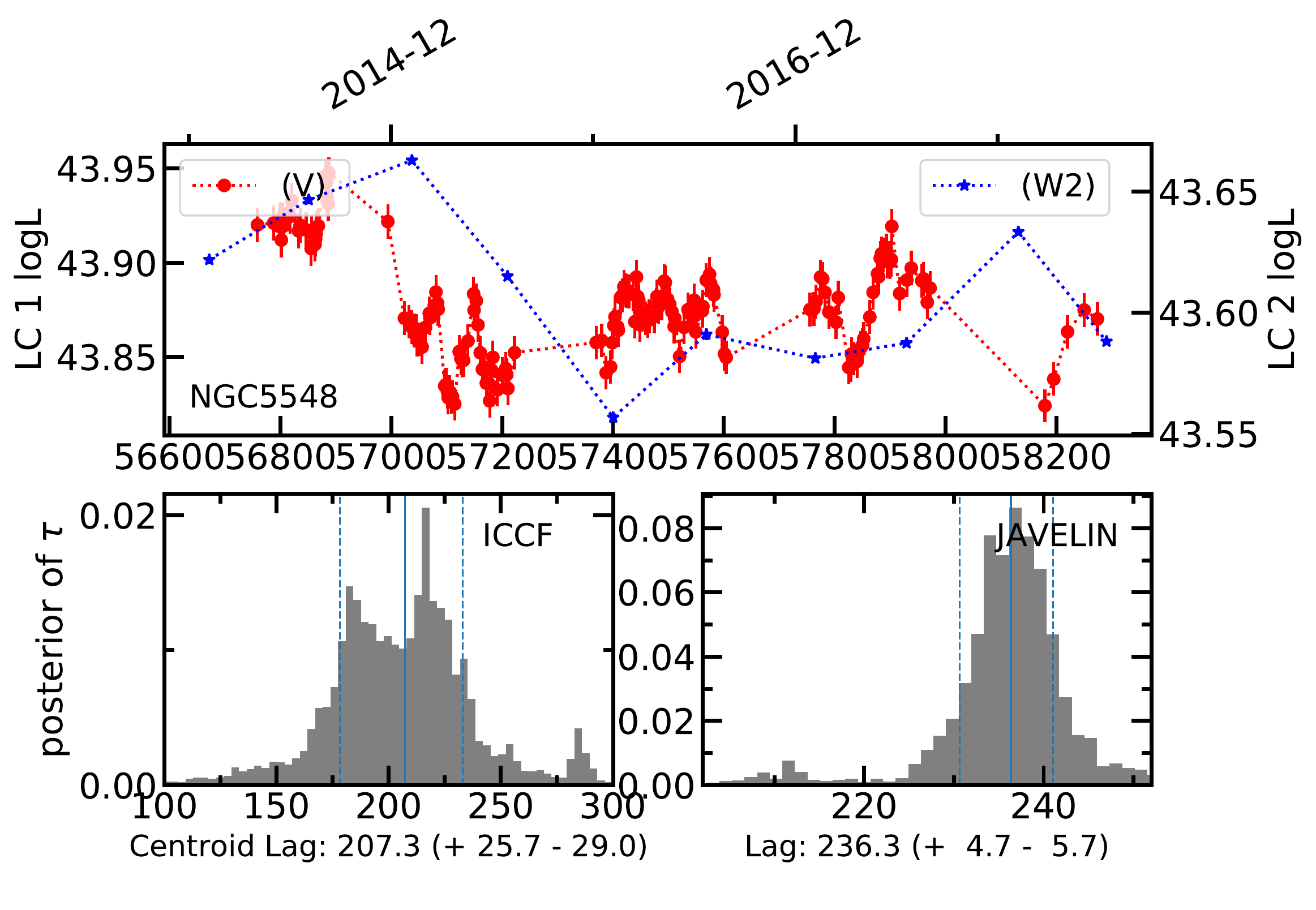} \\    
    };
\end{tikzpicture}
\caption{Mid-IR dust reverberation mapping analysis results for CLAGNs.}

\end{figure}  
\end{document}